\documentclass[12pt]{article}
\usepackage{booktabs}
\usepackage{tabulary}
\usepackage{multirow}
\usepackage{rotating}
\usepackage{epsfig}
\usepackage{psfrag}
\usepackage{latexsym}
\usepackage{indentfirst}
\usepackage{fancyhdr}
\usepackage{dsfont}
\usepackage{adjustbox}
\usepackage{amsmath}
\usepackage{amssymb}
\usepackage{amsfonts}
\usepackage{mathrsfs}
\usepackage{amsthm}
\usepackage{pifont}
\usepackage{dsfont}
\usepackage{multirow}
\usepackage{array}
\usepackage{chngpage}
\usepackage{longtable}
\usepackage{cite}
\usepackage{bbold}
\usepackage{color}
\usepackage{braket}
\usepackage{colordvi}
\usepackage{fancybox}
\usepackage[footnotesize]{caption2}
\usepackage{graphicx}
\usepackage[center,footnotesize,hang]{subfigure}
\usepackage{bm}
\usepackage{bbm}
\usepackage{url}
\usepackage[table]{xcolor}
\usepackage[colorlinks, linkcolor=red, anchorcolor=black, citecolor=green]{hyperref}
\usepackage{multirow}
\usepackage{harpoon}
\usepackage{textcomp}  
\usepackage{enumitem}
\usepackage{mathrsfs}  
\usepackage[normalem]{ulem}

\usepackage{tikz}
\usepackage{diagbox}
\usepackage{enumitem}

\newcommand{\PreserveBackslash}[1]{\let\E^c=\\#1\let\\=\E^c}
\newcolumntype{C}[1]{>{\PreserveBackslash\centering}p{#1}}
\newcolumntype{R}[1]{>{\PreserveBackslash\raggedleft}p{#1}}
\newcolumntype{L}[1]{>{\PreserveBackslash\raggedright}p{#1}}
\allowdisplaybreaks  

\newcommand{\bq}{\begin{eqnarray}}
\newcommand{\nq}{\end{eqnarray}}

\newcommand{\ignore}[1]{}

\numberwithin{equation}{section}

\begin{document}
\title{
\begin{flushright}
\hfill\mbox{\small\tt  UCI-TR-2023-06 } \\[5mm]
\begin{minipage}{0.2\linewidth}
\normalsize
\end{minipage}
\end{flushright}
{\Large \bf
Modular binary octahedral symmetry for flavor structure of Standard Model
\\[2mm]}
\date{}
\author{
Gui-Jun~Ding$^{1}$\footnote{E-mail: {\tt
dinggj@ustc.edu.cn}},  \
Xiang-Gan Liu$^{2}$\footnote{E-mail: {\tt
xianggal@uci.edu}},  \
Jun-Nan Lu$^{1}$\footnote{E-mail: {\tt
junnanlu@ustc.edu.cn}},  \
Ming-Hua Weng$^{3}$\footnote{E-mail: {\tt
mhweng@mju.edu.cn}} \
\\*[20pt]
\centerline{
\begin{minipage}{\linewidth}
\begin{center}
$^1${\it \small Department of Modern Physics, University of Science and Technology of China,\\
Hefei, Anhui 230026, China}\\[2mm]
$^2${\it \small Department of Physics and Astronomy, University of California, Irvine, CA 92697-4575, USA}\\[2mm]
$^3${\it \small College of Physics and Electronic Information Engineering, Minjiang University,\\ Fuzhou 350108, China}
\end{center}
\end{minipage}}
\\[10mm]}}

\maketitle
\thispagestyle{empty}

\begin{abstract}
We have investigated the modular binary octahedral group $2O$ as a flavor symmetry to explain the structure of Standard Model. The vector-valued modular forms in all irreducible representations of this group are constructed. We have classified all possible fermion mass models based on the modular binary octahedral group $2O$. A comprehensive numerical analysis is performed, and we present some benchmark quark/lepton mass models in good agreement with the experimental data. Notably we find a minimal modular invariant model for leptons and quarks, which is able to explain simultaneously the masses and mixing parameters of both quarks and leptons in terms of 14 real free parameters including the modulus $\tau$. The fermion mass hierarchies around the vicinity of the modular fixed points are explored.

\end{abstract}

\clearpage

{\hypersetup{linkcolor=black}
\tableofcontents
}

\section{Introduction}
Symmetries have played a fundamental role in advancing particle physics. Gauge symmetries, for example, have governed the interaction structure of the strong, weak, and electromagnetic forces.  Even in the pursuit of extending the Standard Model (SM), symmetries continue to serve as a guiding principle. For instance, the $\mathrm{U}(1)_{\mathrm{PQ}}$ symmetry has been employed to address the strong CP problem, and it is known that supersymmetry (SUSY) elegantly resolves the gauge hierarchy problem, and so on. In recent years, modular symmetry, inspired by string theory, has been used to address the flavor puzzle of SM~\cite{Feruglio:2017spp}. This can be understood as the nonlinear realization of flavor symmetries~\cite{Feruglio:2019ybq}. In SUSY theories featuring modular invariance, the Yukawa couplings in the superpotential are constrained to be modular forms, which are highly symmetric holomorphic functions of the complex modulus $\tau$.

The modular symmetry approach has been extensively explored from the bottom-up perspective. The model building in this context typically relies on the choice of finite modular groups as well as the representation and modular weight assignments of matter fields.  The family of the finite modular groups has been generalized from the original inhomogeneous finite modular groups $\Gamma_N\equiv\overline{\Gamma}/\overline{\Gamma}(N)$~\cite{Feruglio:2017spp} to homogeneous finite modular groups $\Gamma'_N\equiv\Gamma/\Gamma(N)$~\cite{Liu:2019khw}, further extended to metaplectic finite modular groups~$\widetilde{\Gamma}_N\equiv \widetilde{\Gamma}/\widetilde{\Gamma}(N)$~\cite{Liu:2020msy,Yao:2020zml} and eventually to generalized finite modular groups~\cite{Liu:2021gwa}. Based on these finite modular groups such as $\Gamma_2, \Gamma^{(\prime)}_3,\Gamma^{(\prime)}_4,\widetilde{\Gamma}_4,\Gamma^{(\prime)}_5,\widetilde{\Gamma}_5,\Gamma^{(\prime)}_6,\Gamma^{}_7$ etc, many modular invariant fermion mass models have been constructed in the literature~\cite{Kobayashi:2018vbk,Meloni:2023aru,Feruglio:2017spp,Criado:2018thu,Kobayashi:2018scp,Ding:2019zxk,Penedo:2018nmg,Novichkov:2018ovf,Novichkov:2018nkm,Ding:2019xna,Qu:2021jdy,Liu:2019khw,
Liu:2020akv,Novichkov:2020eep,Liu:2020msy,Yao:2020zml,Wang:2020lxk,Li:2021buv,Ding:2020msi}, and phenomenological implication has been discussed. See recent review~\cite{Kobayashi:2023zzc} for more references therein. Moreover the modular symmetry $\mathrm{SL}(2,\mathbb{Z})$ for single modulus has been generalized to the $\mathrm{Sp}(2g,\mathbb{Z})$ symplectic modular symmetry, which encompasses multiple moduli~\cite{Ding:2020zxw}. The generalized CP symmetry can be consistently combined with (symplectic) modular symmetries~\cite{Novichkov:2019sqv,Baur:2019kwi,Baur:2019iai,Ding:2021iqp}, the coupling constants would be enforced to be real so that the predictive power would be improved further. The modular symmetry framework can also be incorporated into various Grand Unified Theories (GUT)~\cite{deAnda:2018ecu,Kobayashi:2019rzp,Du:2020ylx,Zhao:2021jxg,Chen:2021zty,Ding:2021zbg,Ding:2021eva,Charalampous:2021gmf,Ding:2022bzs,Abe:2023dvr}. Modular symmetry can not only explain the fermion masses and flavor mixing but also address the strong CP problem~\cite{Feruglio:2023uof}. Other applications of modular symmetry to new physics beyond SM were discussed in Refs.~\cite{Nomura:2019jxj,Kobayashi:2021pav,Behera:2022wco,Kobayashi:2022jvy,Abe:2023ylh}.

It is remarkable that the modular invariant models exhibit a remarkable universal scaling behavior near the fixed points~\cite{Feruglio:2023bav,Feruglio:2023mii}, and the fermion mass hierarchies could arise from the proximity of the modulus $\tau$ to a fixed point~\cite{Okada:2020ukr,Feruglio:2021dte,Novichkov:2021evw,Petcov:2022fjf,Petcov:2023vws}. Although the superpotential is highly restricted by modular symmetry, the K\"ahler potential is not under control by modular symmetry~\cite{Chen:2019ewa}. On the other hand, the origin of these modular symmetries have also been explored within top-down approach such as string theory~\cite{Kobayashi:2018bff,Baur:2019kwi,Ohki:2020bpo,Kikuchi:2020frp,Kikuchi:2020nxn,Nilles:2021glx,Almumin:2021fbk,Kikuchi:2022lfv}. In particular, the eclectic flavor symmetry has been identified in heterotic strings on a $\mathbb{T}^2/\mathbb{Z}_k$ orbifold~\cite{Baur:2019iai,Nilles:2020nnc,Nilles:2020kgo,Nilles:2020tdp,Nilles:2020gvu,Kai:2023ivp}. The eclectic flavor group combines traditional flavor symmetry with modular symmetry in a nontrivial manner, allowing for controllability of the K\"ahler potential. Relevant model building examples can be found in Refs.~\cite{Chen:2021prl,Baur:2022hma,Ding:2023ynd,Li:2023dvm}. Moreover, the issue of modulus stabilization has been revisited to dynamically explain the vacuum expectation value (VEV) of the modulus in these modular invariant mass models~\cite{Kobayashi:2019xvz,Ishiguro:2020tmo,Ishiguro:2022pde,Novichkov:2022wvg,Leedom:2022zdm,Knapp-Perez:2023nty}.

Although modular symmetry is increasingly used in particle physics theory,  its potential for unraveling the flavor puzzle remains particularly intriguing. Therefore, one of the primary objectives is to explore different finite modular symmetries and identify realistic models for lepton/quark masses with the fewest free parameters. At present, it has been found that the minimal lepton model can describe all the lepton masses and mixing angles in terms of only six real free parameters~\cite{Ding:2022nzn}.
In this paper, we explore a novel finite modular group, the binary octahedral group denoted as $2O$, which is a double cover group of  $\Gamma_4\cong S_4$ but distinct from $\Gamma'_4\cong S'_4$\footnote{The group $2O$ is the Schur cover of $S_4$ of ``$-$'' type, and $S'_4$ is a double cover of $S_4$ in a broader sense of $S_4\cong S'_4/\mathbb{Z}^{R}_2$.}. We then provide all necessary modular form multiplets for model building. We systematically construct all possible models for lepton masses and quark masses based on this group, successfully identifying a similar lepton mass model with six real input parameters. Our minimal quark model can nicely explain ten flavor observables with nine real parameters. In addition, certain quark model with eight parameters can accommodate the quark data except that $m_{s}/m_{b}$ and $\theta_{13}^{q}$ deviate a bit from their measured values. Furthermore, we present a realistic unified model of leptons and quarks with only 14 input parameters, explaining the masses and mixing parameters of both quark and lepton sectors with a common modulus $\tau$. Finally, we also explore the mass hierarchy patterns that arise naturally in the vicinity of modular fixed points for this group.

The rest of this paper is organized as follows: In section~\ref{sec:VVMF}, we review the modular invariant SUSY theory and vector-valued modular forms (VVMFs). Section~\ref{sec:fermion_models} presents a systematic classification of fermion mass models based on binary octahedral group $2O$. A complete numerical analysis of these lepton and quark models is presented in section~\ref{sec:numerical_analysis}. In section~\ref{sec:lepton_models} and~\ref{sec:quark_models} we showcase some typical minimal lepton models and quark models, respectively. Furthermore, a minimal unified model of leptons and quarks is presented in section~\ref{sec:unified_model}. Finally, we summarize our results and draw the conclusions in section~\ref{sec:conclusion}. The Appendix~\ref{app:123dVVMF} gives the general results for $n$-dimensional ($n$-d) VVMF with $n=1,2,3$. The detailed construction of the 4-d VVMF for $2O$ can be found in Appendix~\ref{app:4dVVMF}. The group theory of binary octahedral group $2O$ is presented in Appendix~\ref{app:2O_group}. Finally, we give in Appendix~\ref{app:fixed-point} all possible mass hierarchy patterns near critical points for $2O$.

\section{\label{sec:VVMF} Modular invariance and vector-valued modular forms }
We consider the $\mathcal{N}=1$ global supersymmetry theory, which includes the matter supermultiplets denoted by $\Phi_I$ and the modulus superfield represented by $\tau$. The most general action can be written as
\begin{equation}
\mathcal{S} = \int d^4 x d^2\theta d^2\bar \theta\, \mathcal{K}(\Phi_I,\bar{\Phi}_I; \tau,\bar{\tau})+\int d^4 x d^2\theta\, \mathcal{W}(\Phi_I,\tau)+\mathrm{h.c.}\,,
\end{equation}
where $\mathcal{K}(\Phi_I,\bar{\Phi}_I; \tau,\bar{\tau})$ is the K\"ahler potential and $\mathcal{W}(\Phi_I,\tau)$ is the superpotential. In order to maintain modular invariance within the theory, it is required that the action $\mathcal{S}$ remains invariant under the following modular transformations of superfields:
\begin{equation}
\label{eq:modular_transform}\left\{\begin{array}{l}
\tau\to \gamma(\tau) = \dfrac{a \tau+b}{c \tau+d}\,,\\ \\
\Phi_I\to (c\tau+d)^{-k_I}\rho_I(\gamma)\Phi_I\,,
\end{array} \right.\quad \text{with}\quad \gamma=\begin{pmatrix}
a  &  b  \\
c  &  d
\end{pmatrix}\in\Gamma\,,
\end{equation}
where the infinite discrete group $\Gamma$ is the full modular group $\mathrm{SL}(2,\mathbb{Z})$, which can be generated by two specific generators, namely $S$ and $T$:
\begin{equation}
S=\begin{pmatrix}
0 &~ 1 \\ -1 &~ 0
\end{pmatrix}\,, \qquad
T=\begin{pmatrix}
1 &~ 1 \\ 0 &~ 1
\end{pmatrix}\,.
\end{equation}
They obeying the following multiplication rule:
\begin{equation}
\label{eq:SL2Zrelation}
S^4=(ST)^3=1\,,\quad S^2 T=T S^2\,.
\end{equation}
The parameter $-k_I\in \mathbb{Z}$ in Eq.~\eqref{eq:modular_transform} represents the modular weight of the matter field $\Phi_I$. Moreover, $\rho_I$ denotes the unitary irreducible representation (irrep) of $\mathrm{SL}(2,\mathbb{Z})$ with finite image. In other words, $\rho_I$ corresponds to the unitary irrep of a finite modular group $\mathcal{G}_f$, which arises as the quotient between $\Gamma$ and a certain normal subgroup~\cite{Liu:2021gwa}.

The minimal K\"ahler potential that satisfies the requirement of modular invariance can be expressed in the following form:
\begin{equation}
\mathcal{K}(\Phi_I,\bar{\Phi}_I; \tau,\bar{\tau}) =-h \log(-i\tau+i\bar\tau)+ \sum_I (-i\tau+i\bar\tau)^{-k_I} |\Phi_I|^2\,,
\end{equation}
where the constant $h>0$. On the other hand, the modular invariant superpotential $\mathcal{W}(\Phi_I,\tau)$ can be expanded in power series of the supermultiplets $\Phi_I$:
\begin{equation}
\mathcal{W}(\Phi_I,\tau) =\sum_n Y_{I_1...I_n}(\tau)~ \Phi_{I_1}... \Phi_{I_n}\,,
\end{equation}
where Yukawa couplings $Y_{I_1...I_n}(\tau)$ are the modular form multiplets of weight $k_Y$ in representation $\rho_Y$, in accordance with modular invariance. Hence, they satisfy the following equation:
\begin{equation}
\label{eq:modularity}
Y_{I_1...I_n}(\tau)\to Y_{I_1...I_n}(\gamma \tau)=(c\tau+d)^{k_Y}\rho_{Y}(\gamma)Y_{I_1...I_n}(\tau)\,,
\end{equation}
with
\begin{equation}
k_Y=k_{I_1}+...+k_{I_n}\,,\qquad \rho_{Y}\otimes\rho_{I_1}\otimes...\otimes\rho_{I_n} \supset \mathbf{1}\,.
\end{equation}
In a bottom-up model, the modular weights $-k_I$ and representations $\rho_I$ are freely assigned, uniquely determining all possible Yukawa couplings $Y_{I_1...I_n}(\tau)$ in the model. It is important to note that since modular form spaces are typically finite-dimensional, the model contains only a finite number of possible Yukawa couplings. As a result, the model exhibits significant predictive power. For this reason, we apply modular symmetry to a specific minimal supersymmetric standard model (MSSM) to gain insights into fermion masses and flavor mixing.

Furthermore, it is worth emphasizing that the scalar component of the modulus field $\tau$ corresponds geometrically to the complex structure modulus of a two-dimensional torus $\mathbb{T}^2$, which characterizes the torus shape. Consequently, the moduli space spanned by the modulus is known as the upper-half plane $\mathcal{H}=\{\tau\in\mathbb{C}~|~\text{Im} \tau >0\}$,  Additionally, due to the modular symmetry, the inequivalent moduli vacua are described by the fundamental domain $\mathcal{F}\equiv \mathcal{H}/\Gamma$:
\begin{equation}
\mathcal{F}= \left\{\tau\in\mathcal{H} ~\Big|~ |\tau|\geq 1\,,~ -1/2\leq\text{Re}{\tau}\leq 0\right\}\cup  \left\{\tau\in\mathcal{H} ~\Big|~ |\tau|>1 \,,~ 0 <\text{Re}{\tau}< 1/2\right\}\,.
\end{equation}

As the key role of the modular invariant theory, the modular form multiplets (also known as the VVMFs in modern mathematics) possess an elegant mathematical theory, which we will briefly introduce below.

A vector-valued holomorphic function in $d$ dimension is considered as a VVMF if it satisfies Eq.~\eqref{eq:modularity}.  All VVMFs in the irrep $\rho$ constitute a free module denoted by $\mathcal{M}(\rho)$ over the ring $\mathcal{M}(\mathbf{1})=\mathbb{C}[E_4,E_6]$~\footnote{The VVMFs of $\mathrm{SL}(2,\mathbb{Z})$ in one-dimensional irreps include Eisenstein series  $E_4(\tau)$, $E_6(\tau)$ and eta products~\cite{cohen2017modular}:
\begin{align}
\label{eq:E4-E6-eta}
\nonumber
E_4(\tau)=1+240\sum_{n=1}^{\infty}\sigma_3(n)q^n \,,\quad
E_6(\tau)=1-504\sum_{n=1}^{\infty}\sigma_5(n)q^n\,,\quad \eta^{2p}(\tau)=q^{p/12}\prod_{n=1}^\infty \left(1-q^n \right)^{2p}\,,
\end{align}
where $q\equiv e^{2 \pi i\tau}$, $\sigma_k(n) = \sum_{d|n} d^k$ is the sum of the $k$-th power of the divisors of $n$,  $p\in \mathbb{N}^+$ and $\eta(\tau)$ is the Dedekind-eta function.}. The rank of this module is equal to the dimension of the irrep $\rho$ which is denoted as $d\equiv \dim\rho$. Typically, a basis for this module $\mathcal{M}(\rho)$ can be obtained by applying modular differential operators $D^n_k$ to the VVMFs of minimal weight. The modular differential operators $D^n_k$ are defined as follows:
\begin{equation}
D^{n}_k\equiv D_{k+2(n-1)}\circ D_{k+2(n-2)}\circ\dots \circ D_k\,,
\end{equation}
where
\begin{equation}
D_k\equiv \frac{1}{2\pi i} \frac{d}{d\tau}- \frac{kE_2(\tau)}{12}\,,\quad k\in\mathbb{N}^+\,,
\end{equation}
and $E_2(\tau)$ is the well-known quasi-modular Eisenstein series~\cite{cohen2017modular}
\begin{equation}
\label{eq:E2}
E_2(\tau)=1-24\sum_{n=1}^{\infty}\sigma_1(n)q^n \,.
\end{equation}
The action of $D^n_k$ on a VVMF $Y(\tau)$ of weight $k$ yields a VVMF $D^n_kY(\tau)$ in the same representation but of higher weights $k+2n$. If the module $\mathcal{M}(\rho)$ is cyclic, the basis vectors of $\mathcal{M}(\rho)$ could be chosen as  $\{Y^{(k_0)},D_{k_0} Y^{(k_0)},\dots,D_{k_0}^{d-1} Y^{(k_0)}\}$, where $Y^{(k_0)}$ is the VVMF of minimal weight $k_0$. The modular form multiplets $D_{k_0}^{d} Y^{(k_0)}$  of higher weight $k_0+2d$ can always be expressed as linear combinations of these bases over the ring $\mathcal{M}(\mathbf{1})=\mathbb{C}[E_4,E_6]$:
\begin{equation}
\label{eq:MLDE}
(D^d_{k_0}+M_4D_{k_0}^{d-2}+\dots+M_{2(d-1)}D_{k_0}+M_{2d})Y^{(k_0)}=0\,,
\end{equation}
where  $M_k\in \mathbb{C}[E_4,E_6]$ is the scalar modular form of weight $k$. From another point of view, the above equation is actually the modular linear differential equation (MLDE) satisfied by the VVMF of minimal weight, and its solution provides us with the specific form of $Y^{(k_0)}$. Notably, $\mathcal{M}(\rho)$ is fully determined by the irrep $\rho$. If $\rho$ is a 1-d, 2-d, or 3-d irrep of $\mathrm{SL}(2,\mathbb{Z})$, the solutions $Y^{(k_0)}$ can be expressed either as eta products or generalized hypergeometric series.
A comprehensive overview of VVMFs theory and related results can be found in Ref.~\cite{Liu:2021gwa}.

The finite modular groups can generally be expressed as the quotient of $\mathrm{SL}(2,\mathbb{Z})$ over its normal subgroup with finite index, and they include other intriguing groups besides the known $\Gamma_N$ and $\Gamma'_N$. All the finite modular group up to order 72 have been listed in~\cite{Liu:2021gwa}. The homogeneous finite modular groups $T'$ and $A'_5$ are the binary tetrahedral group and the binary icosahedral group respectively, they have been investigated in the context of modular flavor symmetry~\cite{Liu:2019khw,Lu:2019vgm,Ding:2022aoe,Ding:2023ynd,Yao:2020zml,Wang:2020lxk,Behera:2021eut}.
This paper will focus on the binary octahedral group $2O$. In the subsequent sections, we construct all the VVMFs in irreps of binary octahedral group $2O$. The detailed group theory for $2O$ is provided in Appendix~\ref{app:2O_group}.
\subsection{\label{sec:VVMF-modules-2O}VVMFs in representations of binary octahedral group}
Since all the VVMFs (denoted by $\mathcal{M}(2O)$) in the representations of the binary octahedral group $2O$ are simply the set of each irreducible VVMF module, the whole module $\mathcal{M}(2O)$ can be organized as the direct sum of eight irreducible VVMF modules:
\begin{equation}
\mathcal{M}(2O)=\mathcal{M}(\mathbf{1})\oplus\mathcal{M}(\mathbf{1'})\oplus\mathcal{M}(\mathbf{2})\oplus \mathcal{M}(\widehat{\mathbf{2}}) \oplus \mathcal{M}(\widehat{\mathbf{2}}') \oplus \mathcal{M}(\mathbf{3}) \oplus \mathcal{M}(\mathbf{3'})\oplus \mathcal{M}(\widehat{\mathbf{4}})\,.
\end{equation}
From the general theory of VVMFs~\cite{Liu:2021gwa}, we can determine the basis of each module and identify the VVMF of minimal weight within them. Specifically, each irreducible VVMF module can be generated by the following modular multiplets over the ring $\mathcal{M}(\mathbf{1})=\mathbb{C}[E_4,E_6]$:
\begin{align}
\nonumber\mathcal{M}(\mathbf{1})&=\langle 1 \rangle\,,\\
\nonumber\mathcal{M}(\mathbf{1'})&=\langle Y^{(6)}_{\mathbf{1'}} \rangle\,,\\
\nonumber\mathcal{M}(\mathbf{2})&=\langle Y^{(2)}_{\mathbf{2}},~~D_2 Y^{(2)}_{\mathbf{2}} \rangle\,,\\
\nonumber\mathcal{M}(\widehat{\mathbf{2}})&=\langle Y^{(5)}_{\widehat{\mathbf{2}}}, ~~D_5 Y^{(5)}_{\widehat{\mathbf{2}}}\rangle\,,\\
\nonumber\mathcal{M}(\widehat{\mathbf{2}}')&=\langle Y^{(5)}_{\widehat{\mathbf{2}}'}, ~~D_5Y^{(5)}_{\widehat{\mathbf{2}}'} \rangle\,,\\
\nonumber\mathcal{M}(\mathbf{3})&=\langle Y^{(2)}_{\mathbf{3}}, ~~D_2Y^{(2)}_{\mathbf{3}}, ~~D^2_2Y^{(2)}_{\mathbf{3}}\rangle\,,\\
\nonumber\mathcal{M}(\mathbf{3'})&=\langle Y^{(4)}_{\mathbf{3'}}, ~~D_4 Y^{(4)}_{\mathbf{3'}}, ~~D_4^2 Y^{(4)}_{\mathbf{3'}} \rangle \,,\\
\mathcal{M}(\widehat{\mathbf{4}})&=\langle Y^{(3)}_{\widehat{\mathbf{4}}}, ~~D_3 Y^{(3)}_{\widehat{\mathbf{4}}},~~D_3^2 Y^{(3)}_{\widehat{\mathbf{4}}}, ~~D_3^3 Y^{(3)}_{\widehat{\mathbf{4}}} \rangle \,,
\label{eq:modules-2O}
\end{align}
where all the 1-d, 2-d and 3-d VVMFs of minimum weight can be obtained relatively directly from the corresponding MLDEs. These solutions are consistently expressed in the form of eta products or generalized hypergeometric functions, as demonstrated in Appendix~\ref{app:123dVVMF}. However, in the 4-d case, the solution of MLDE does not possess a compact analytic expression. Instead, we can solve for the $q$-expansion of the corresponding VVMF through a recursive method. The detailed procedure is described in Appendix~\ref{app:4dVVMF}. The VVMFs of minimal weight in each irreps of $2O$ are determined to be
\begin{align}
\label{eq:MinimalWeightVVMF2O}
\nonumber&Y^{(6)}_{\mathbf{1'}}(\tau)=\eta^{12}(\tau)\,,\\
\nonumber
&Y^{(2)}_{\mathbf{2}}(\tau)=\begin{pmatrix}
        \eta^4(\tau)(\frac{K(\tau)}{1728})^{-\frac{1}{6}}~ {}_2F_1(-\frac{1}{6},\frac{1}{6};\frac{1}{2};K(\tau)) \\
        8\sqrt{3}\eta^4(\tau)(\frac{K(\tau)}{1728})^{\frac{1}{3}}~ {}_2F_1(\frac{1}{3},\frac{2}{3};\frac{3}{2};K(\tau)) \\
\end{pmatrix}\,,\\
\nonumber
&Y^{(5)}_{\widehat{\mathbf{2}}}(\tau)=\begin{pmatrix}
         \eta^{10}(\tau)(\frac{K(\tau)}{1728})^{-\frac{1}{24}}~ {}_2F_1(-\frac{1}{24},\frac{7}{24};-\frac{3}{4};K(\tau)) \\
        2\eta^{10}(\tau)(\frac{K(\tau)}{1728})^{\frac{5}{24}}~ {}_2F_1(\frac{5}{24},\frac{13}{24};\frac{5}{4};K(\tau)) \\
\end{pmatrix}\,,\\
\nonumber
&Y^{(5)}_{\widehat{\mathbf{2}}'}(\tau)=\begin{pmatrix}
        \eta^{10}(\tau)(\frac{K(\tau)}{1728})^{-\frac{7}{24}}~ {}_2F_1(-\frac{7}{24},\frac{1}{24};\frac{1}{4};K(\tau)) \\
        56\eta^{10}(\tau)(\frac{K(\tau)}{1728})^{\frac{29}{48}}~ {}_2F_1(\frac{29}{48},\frac{15}{16};\frac{49}{24};K(\tau)) \\
\end{pmatrix}\,,\\
\nonumber
&Y^{(2)}_{\mathbf{3}}(\tau)=\begin{pmatrix}
        \eta^4(\tau)(\frac{K(\tau)}{1728})^{-\frac{1}{6}}~ {}_3F_2(-\frac{1}{6},\frac{1}{6},\frac{1}{2};\frac{3}{4},\frac{1}{4};K(\tau)) \\
        -4\sqrt{2} \eta^4(\tau)(\frac{K(\tau)}{1728})^{\frac{1}{12}}~ {}_3F_2(\frac{1}{12},\frac{5}{12},\frac{3}{4};\frac{1}{2},\frac{5}{4};K(\tau)) \\
        -16\sqrt{2} \eta^4(\tau)(\frac{K(\tau)}{1728})^{\frac{7}{12}}~ {}_3F_2(\frac{7}{12},\frac{11}{12},\frac{5}{4};\frac{7}{4},\frac{3}{2};K(\tau))
\end{pmatrix}\,,\\
\nonumber &Y^{(4)}_{\mathbf{3'}}(\tau)=\begin{pmatrix}
        2\sqrt{2}\eta^8(\tau)(\frac{K(\tau)}{1728})^{\frac{1}{6}}~ {}_3F_2(\frac{1}{6},\frac{1}{2},\frac{5}{6};\frac{3}{4},\frac{5}{4};K(\tau)) \\
        4 \eta^8(\tau)(\frac{K(\tau)}{1728})^{\frac{5}{12}}~ {}_3F_2(\frac{5}{12},\frac{3}{4},\frac{13}{12};\frac{3}{2},\frac{5}{4};K(\tau)) \\
        - \eta^8(\tau)(\frac{K(\tau)}{1728})^{-\frac{1}{12}}~ {}_3F_2(-\frac{1}{12},\frac{1}{4},\frac{7}{12};\frac{3}{4},\frac{1}{2};K(\tau))
\end{pmatrix}\,, \\
&Y^{(3)}_{\widehat{\mathbf{4}}}(\tau)=\begin{pmatrix}
4\sqrt{3}q^{5/8} \left(1-q-4 q^2+3 q^3+q^4+3 q^5+13q^6+\dots\right) \\
2\sqrt{3}q^{3/8} \left(1+q-7 q^2-6 q^3+16q^4+9 q^5-6q^6\dots\right) \\
-8 q^{7/8} \left(1-3q+ 3q^2-4q^3+3q^4+6q^5-3q^6+\dots\right) \\
q^{1/8} \left(1+3q-6q^2-23q^3+12q^4+66q^5-15q^6+\dots\right)
\end{pmatrix}\,,
\end{align}
where ${}_2F_1$ and ${}_3F_2$ are the generalized hypergeometric series.
The $q$-expansions for above VVMFs are given as follows
\begin{align}
\label{eq:q-expansion}
\nonumber&Y^{(6)}_{\mathbf{1'}}(\tau)=q^{1/2}\left(1-12 q+54 q^2-88 q^3-99 q^4+540 q^5-418 q^6 +\dots \right)\,,\\
\nonumber
&Y^{(2)}_{\mathbf{2}}(\tau)=\begin{pmatrix}
 1 + 24 q + 24 q^2 + 96 q^3 + 24 q^4 + 144 q^5 + 96 q^6+\dots \\
 8 \sqrt{3} q^{1/2} \left(1 + 4 q + 6 q^2 + 8 q^3 + 13 q^4 + 12 q^5 + 14 q^6 +\dots \right)\\
\end{pmatrix}\,,\\
\nonumber
&Y^{(5)}_{\widehat{\mathbf{2}}}(\tau)=\begin{pmatrix}
 q^{3/8} \left(1 - 7 q + 9 q^2 + 42 q^3 - 112 q^4 - 63 q^5 + 378 q^6 + \dots\right) \\
 2 q^{5/8} \left(1 - 9 q + 28 q^2 - 21 q^3 - 63 q^4 + 123 q^5 - 35 q^6+\dots\right)\\
\end{pmatrix}\,,\\
\nonumber
&Y^{(5)}_{\widehat{\mathbf{2}}'}(\tau)=\begin{pmatrix}
q^{1/8} \left(1 + 123 q + 378 q^2 - 191 q^3 - 1428 q^4 - 1134 q^5 - 735 q^6+\dots\right) \\
8 q^{7/8} \left(7 + 51 q - 27 q^2 - 28 q^3 - 459 q^4 + 378 q^5 - 357 q^6+\dots\right)\\
\end{pmatrix}\,,\\
\nonumber
&Y^{(2)}_{\mathbf{3}}(\tau)=\begin{pmatrix}
 1 - 8 q + 24 q^2 - 32 q^3 + 24 q^4 - 48 q^5 + 96 q^6+\dots\\
-4 \sqrt{2} q^{1/4} \left(1 + 6 q + 13 q^2 + 14 q^3 + 18 q^4 + 32 q^5 + 31 q^6 +\dots\right)\\
-16 \sqrt{2} q^{3/4} \left(1 + 2 q + 3 q^2 + 6 q^3 + 5 q^4 + 6 q^5 + 10 q^6 +\dots \right)
\end{pmatrix}\,, \\
&Y^{(4)}_{\mathbf{3'}}(\tau)=\begin{pmatrix}
2 \sqrt{2}q^{1/2} \left(1 - 4 q - 2 q^2 + 24 q^3 - 11 q^4 - 44 q^5 + 22 q^6+\dots\right) \\
 4 q^{3/4} \left(1 - 6 q + 11 q^2 - 2 q^3 - 11 q^4 + 14 q^5 - 38 q^6 +\dots\right) \\
 q^{1/4}\left(-1 + 2 q + 11 q^2 - 22 q^3 - 50 q^4 + 96 q^5 + 121 q^6 +\dots\right)
\end{pmatrix}\,.
\end{align}
The linearly independent modular multiplets of $2O$ at each permissible weight can be straightforwardly obtained by multiplying the polynomial of $E_4,E_6$ basis vectors of modules in Eq.~\eqref{eq:modules-2O}:
\begin{align}
\label{eq:MF-multiplets-2O}
\nonumber&k=2:~~ Y^{(2)}_{\mathbf{2}}\,,\quad Y^{(2)}_{\mathbf{3}}\,,\\
\nonumber&k=3:~~ Y^{(3)}_{\widehat{\mathbf{4}}}\,,\\
\nonumber&k=4:~~ Y^{(4)}_{\mathbf{1}}\equiv E_4\,, \quad Y^{(4)}_{\mathbf{2}}\equiv 6D_2 Y^{(2)}_{\mathbf{2}}\,,\quad Y^{(4)}_{\mathbf{3}}\equiv -6 D_2 Y^{(2)}_{\mathbf{3}}\,,\quad Y^{(4)}_{\mathbf{3'}}, \\
\nonumber&k=5:~~ Y^{(5)}_{\widehat{\mathbf{2}}}\,, \quad Y^{(5)}_{\widehat{\mathbf{2}}'}\,,\quad  Y^{(5)}_{\widehat{\mathbf{4}}}\equiv -8D_3 Y^{(3)}_{\widehat{\mathbf{4}}}\,,\\
\nonumber&k=6:~~ Y^{(6)}_{\mathbf{1}}\equiv E_6\,,\quad Y^{(6)}_{\mathbf{1'}}\,,\quad Y^{(6)}_{\mathbf{2}}\equiv E_4Y^{(2)}_{\mathbf{2}}\,,\quad Y^{(6)}_{\mathbf{3}I}\equiv E_4Y^{(2)}_{\mathbf{3}}\,,\quad Y^{(6)}_{\mathbf{3}II}\equiv -18\sqrt{2}D^2_2\ Y^{(2)}_{\mathbf{3}}\,,\\
\nonumber& ~~~~~~~~~~~~Y^{(6)}_{\mathbf{3'}}\equiv 12D_4 Y^{(4)}_{\mathbf{3'}}\,,\\
\nonumber&k=7:~~ Y^{(7)}_{\widehat{\mathbf{2}}}\equiv 24D_5 Y^{(5)}_{\widehat{\mathbf{2}}}\,,\quad Y^{(7)}_{\widehat{\mathbf{2}}'}\equiv -\frac{24}{7}D_5Y^{(5)}_{\widehat{\mathbf{2}}'}\,,\quad Y^{(7)}_{\mathbf{4}I}\equiv E_4Y^{(3)}_{\widehat{\mathbf{4}}}\,,\quad Y^{(7)}_{\widehat{\mathbf{4}}II}\equiv -192 D^2_3 Y^{(3)}_{\widehat{\mathbf{4}}}\,,\\
\nonumber&k=8:~~ Y^{(8)}_{\mathbf{1}}\equiv E_4^2\,,\quad Y^{(8)}_{\mathbf{2}I}\equiv E_6Y^{(2)}_{\mathbf{2}}\,,\quad Y^{(8)}_{\mathbf{2}II}\equiv 6E_4 D_2Y^{(2)}_{\mathbf{2}}\,,\quad Y^{(8)}_{\mathbf{3}I}\equiv E_6 Y^{(2)}_{\mathbf{3}}\,,\\
\nonumber& ~~~~~~~~~~~~Y^{(8)}_{\mathbf{3}II}\equiv -6E_4 D_2Y^{(2)}_{\mathbf{3}}\,,\quad Y^{(8)}_{\mathbf{3'}I}\equiv E_4 Y^{(4)}_{\mathbf{3'}}\,,\quad Y^{(8)}_{\mathbf{3'}II}\equiv 48E_4 D_4^2Y^{(4)}_{\mathbf{3'}}\,,\\
\nonumber&k=9:~~ Y^{(9)}_{\widehat{\mathbf{2}}}\equiv E_4 Y^{(5)}_{\widehat{\mathbf{2}}}\,,\quad Y^{(9)}_{\widehat{\mathbf{2}}'}\equiv E_4 Y^{(5)}_{\widehat{\mathbf{2}}'}\,,\quad Y^{(9)}_{\widehat{\mathbf{4}}I}\equiv E_6 Y^{(3)}_{\widehat{\mathbf{4}}}\,,\quad Y^{(9)}_{\widehat{\mathbf{4}}II}\equiv -8E_4 D_3Y^{(3)}_{\widehat{\mathbf{4}}}\,,\\
\nonumber& ~~~~~~~~~~~~ Y^{(9)}_{\widehat{\mathbf{4}}III}\equiv -4608 D_3^3 Y^{(3)}_{\widehat{\mathbf{4}}}\,,\\
\nonumber&k=10:~~ Y^{(10)}_{\mathbf{1'}}\equiv E_4 Y^{(6)}_{\mathbf{1'}}\,,\quad Y^{(10)}_{\mathbf{2}I}\equiv E_4^2 Y^{(2)}_{\mathbf{2}}\,,\quad Y^{(10)}_{\mathbf{2}II}\equiv 6E_6 D_2Y^{(2)}_{\mathbf{2}}\,,\quad Y^{(10)}_{\mathbf{3}I}\equiv E_4^2 Y^{(2)}_{\mathbf{3}}\,,\\
\nonumber& ~~~~~~~~~~~~~Y^{(10)}_{\mathbf{3}II}\equiv -6E_6 D_2 Y^{(2)}_{\mathbf{3}}\,,\quad Y^{(10)}_{\mathbf{3}III}\equiv -18\sqrt{2}E_4 D_2^2 Y^{(2)}_{\mathbf{3}}\,,\quad Y^{(10)}_{\mathbf{3'}I}\equiv E_6 Y^{(4)}_{\mathbf{3'}}\,,\\
& ~~~~~~~~~~~~~ Y^{(10)}_{\mathbf{3'}II}\equiv 12E_4 D_4Y^{(4)}_{\mathbf{3'}}\,.
\end{align}
Notice that there is no VVMFs of weight $k=1$ for $2O$. Additionally, all the modular form multiplets mentioned above of weights greater than $3$, can also be directly obtained by tensor product from the modular form multiplets of weights $2$ and $3$.

\begin{table}[t!]
\centering
\begin{tabular}{|c|c|}
\hline  \hline

Modular weight $k$ & Modular form $Y^{(k)}_{\bm{r}}$ \\ \hline
$k=2$ & $Y^{(2)}_{\mathbf{2}},Y^{(2)}_{\mathbf{3}}$\\ \hline
$k=3$ & $Y^{(3)}_{\mathbf{\widehat{4}}}$\\ \hline
$k=4$ & $Y^{(4)}_{\mathbf{1}}, Y^{(4)}_{\mathbf{2}}, Y^{(4)}_{\mathbf{3}}, Y^{(4)}_{\mathbf{3'}}$\\ \hline
$k=5$ & $Y^{(5)}_{\mathbf{\widehat{2}}}, Y^{(5)}_{\mathbf{\widehat{2}'}}, Y^{(5)}_{\mathbf{\widehat{4}}}$\\ \hline
  $k=6$ & $Y^{(6)}_{\mathbf{1}}, Y^{(6)}_{\mathbf{1'}}, Y^{(6)}_{\mathbf{2}}, Y^{(6)}_{\mathbf{3}I}, Y^{(6)}_{\mathbf{3}II}, Y^{(6)}_{\mathbf{3}'}$\\ \hline
  $k=7$ & $Y^{(7)}_{\mathbf{\widehat{2}}},Y^{(7)}_{\mathbf{\widehat{2}'}}, Y^{(7)}_{\mathbf{\widehat{4}}I}, Y^{(7)}_{\mathbf{\widehat{4}}II}$\\ \hline
  $k=8$ & $Y^{(8)}_{\mathbf{1}}, Y^{(8)}_{\mathbf{2}I}, Y^{(8)}_{\mathbf{2}II},Y^{(8)}_{\mathbf{3}I}, Y^{(8)}_{\mathbf{3}II}, Y^{(8)}_{\mathbf{3'}I}, Y^{(8)}_{\mathbf{3'}II}$\\ \hline
  $k=9$ & $Y^{(9)}_{\mathbf{\widehat{2}}}, Y^{(9)}_{\mathbf{\widehat{2}'}},Y^{(9)}_{\mathbf{\widehat{4}}I}, Y^{(9)}_{\mathbf{\widehat{4}}II}, Y^{(9)}_{\mathbf{\widehat{4}}III}$\\ \hline
  $k=10$ & $Y^{(10)}_{\mathbf{1'}}, Y^{(10)}_{\mathbf{2}I},Y^{(10)}_{\mathbf{2}II}, Y^{(10)}_{\mathbf{3}I}, Y^{(10)}_{\mathbf{3}II}, Y^{(10)}_{\mathbf{3}III}, Y^{(10)}_{\mathbf{3'}I}, Y^{(10)}_{\mathbf{3'}II}$\\ \hline
  \hline
\end{tabular}
\caption{\label{Tab:2O_MM}Summary of modular form multiplets of finite modular group $2O$ up to weight 10, the subscript $\mathbf{r}$ denotes the transformation property under $2O$. Here $Y^{(k)}_{\mathbf{r}I,II,III}$ stand for linearly independent weight-$k$ modular form multiplets transforming in the representation $\mathbf{r}$ of $2O$. }
\end{table}

\section{\label{sec:fermion_models}Fermion models based on $2O$ modular symmetry}
In the following, we shall present a systematic analysis of fermions (leptons and quarks) mass models based on the $2O$ modular symmetry. First of all, we will clarify the assignments of representations and modular weights of both matter fields and Higgs fields under $2O$ modular group.

Regarding the representation assignments, the matter fields can transform as either reducible or irreducible representations of $2O$. As can be seen from Appendix~\ref{app:2O_group}, the $2O$ group possesses two one-dimensional representations $\mathbf{1}$, $\mathbf{1'}$, three two-dimensional representations $\mathbf{2}$, $\widehat{\mathbf{2}}$, $\widehat{\mathbf{2}}'$, two three-dimensional representations $\mathbf{3}$, $\mathbf{3}'$ and one four-dimensional representations $\widehat{\mathbf{4}}$. Thus the three generations of fermions can transform as a triplet or a direct sum of one-dimensional and two-dimensional representations or a direct sum of  three one-dimensional representations of $2O$, i.e.,
\begin{eqnarray} \label{eq:assign_L}
\hskip-0.3in &&\psi\equiv\begin{pmatrix}
\psi_1 \\
\psi_2 \\
\psi_3 \\
\end{pmatrix}\sim\mathbf{3}/\mathbf{3'}\,,~~\text{or}~~\psi_D\equiv\begin{pmatrix}
\psi_1 \\
\psi_2
\end{pmatrix}\sim\mathbf{2}/\widehat{\mathbf{2}}/\widehat{\mathbf{2}}',\quad \psi_3\sim \mathbf{1}/\mathbf{1'}\,,~~\text{or}~~
\psi_{1,2,3}\sim \mathbf{1}/\mathbf{1'}\,,\\
\hskip-0.3in &&\psi^{c}\equiv\begin{pmatrix}
\psi_{1}^{c} \\
\psi_{2}^{c} \\
\psi_{3}^{c} \\
\end{pmatrix}\sim\mathbf{3}/\mathbf{3'}\,,~~\text{or}~~\psi^{c}_D\equiv\begin{pmatrix}
\psi_{1}^{c} \\
\psi_{2}^{c}
\end{pmatrix}\sim\mathbf{2}/\widehat{\mathbf{2}}/\widehat{\mathbf{2}}',\quad \psi_{3}^{c}\sim \mathbf{1}/\mathbf{1'}\,,~~\text{or}~~
\psi_{1,2,3}^{c}\sim\mathbf{1}/\mathbf{1'}\,.
\end{eqnarray}
Here $\psi$ represents the left-handed (LH) fermions which can be the lepton $SU(2)$ doublet $L=(L_{1},L_{2},L_{3})^{T}$ or the quark $SU(2)$ doublet $Q=(Q_{1},Q_{2},Q_{3})^{T}$. $\psi^{c}$ represents the right-handed (RH) fermions which can be the $SU(2)$ singlet $E^{c}=(e^{c},\mu^{c},\tau^{c})^{T}$ or $U^{c}=(u^{c},c^{c},t^{c})^{T}$ or $D^{c}=(d^{c},s^{c},b^{c})^{T}$. The subscript $i$ of  $\psi/\psi^{c}$ with $i=1,2,3$ indicates the $i$-th generation of $\psi/\psi^{c}$. We denote the modular weights of $\psi$, $\psi_{D}$, $\psi_{1,2,3}$, $\psi^{c}$, $\psi_{D}^{c}$ and $\psi_{1,2,3}^{c}$ as $k_{\psi}$, $k_{\psi_{D}}$, $k_{\psi_{1,2,3}}$, $k_{\psi^{c}}$, $k_{\psi^{c}_{D}}$ and $k_{\psi^{c}_{1,2,3}}$ respectively. Note that one can permute the three generations of fermions for the above representation assignments. This amounts to multiplying the corresponding fermion mass matrix on the left and/or right by permutation matrices. However, this does not affect the fermion masses and flavor mixing. In this work, we will not consider the case in which all fermion fields are all singlets of $2O$, since a large number of free parameters would have to be introduced to explain the experimental results of fermion masses and mixing. Without loss of generality, the Higgs doublets $H_{u}$ and $H_{d}$ are assumed to transform trivially under $2O$ with vanishing modular weight $k_{H_{u}}=k_{H_{d}}=0$.

To carry out a full analysis of fermion models in the frame of $2O$ modular symmetry, we will discuss all possible assignments of fermion fields as well as the resulting fermion mass matrices. The neutrinos are assumed to be Majorana particles. In the following, we will first focus on the Dirac fermions and then turn to consider the case of Majorana neutrinos.

\subsection{Dirac fermions\label{sec:Dirac_fermion}}

We start with the analysis of the Dirac fermions. The most general form of the Yukawa superpotential for the Dirac fermion masses can be written as
\begin{equation}
\mathcal{W}_{\psi}= g\left(Y^{(k_{\psi^{c}}+k_{\psi})}\psi^{c}\psi  H_{u/d}\right)_{\mathbf{1}}\,,
\end{equation}
where all independent $2O$ contractions should be considered and different $2O$ invariant contractions are associated with distinct Yukawa coupling constants. Modular invariance allows us to fix the modular form multiplets $Y^{(k_{\psi^{c}}+k_{\psi})}$ once the weight and representation assignments of the fields $\psi$, $\psi^{c}$ and $H_{u/d}$ are specified. In table~\ref{Tab:2O_MM}, we summarize the possible modular form multiplets of $2O$ up to weight $10$.
However, we will only consider the modular forms involved in fermion models up to weight $6$ in this work. Since the higher weight modular forms generally lead to more free parameters which will weaken the predictive power of models.

In order to obtain all possible fermion mass matrices, we first consider the independent representation assignments for the LH and RH fermions as well as the resulting sub-matrix of fermion mass matrix. In table~\ref{tab:fermion_submatrix}, we list all independent pairs of the representations of fermions $(\rho_{\psi^{c}},\rho_{\psi})$ and the corresponding sub-matrix of fermion mass matrix which are obtained using the Kronecker products and Clebsch-Gordan (CG) coefficients of $2O$ given in table~\ref{tab:2O_CG-1st} of Appendix~\ref{app:2O_group}.
The obtained sub-matrices are $n_{R}\times n_{L}$ matrices, with $n_{R}$ and $n_{L}$ are the number of generations of the involved RH and LH fermions respectively. Note that in table~\ref{tab:fermion_submatrix}, we do not discuss explicitly the cases where the assignments of the LH fermions and the RH fermions under $2O$ are switched, since in this case we only need to transpose the original sub-matrix to obtain the new sub-matrix. In concrete models, once the assignments of representations and modular weights of matter fields are fixed, we can easily read out the explicit form of fermion mass matrices from table~\ref{tab:fermion_submatrix}. We now consider distinct assignments of Dirac fermion fields as well as the resulting mass matrices.

\begin{itemize}
\item{$\psi\equiv (\psi_{1},\psi_{2},\psi_{3})^{T}\sim \mathbf{3}/\mathbf{3'},~~\psi^{c}\equiv (\psi^{c},\psi^{c},\psi^{c})^{T}\sim \mathbf{3}/\mathbf{3'}$}

  In the case that both the LH fermions $\psi$ and RH fermions $\psi^{c}$ transform as three-dimensional representations of $2O$, the general effective Yukawa terms for the Dirac fermions in the superpotential are given by
\begin{equation}
\label{eq:W_E_1}
\mathcal{W}_{\psi}=\sum_{\mathbf{r}, a}g^{\psi}_{\mathbf{r}a}\left[Y^{(k_{\psi}+k_{\psi^{c}})}_{\mathbf{r}a}\psi^{c}\psi \right]_{\mathbf{1}}H_{u/d}\,,
\end{equation}
where $g^{\psi}_{\mathbf{r}a}$ are coupling constants, with $a$ possibly labelling linearly independent multiplets of the same type and $\mathbf{r}$ labelling distinct $2O$ contractions into the $2O$ invariant singlet. From table~\ref{tab:fermion_submatrix}, we can find the corresponding Dirac fermion mass matrix is given as
\begin{equation}
  M_{\psi}=M_{33}^{\alpha}\,,
\end{equation}
where $M_{33}^{\alpha}$ is presented in table~\ref{tab:fermion_submatrix}. The superscript $\alpha=1$ corresponds to $\rho_{\psi}=\rho_{\psi^{c}}=\mathbf{3}/\mathbf{3'}$, and $\alpha=2$ if $\rho_{\psi}\neq\rho_{\psi^{c}}$ and $\rho_{\psi},\rho_{\psi^{c}}\in \{\mathbf{3},\mathbf{3'}\}$.

\item{$\psi\equiv (\psi_{1},\psi_{2},\psi_{3})^{T}\sim \mathbf{3}/\mathbf{3'},~~\psi^{c}_{D}\equiv (\psi_{1}^{c},\psi_{2}^{c})^{T}\sim \mathbf{2}/\widehat{\mathbf{2}}/\widehat{\mathbf{2}}',~~\psi_{3}^{c}\sim \mathbf{1}/\mathbf{1'}$}

If the LH fermions transform as a triplet of $2O$, and the RH fermions transform as a direct sum of one- and two-dimensional representations of $2O$, the superpotential terms relevant for Dirac fermion mass generation are
\begin{equation}\label{eq:W_E_2}
\mathcal{W}_{\psi }=\sum_{\mathbf{r},a,b}g^{\psi}_{1\mathbf{r}a}\left[Y^{(k_{\psi}+k_{\psi_{D}^{c}})}_{\mathbf{r}a}\psi_{D}^{c}\psi\right]_{\mathbf{1}}H_{u/d}+g^{\psi}_{2\mathbf{r}b}\left[Y^{(k_{\psi}+k_{\psi_{3}^{c}})}_{\mathbf{r}b}\psi_{3}^{c}\psi \right]_{\mathbf{1}}H_{u/d}\,,
\end{equation}
where $g^{\psi}_{1\mathbf{r}a}$, $g^{\psi}_{2\mathbf{r}b}$ are coupling constants. In this case, the $3\times 3$ fermion mass matrix can be divided into a $2\times 3$ and a $1\times 3$ sub-matrices. Using the convention presented in table~\ref{tab:fermion_submatrix}, we can write down the explicit form of the Dirac fermion mass matrix as
\begin{equation}
  M_{\psi}=\begin{pmatrix} M_{23}^{\alpha} \\ M_{13}^{\beta}\end{pmatrix}\,,
\end{equation}
where the values of $\alpha$ and $\beta$ depend on the choices of the representations of $\psi$, $\psi^{c}_{D}$ and $\psi^{c}_{3}$.

\item{$\psi \equiv (\psi_{1},\psi_{2},\psi_{3})^{T}\sim \mathbf{3}/\mathbf{3'},~~\psi^{c}_{i}\sim \mathbf{1}/\mathbf{1'}$}

  With all RH fermions transforming as singlets of $2O$, and LH fermions forming a triplet of $2O$, the Yukawa terms for fermion masses are
\begin{equation}
\label{eq:W_E_3}
  \mathcal{W}_{\psi}=\sum_{i=1}^{3}\sum_{a}g^{\psi}_{i \mathbf{r}a}\left[Y_{\mathbf{r}a}^{(k_{\psi}+k_{\psi^{c}_{i}})}\psi_{i}^{c}\psi \right]_{\mathbf{1}}H_{u/d}\,,
\end{equation}
where $g^{\psi}_{i \mathbf{r}a}$ are coupling constants. The $i$-th row of the corresponding mass matrix can be read out from table~\ref{tab:fermion_submatrix} as
\begin{eqnarray} \nonumber
  (\rho_{\psi^{c}_{i}},\rho_{\psi})=(\mathbf{1},\mathbf{3}')/(\mathbf{1}',\mathbf{3})\,,\quad i\text{-th row of } M_{\psi}=M_{13}^{1}\,,\\
  (\rho_{\psi^{c}_{i}},\rho_{\psi})=(\mathbf{1}',\mathbf{3}')/(\mathbf{1},\mathbf{3})\,,\quad i\text{-th row of } M_{\psi}=M_{13}^{2}\,.
\end{eqnarray}

\item{$\psi_{D}\equiv (\psi_{1},\psi_{2})^{T}\sim \mathbf{2}/\widehat{\mathbf{2}}/\widehat{\mathbf{2}}',~~\psi_{3}\sim \mathbf{1}/\mathbf{1}',~~\psi_{D}^{c}\equiv (\psi_{1}^{c},\psi_{2}^{c})^{T}\sim \mathbf{2}/\widehat{\mathbf{2}}/\widehat{\mathbf{2}}',~~\psi^{c}_{3}\sim \mathbf{1}/\mathbf{1}'$}

In the case that both the LH and RH fermions transform as direct sums of one- and two-dimensional representations of $2O$, the superpotential $\mathcal{W}_{\psi}$ is given as
\begin{eqnarray}\nonumber\mathcal{W}_{\psi}&=&\sum_{\mathbf{r},a,b,c,d}g^{\psi}_{1\mathbf{r}a}\left[Y^{(k_{\psi_{D}}+k_{\psi_{D}^{c}})}_{\mathbf{r}a}\psi_{D}^{c}\psi_{D}\right]_{\mathbf{1}}H_{u/d}+g^{\psi}_{2\mathbf{r}b}\left[Y^{(k_{\psi_{3}}+k_{\psi_{D}^{c}})}_{\mathbf{r}b}\psi_{D}^{c}\psi_{3}\right]_{\mathbf{1}}H_{u/d}\\ \label{eq:W_E_5}
&~&~~+g^{\psi}_{3\mathbf{r}c}\left[Y^{(k_{\psi_{D}}+k_{\psi_{3}^{c}})}_{\mathbf{r}c}\psi_{3}^{c}\psi_{D}\right]_{\mathbf{1}}H_{u/d}+g^{\psi}_{4\mathbf{r}d}\left[Y^{(k_{\psi_{3}}+k_{\psi_{3}^{c}})}_{\mathbf{r}d}\psi_{3}^{c}\psi_{3}\right]_{\mathbf{1}}H_{u/d}\,.
\end{eqnarray}
The corresponding Dirac fermion mass matrix can be divided into four blocks which correspond to $2\times 2$, $2\times 1$, $1\times 2$ and $1\times 1$ sub-matrices,
\begin{equation}
  M_{\psi}=\begin{pmatrix} M_{22}^{\alpha}& M_{21}^{\beta} \\ M_{12}^{\gamma} & M_{11}^{\delta} \end{pmatrix}\,,
\end{equation}
where $M_{22}^{\alpha},~M_{21}^{\beta},~M_{12}^{\gamma}$ and $M_{11}^{\delta}$ are given in table~\ref{tab:fermion_submatrix} and the values of $\alpha$, $\beta$, $\gamma$ and $\delta$ can be fixed by the representations of the LH and RH fermions.

\item{$\psi_{D}\equiv (\psi_{1},\psi_{2})^{T}\sim \mathbf{2}/\widehat{\mathbf{2}}/\widehat{\mathbf{2}}',~~\psi_{3}\sim \mathbf{1}/\mathbf{1}',~~~~\psi^{c}_{i}\sim \mathbf{1}/\mathbf{1}'$}

In this case, the modular invariant superpotential $\mathcal{W}_{\psi}$ is now given by
\begin{equation}\label{eq:We_5}\mathcal{W}_{\psi}=\sum_{i=1}^{3}\sum_{\mathbf{r},a,b}g^{\psi}_{i\mathbf{r}a}\left[Y_{\mathbf{r}a}^{(k_{\psi_{D}}+k_{\psi^{c}_{i}})}\psi_{i}^{c}\psi_{D}\right]_{\mathbf{1}}H_{u/d}+g'^{\psi}_{i\mathbf{r}b}\left[Y_{\mathbf{r}b}^{(k_{\psi_{3}}+k_{\psi^{c}_{i}})}\psi_{i}^{c}\psi_{3}\right]_{\mathbf{1}}H_{u/d}\,.
\end{equation}
The $i$-th row of $M_{\psi}$ can be expressed as
\begin{equation}
  \left(M_{12}^{\alpha}~~M_{11}^{\beta} \right),
\end{equation}
and the values of $\alpha$ and $\beta$ can be read out from table~\ref{tab:fermion_submatrix} once $\rho_{\psi_{D}}$, $\rho_{\psi_{3}}$ and $\rho_{\psi^{c}_{i}}$ are fixed.
\end{itemize}

\begin{table}[ht!]
 \centering
 \resizebox{1.0\textwidth}{!}{
\begin{tabular}{|c|c|c|} \hline\hline
$(\rho_{\psi^{c}},\rho_{\psi})$ & Case & Mass matrix/sub-matrix \\ \hline $(\mathbf{3},\mathbf{3})/(\mathbf{3'},\mathbf{3'})$ & $M_{33}^{1}$ &$g_{\mathbf{1}a}\left(
\begin{array}{ccc}
 Y_{\mathbf{1}a}^{(k)} & 0 & 0 \\
 0 & 0 & Y_{\mathbf{1}a}^{(k)} \\
 0 & Y_{\mathbf{1}a}^{(k)} & 0 \\
\end{array}
\right)+g_{\mathbf{2}b}\left(
\begin{array}{ccc}
 -2 Y_{\mathbf{2}b,1}^{(k)} & 0 & 0 \\
 0 & \sqrt{3} Y_{\mathbf{2}b,2}^{(k)} & Y_{\mathbf{2}b,1}^{(k)} \\
 0 & Y_{\mathbf{2}b,1}^{(k)} & \sqrt{3} Y_{\mathbf{2}b,2}^{(k)} \\
\end{array}
\right)+g_{\mathbf{3}c}\left(
\begin{array}{ccc}
 0 & Y_{\mathbf{3}c,3}^{(k)} & -Y_{\mathbf{3}c,2}^{(k)} \\
 -Y_{\mathbf{3}c,3}^{(k)} & 0 & Y_{\mathbf{3}c,1}^{(k)} \\
 Y_{\mathbf{3}c,2}^{(k)} & -Y_{\mathbf{3}c,1}^{(k)} & 0 \\
\end{array}
\right)+g_{\mathbf{3'}d}\left(
\begin{array}{ccc}
 0 & -Y_{\mathbf{3'}d,2}^{(k)} & Y_{\mathbf{3'}d,3}^{(k)} \\
 -Y_{\mathbf{3'}d,2}^{(k)} & -Y_{\mathbf{3'}d,1}^{(k)} & 0 \\
 Y_{\mathbf{3'}d,3}^{(k)} & 0 & Y_{\mathbf{3'}d,1}^{(k)} \\
\end{array}
\right)$\\\hline$(\mathbf{3'},\mathbf{3})$ & $M_{33}^{2}$ &$g_{\mathbf{1}a}\left(
\begin{array}{ccc}
 Y_{\mathbf{1'}}^{(k)} & 0 & 0 \\
 0 & 0 & Y_{\mathbf{1'}}^{(k)} \\
 0 & Y_{\mathbf{1'}}^{(k)} & 0 \\
\end{array}
\right)+g_{\mathbf{2}b}\left(
\begin{array}{ccc}
 2 Y_{\mathbf{2},2}^{(k)} & 0 & 0 \\
 0 & \sqrt{3} Y_{\mathbf{2},1}^{(k)} & -Y_{\mathbf{2},2}^{(k)} \\
 0 & -Y_{\mathbf{2},2}^{(k)} & \sqrt{3} Y_{\mathbf{2},1}^{(k)} \\
\end{array}
\right)+g_{\mathbf{3}c}\left(
\begin{array}{ccc}
 0 & -Y_{\mathbf{3},2}^{(k)} & Y_{\mathbf{3},3}^{(k)} \\
 -Y_{\mathbf{3},2}^{(k)} & -Y_{\mathbf{3},1}^{(k)} & 0 \\
 Y_{\mathbf{3},3}^{(k)} & 0 & Y_{\mathbf{3},1}^{(k)} \\
\end{array}
\right)+g_{\mathbf{3}'d}\left(
\begin{array}{ccc}
 0 & Y_{\mathbf{3'}d,3}^{(k)} & -Y_{\mathbf{3'}d,2}^{(k)} \\
 -Y_{\mathbf{3'}d,3}^{(k)} & 0 & Y_{\mathbf{3'}d,1}^{(k)} \\
 Y_{\mathbf{3'}d,2}^{(k)} & -Y_{\mathbf{3'}d,1}^{(k)} & 0 \\
\end{array}
\right)$\\\hline$(\mathbf{2},\mathbf{3})$ & $M_{23}^{1} $ & $g_{\mathbf{3}a}\left(
\begin{array}{ccc}
 -2 Y_{\mathbf{3}a,1}^{(k)} & Y_{\mathbf{3}a,3}^{(k)} & Y_{\mathbf{3}a,2}^{(k)} \\
 0 & \sqrt{3} Y_{\mathbf{3}a,2}^{(k)} & \sqrt{3} Y_{\mathbf{3}a,3}^{(k)} \\
\end{array}
\right)+g_{\mathbf{3}'b}\left(
\begin{array}{ccc}
 0 & -\sqrt{3} Y_{\mathbf{3'}b,2}^{(k)} & -\sqrt{3} Y_{\mathbf{3'}b,3}^{(k)} \\
 -2 Y_{\mathbf{3'}b,1}^{(k)} & Y_{\mathbf{3'}b,3}^{(k)} & Y_{\mathbf{3'}b,2}^{(k)} \\
\end{array}
\right)$\\\hline$(\mathbf{\widehat{2}},\mathbf{3})$ & $M_{23}^{2} $ &$g_{\mathbf{\widehat{2}}a}\left(
\begin{array}{ccc}
 -Y_{\mathbf{\widehat{2}}a,2}^{(k)} & \sqrt{2} Y_{\mathbf{\widehat{2}}a,1}^{(k)} & 0 \\
 -Y_{\mathbf{\widehat{2}}a,1}^{(k)} & 0 & -\sqrt{2} Y_{\mathbf{\widehat{2}}a,2}^{(k)} \\
\end{array}
\right)+g_{\widehat{\mathbf{4}}b}\left(
\begin{array}{ccc}
 -\sqrt{2} Y_{\widehat{\mathbf{4}}b,1}^{(k)} & Y_{\widehat{\mathbf{4}}b,2}^{(k)} & \sqrt{3} Y_{\widehat{\mathbf{4}}b,3}^{(k)} \\
 \sqrt{2} Y_{\widehat{\mathbf{4}}b,2}^{(k)} & -\sqrt{3} Y_{\widehat{\mathbf{4}}b,4}^{(k)} & Y_{\widehat{\mathbf{4}}b,1}^{(k)} \\
\end{array}
\right)$\\\hline$(\mathbf{\widehat{2}'},\mathbf{3})$ & $M_{23}^{3} $ &$g_{\mathbf{\widehat{2}'}a}\left(
\begin{array}{ccc}
 -Y_{\mathbf{\widehat{2}'}a,2}^{(k)} & 0 & \sqrt{2} Y_{\mathbf{\widehat{2}'}a,1}^{(k)} \\
 -Y_{\mathbf{\widehat{2}'}a,1}^{(k)} & -\sqrt{2} Y_{\mathbf{\widehat{2}'}a,2}^{(k)} & 0 \\
\end{array}
\right)+g_{\widehat{\mathbf{4}}b}\left(
\begin{array}{ccc}
 -\sqrt{2} Y_{\widehat{\mathbf{4}}b,3}^{(k)} & \sqrt{3} Y_{\widehat{\mathbf{4}}b,1}^{(k)} & Y_{\widehat{\mathbf{4}}b,4}^{(k)} \\
 \sqrt{2} Y_{\widehat{\mathbf{4}}b,4}^{(k)} & Y_{\widehat{\mathbf{4}}b,3}^{(k)} & -\sqrt{3} Y_{\widehat{\mathbf{4}}b,2}^{(k)} \\
\end{array}
\right)$\\\hline$(\mathbf{2},\mathbf{3'})$ & $M_{23}^{4} $ &$g_{\mathbf{3}a}\left(
\begin{array}{ccc}
 0 & -\sqrt{3} Y_{\mathbf{3}a,2}^{(k)} & -\sqrt{3} Y_{\mathbf{3}a,3}^{(k)} \\
 -2 Y_{\mathbf{3}a,1}^{(k)} & Y_{\mathbf{3}a,3}^{(k)} & Y_{\mathbf{3}a,2}^{(k)} \\
\end{array}
\right)+g_{\mathbf{3'}b}\left(
\begin{array}{ccc}
 -2 Y_{\mathbf{3'}b,1}^{(k)} & Y_{\mathbf{3'}b,3}^{(k)} & Y_{\mathbf{3'}b,2}^{(k)} \\
 0 & \sqrt{3} Y_{\mathbf{3'}b,2}^{(k)} & \sqrt{3} Y_{\mathbf{3'}b,3}^{(k)} \\
\end{array}
\right)$\\\hline$(\mathbf{\widehat{2}},\mathbf{3'})$ & $M_{23}^{5} $ &$g_{\mathbf{\widehat{2}'}a}\left(
\begin{array}{ccc}
 Y_{\mathbf{\widehat{2}'}a,1}^{(k)} & \sqrt{2} Y_{\mathbf{\widehat{2}'}a,2}^{(k)} & 0 \\
 -Y_{\mathbf{\widehat{2}'}a,2}^{(k)} & 0 & \sqrt{2} Y_{\mathbf{\widehat{2}'}a,1}^{(k)} \\
\end{array}
\right)+g_{\widehat{\mathbf{4}}b}\left(
\begin{array}{ccc}
 \sqrt{2} Y_{\widehat{\mathbf{4}}b,4}^{(k)} & Y_{\widehat{\mathbf{4}}b,3}^{(k)} & -\sqrt{3} Y_{\widehat{\mathbf{4}}b,2}^{(k)} \\
 \sqrt{2} Y_{\widehat{\mathbf{4}}b,3}^{(k)} & -\sqrt{3} Y_{\widehat{\mathbf{4}}b,1}^{(k)} & -Y_{\widehat{\mathbf{4}}b,4}^{(k)} \\
\end{array}
\right)$\\\hline$(\mathbf{\widehat{2}'},\mathbf{3'})$ & $M_{23}^{6} $ &$g_{\mathbf{\widehat{2}}a}\left(
\begin{array}{ccc}
 Y_{\mathbf{\widehat{2}}a,1}^{(k)} & 0 & \sqrt{2} Y_{\mathbf{\widehat{2}}a,2}^{(k)} \\
 -Y_{\mathbf{\widehat{2}}a,2}^{(k)} & \sqrt{2} Y_{\mathbf{\widehat{2}}a,1}^{(k)} & 0 \\
\end{array}
\right)+g_{\widehat{\mathbf{4}}b}\left(
\begin{array}{ccc}
 \sqrt{2} Y_{\widehat{\mathbf{4}}b,2}^{(k)} & -\sqrt{3} Y_{\widehat{\mathbf{4}}b,4}^{(k)} & Y_{\widehat{\mathbf{4}}b,1}^{(k)} \\
 \sqrt{2} Y_{\widehat{\mathbf{4}}b,1}^{(k)} & -Y_{\widehat{\mathbf{4}}b,2}^{(k)} & -\sqrt{3} Y_{\widehat{\mathbf{4}}b,3}^{(k)} \\
\end{array}
\right)$\\\hline$(\mathbf{1},\mathbf{3'})/(\mathbf{1'},\mathbf{3})$ & $M_{13}^{1} $ &$g_{\mathbf{3'}a}\left(
\begin{array}{ccc}
 Y_{\mathbf{3'}a,1}^{(k)} ~& ~Y_{\mathbf{3'}a,3}^{(k)} ~&~ Y_{\mathbf{3'}a,2}^{(k)} \\
\end{array}
\right)$\\\hline$(\mathbf{1'},\mathbf{3'})/(\mathbf{1},\mathbf{3})$ & $M_{13}^{2} $ &$g_{\mathbf{3}b}\left(
\begin{array}{ccc}
 Y_{\mathbf{3}b,1}^{(k)} ~&~ Y_{\mathbf{3}b,3}^{(k)} ~&~ Y_{\mathbf{3}b,2}^{(k)} \\
\end{array}
\right)$\\\hline$(\mathbf{2},\mathbf{2})$ & $M_{22}^{1} $&$g_{\mathbf{1}a}\left(
\begin{array}{cc}
 Y_{\mathbf{1}a}^{(k)} & 0 \\
 0 & Y_{\mathbf{1}a}^{(k)} \\
\end{array}
\right)+g_{\mathbf{1'}b}\left(
\begin{array}{cc}
 0 & Y_{\mathbf{1'}b}^{(k)} \\
 -Y_{\mathbf{1'}b}^{(k)} & 0 \\
\end{array}
\right)+g_{\mathbf{2}c}\left(
\begin{array}{cc}
 -Y_{\mathbf{2},1}^{(k)} & Y_{\mathbf{2},2}^{(k)} \\
 Y_{\mathbf{2},2}^{(k)} & Y_{\mathbf{2},1}^{(k)} \\
\end{array}
\right)$\\\hline$(\mathbf{\widehat{2}},\mathbf{2})$ & $M_{22}^{2} $&$g_{\widehat{\mathbf{4}}a}\left(
\begin{array}{cc}
 Y_{\widehat{\mathbf{4}}a,1}^{(k)} & -Y_{\widehat{\mathbf{4}}a,4}^{(k)} \\
 Y_{\widehat{\mathbf{4}}a,2}^{(k)} & Y_{\widehat{\mathbf{4}}a,3}^{(k)} \\
\end{array}
\right)$\\\hline$(\mathbf{\widehat{2}'},\mathbf{2})$ & $M_{22}^{3} $&$g_{\widehat{\mathbf{4}}a}\left(
\begin{array}{cc}
 Y_{\widehat{\mathbf{4}}a,3}^{(k)} & -Y_{\widehat{\mathbf{4}}a,2}^{(k)} \\
 Y_{\widehat{\mathbf{4}}a,4}^{(k)} & Y_{\widehat{\mathbf{4}}a,1}^{(k)} \\
\end{array}
\right)$\\\hline$(\mathbf{\widehat{2}},\mathbf{\widehat{2}})$ & $M_{22}^{4} $ &$g_{\mathbf{1}a}\left(
\begin{array}{cc}
 0 & Y_{\mathbf{1}a}^{(k)} \\
 -Y_{\mathbf{1}a}^{(k)} & 0 \\
\end{array}
\right)+g_{\mathbf{3}b}\left(
\begin{array}{cc}
 -\sqrt{2} Y_{\mathbf{3}b,2}^{(k)} & Y_{\mathbf{3}b,1}^{(k)} \\
 Y_{\mathbf{3}b,1}^{(k)} & \sqrt{2} Y_{\mathbf{3}b,3}^{(k)} \\
\end{array}
\right)$\\\hline$(\mathbf{\widehat{2}'},\mathbf{\widehat{2}})$ & $M_{22}^{5} $&$g_{\mathbf{1'}a}\left(
\begin{array}{cc}
 Y_{\mathbf{1'}a}^{(k)} & 0 \\
 0 & Y_{\mathbf{1'}a}^{(k)} \\
\end{array}
\right)+g_{\mathbf{3'}b}\left(
\begin{array}{cc}
 Y_{\mathbf{3'}b,1}^{(k)} & \sqrt{2} Y_{\mathbf{3'}b,3}^{(k)} \\
 \sqrt{2} Y_{\mathbf{3'}b,2}^{(k)} & -Y_{\mathbf{3'}b,1}^{(k)} \\
\end{array}
\right)$\\\hline$(\mathbf{\widehat{2}'},\mathbf{\widehat{2}'})$ & $M_{22}^{6}$&$g_{\mathbf{1}a}\left(
\begin{array}{cc}
 0 & Y_{\mathbf{1}a}^{(k)} \\
 -Y_{\mathbf{1}a}^{(k)} & 0 \\
\end{array}
\right)+g_{\mathbf{3}b}\left(
\begin{array}{cc}
 -\sqrt{2} Y_{\mathbf{3}b,3}^{(k)} & Y_{\mathbf{3}b,1}^{(k)} \\
 Y_{\mathbf{3}b,1}^{(k)} & \sqrt{2} Y_{\mathbf{3}b,2}^{(k)} \\
\end{array}
\right)$\\\hline$(\mathbf{1},\mathbf{2})$ & $M_{12}^{1} $ & $g_{\mathbf{2}a}\left(
\begin{array}{cc}
 Y_{\mathbf{2}a,1}^{(k)} ~&~ Y_{\mathbf{2}a,2}^{(k)} \\
\end{array}
\right)$\\\hline$(\mathbf{1},\mathbf{\widehat{2}})$ & $M_{12}^{2} $ & $g_{\mathbf{\widehat{2}}a}\left(
\begin{array}{cc}
 -Y_{\mathbf{\widehat{2}}a,2}^{(k)} ~&~ Y_{\mathbf{\widehat{2}}a,1}^{(k)} \\
\end{array}
\right)$\\\hline$(\mathbf{1},\mathbf{\widehat{2}'})$ & $M_{12}^{3} $ & $g_{\mathbf{\widehat{2}'}a}\left(
\begin{array}{cc}
 -Y_{\mathbf{\widehat{2}'}a,2}^{(k)} ~&~ Y_{\mathbf{\widehat{2}'}a,1}^{(k)} \\
\end{array}
\right)$\\\hline$(\mathbf{1'},\mathbf{2})$ &$M_{12}^{4} $ &$g_{\mathbf{2}a}\left(
\begin{array}{cc}
 -Y_{\mathbf{2}a,2}^{(k)} ~&~ Y_{\mathbf{2}a,1}^{(k)} \\
\end{array}
\right)$\\\hline$(\mathbf{1'},\mathbf{\widehat{2}})$ & $M_{12}^{5} $ & $g_{\mathbf{\widehat{2}'}a}\left(
\begin{array}{cc}
 Y_{\mathbf{\widehat{2}'}a,1}^{(k)} ~&~ Y_{\mathbf{\widehat{2}'}a,2}^{(k)} \\
\end{array}
\right)$\\\hline$(\mathbf{1'},\mathbf{\widehat{2}'})$ & $M_{12}^{6} $ & $g_{\mathbf{\widehat{2}}a}\left(
\begin{array}{cc}
 Y_{\mathbf{\widehat{2}}a,1}^{(k)} ~&~ Y_{\mathbf{\widehat{2}}a,2}^{(k)} \\
\end{array}
\right)$\\\hline$(\mathbf{1},\mathbf{1})/(\mathbf{1'},\mathbf{1'})$ &$M_{11}^{1} $ &$g_{\mathbf{1}a}
 Y_ {\mathbf{1}a}^{(k)} $\\ \hline$(\mathbf{1'},\mathbf{1})$ &$M_{11}^{2}$ &$g_{\mathbf{1'}a}
 Y_ {\mathbf{1'}a}^{(k)}$ \\\hline \hline \end{tabular}}
\caption{\label{tab:fermion_submatrix}Textures for the submatrix of fermion mass matrix, where $\rho_{\psi}$ and $\rho_{\psi^{c}}$ are the representations of the involved LH and RH fermions under the modular symmetry $2O$ respectively. The $Y_{\mathbf{r}a}^{(k)}$ denotes the modular form multiplet of weight $k$ and representation $\mathbf{r}$ with $a$ possibly labelling linearly independent multiplets of the same type. The $g_{\mathbf{r}a}$ are coupling constants and the repeated indices are implicitly summed over. Here we drop the VEVs $v_u\equiv\langle H_u\rangle$ and $v_d\equiv\langle H_d\rangle$ of the Higgs fields $H_{u}$ and $H_{d}$ in the mass matrices. The fermion mass matrices are read out in the $RL$ basis.}
\end{table}

\subsection{Majorana neutrinos\label{sec:Majorana_fermion}}

We now turn to analyze the assignments and the resulting mass matrices of Majorana neutrinos. We consider two distinct neutrino mass generation mechanisms: the effective Weinberg operator and the type-I seesaw mechanism. If neutrino masses are generated via the Weinberg operator, the modular-invariant Weinberg operator of neutrino masses can be written as
\begin{equation}
\mathcal{W}_{\nu}= \frac{g^{W}}{\Lambda}\left(Y^{(2k_{L})}  LLH_{u}H_{u}\right)_{\mathbf{1}}\,,
\end{equation}
where $\Lambda$ denotes the lepton number breaking scale. In this case, the effective neutrino mass matrix $M_{\nu}$ only depends on the $2O$ representation assignments and modular weights of the LH lepton fields $L$. Similar to what we have done in section~\ref{sec:Dirac_fermion}, we can discuss all possible assignments of $L$ and the corresponding $M_{\nu}$. For different assignments of the LH leptons, the resulting matrices/sub-matrices of $M_{\nu}$ can be obtained from table~\ref{tab:fermion_submatrix} with $\rho_{\psi}=\rho_{\psi^{c}}=\rho_{L}$. Besides, the anti-symmetric terms of the neutrino mass matrix $M_{\nu}$ should be removed, since the Majorana neutrino mass matrix is symmetric.

For the case that neutrino masses are generated through the type-I seesaw mechanism, the general form of the superpotential in neutrino sector is
\begin{equation}\label{eq:W_ss}
  W_{\nu}= g^{D}\left(Y^{(k_{L}+k_{N^{c}})} N^{c}L  H_{u}\right)_{\mathbf{1}} + g^{N}\Lambda\left(Y^{(2k_{N^{c}})} N^{c}N^{c} \right)_{\mathbf{1}}\,,
\end{equation}
where $N^{c}\equiv (N_{1}^{c},N_{2}^{c},N_{3}^{c})$ denotes three generations of RH neutrinos, and the first and second terms in the RH part of Eq~\eqref{eq:W_ss} correspond to the Dirac and Majorana neutrino mass terms respectively. The relevant analysis of Dirac neutrino mass matrix $M_D$ which involves $N^c$ and $L$ is similar to the analysis of Dirac fermions presented in section~\ref{sec:Dirac_fermion}. The Majorana neutrino mass term for the RH neutrinos only depends on $N^{c}$. We can also use table~\ref{tab:fermion_submatrix} to read out the explicit form of the RH neutrino mass matrix $M_{N}$ with $\rho_{\psi}=\rho_{\psi^{c}}=\rho_{N^{c}}$. Similarly, we need to drop the anti-symmetric terms in $M_{N}$. After integrating out the heavy neutrinos $N^{c}$,  the effective neutrino mass matrix $M_{\nu}$ can be given by the seesaw formula
\begin{equation}\label{eq:seesaw_formula}
  M_{\nu}=-M_{D}^{T}M_{N}^{-1}M_{D}\,.
\end{equation}

\subsection{Fermion models}

In section~\ref{sec:Dirac_fermion} and section~\ref{sec:Majorana_fermion}, we have discussed the assignments and the resulting mass matrices of Dirac fermions and Majorana neutrinos. All independent pairs of the representations of Dirac fermions $(\rho_{\psi^{c}},\rho_{\psi})$ and the corresponding sub-matrices of Dirac fermion mass matrix are presented in table~\ref{tab:fermion_submatrix}. The Majorana neutrinos mass matrix can also be obtained from table~\ref{tab:fermion_submatrix} by demanding $\rho_{\psi}=\rho_{\psi^{c}}$ and dropping the anti-symmetric terms. In concrete fermion models, once the assignments of representations and modular weights of fermions are fixed, we can easily read out the explicit form of fermion mass matrices from table~\ref{tab:fermion_submatrix}. Generally, the modular weights of the LH and RH fermions are arbitrary integers. As mentioned in section~\ref{sec:Dirac_fermion}, we will be concerned with the modular forms up to weight 6. As a result, for Dirac fermions, the modular weights $k_{\psi}$ and $k_{\psi^{c}}$ should satisfy $k_{\psi}+k_{\psi^{c}}\leq 6$ and for Majorana neutrinos $2k_{L}\leq 6$ or $2k_{N^{c}}\leq 6$. The higher weight modular forms can be discussed in a similar way, and the higher weight modular forms generally contain more modular multiplets which lead to more Yukawa coupling constants.

In the following, we are interested in those fermion models which are, in some sense, predictive. In lepton and quark sectors, there are $9$ and $10$ experimentally-measured observables respectively,
\begin{eqnarray}
  \text{Lepton sector}:&&~~ m_{e},~m_{\mu},~m_{\tau},~\Delta m_{21}^{2},~\Delta m_{31}^{2},~\theta^{l}_{12},~\theta^{l}_{13},~\theta^{l}_{23},~\delta^{l}_{CP}\,,\\
  \text{Quark sector}:&&~~ m_{u},~m_{c},~m_{t},~m_{d},~m_{s},~m_{b},~\theta_{12}^{q},~\theta_{13}^{q},~\theta_{23}^{q},~\delta_{CP}^{q}\,,
\end{eqnarray}
where $m_{e,\mu,\tau}$ are three charged-lepton masses, $\Delta m_{21}^{2}\equiv m_{2}^{2}-m_{1}^{2}$, $\Delta m_{31}^{2}\equiv m_{3}^{2}-m_{1}^{2}$ are two neutrino mass-squared differences, $\theta^{l}_{12}$, $\theta^{l}_{13}$, $\theta^{l}_{23}$ are three lepton mixing angles and $\delta^{l}_{CP}$ is the Dirac CP-violation phase in the lepton sector, $m_{u,c,t}$, $m_{d,s,b}$ are six quark masses, $\theta_{12}^{q}$, $\theta_{13}^{q}$, $\theta_{23}^{q}$ are three quark mixing angles and $\delta_{CP}^{q}$ is quark CP-violation phase. The fermion model will not be attractive if too many free parameters are required to fit the measured observables. Hence we only focus on the lepton models and quark models which contain no more than 9 and 10 free real parameters respectively.

It can be observed from table~\ref{tab:fermion_submatrix} that the fermion mass matrices consist of the Yukawa coupling constants and modular forms. The Yukawa coupling constants within the fermion model are generally complex numbers and the modular forms are holomorphic functions of a complex modulus $\tau$. However, there are some unphysical complex phases in Yukawa coupling constants which can be removed by rephasing fermion fields. In the process of counting the number of real input parameters, the unphysical complex phases should be omitted. A common approach to further constrain the number of input parameters is to include the generalized CP symmetry (gCP) in fermion models.

It has been established that the action of gCP on the complex modulus is $\tau\xrightarrow[]{\mathcal{CP}} -\tau^{*}$, up to modular transformations~\cite{Acharya:1995ag,Dent:2001cc,Giedt:2002ns,Baur:2019kwi,Novichkov:2019sqv}. Under the action of gCP, the VVMF transforms in the same way as the matter field. The coupling constants in models would be enforced to be real by gCP if both modular generators $S$ and $T$ are represented by symmetric and unitary matrices in all irreps and the CG coefficients are real. We are indeed working in the symmetric basis of the $2O$ group with real CG coefficients, as shown in Appendix~\ref{app:2O_group}. In this paper, we consider all fermion models with modular symmetry $2O$ and gCP. Consequently, all Yukawa coupling constants become real and modular flavor symmetry breaking as well as CP violation solely originate from the VEV of $\tau$.

Considering the previous convention of assignments of representations and modular weights of lepton fields, and demanding the lepton model to contain no more than $9$ real free input parameters, we can realize over two thousands kinds of lepton models in the case of neutrino masses generated via the Weinberg operator. If neutrino masses are generated via the see-saw mechanism, we can obtain nearly four thousands lepton models. For quark models, we find approximate thirty thousands kinds of quark models with $10$ or less real free input parameters. With these constructed fermion models at our disposal, we will perform numerical analysis of these lepton models and the quark models in the next section.

\section{Numerical analysis}
\label{sec:numerical_analysis}

We have systematically constructed the lepton and quark models based on $2O$ modular symmetry. In this section, we will perform numerical analysis of these fermion models and give predictions of fermion observables. For each fermion model, we need to check if it can reproduce the input data within experimental errors. In order to do it, we perform a $\chi^2$ analysis of all fermion models. The $\chi^{2}$ function is taken as usual form
\begin{equation}
\chi^2=\sum^{n}_{i=1}\left(\frac{P_i(x)-\mu_i}{\sigma_i}\right)^2\,,
\end{equation}
where the vector $x$ contains the model parameters, $P_{i}(x)$ are the model predictions for the observables, $\mu_{i}$ and $\sigma_{i}$ denote the central values and standard deviations of the corresponding quantities obtained from experimental data -- see table~\ref{Tab:parameter_values}. For lepton models, we fit seven dimensionless physical observables: $\theta_{12}$, $\theta_{13}$, $\theta_{23}$, $\delta_{CP}$, $m_{e}/m_{\mu}$, $m_{\mu}/m_{\tau}$ and $\Delta m_{21}^{2}/\Delta m_{31}^{2}$. The mass of tau $m_{\tau}$ and the solar neutrino mass squared difference $\Delta m_{21}^{2}$ can be fixed by the overall mass scale parameters of the charged lepton and neutrino mass matrices respectively. For quark models, physical observables are chosen as  $\theta_{12}^{q}$, $\theta_{13}^{q}$, $\theta_{23}^{q}$, $\delta_{CP}^{q}$, $m_{u}/m_{c}$, $m_{c}/m_{t}$, $m_{d}/m_{s}$ and $m_{s}/m_{b}$. Similarly, we can use the overall parameters of up- and down-type quark mass matrices to fix $m_{t}$ and $m_{b}$.

The experimental results of lepton and quark observables are summarized in table~\ref{Tab:parameter_values}, where the charged fermion masses and the quark mixing parameters have been extrapolated to the GUT scale~\cite{Antusch:2013jca}. The neutrino masses and mixings are taken from NuFIT 5.2~\cite{Esteban:2020cvm}. Since the normal ordering neutrino masses are favored at $2.7\sigma$ level if the atmospheric neutrino data of Super-Kamiokande is taken into account~\cite{Esteban:2020cvm}. We only focus on the normal ordering spectrum of neutrino masses, $m_1 < m_2 < m_3$ in this work. The renormalization group effects of the neutrino mass ratios and the mixing angles are known to be negligible to a good approximation for the normal hierarchical spectrum~\cite{Criado:2018thu}. Therefore, for neutrino oscillation parameters, we ignore the effects of the evolution from the low energy scale to the GUT scale. Note that the present statistical significance of the leptonic CP phase $\delta^{l}_{CP}$ measurement is rather weak~\cite{Esteban:2020cvm}. For normal ordering neutrino mass spectrum,  the $3\sigma$ region of $\delta^{l}_{CP}$ is $[ 144^{\circ},350^{\circ}]$, taken from NuFIT 5.2~\cite{Esteban:2020cvm} with Super-Kamiokande atmospheric data. In this paper, we impose gCP symmetry on the fermion mass models, consequently all CP violation phases arise from $\text{Re}\braket{\tau}$, which represents the real part of modulus $\braket{\tau}$ the VEV of the modulus $\tau$. As a result, it is worth to include $\delta^{l}_{CP}$ in $\chi^2$ analysis of lepton sector to constrain the allowed region of $\tau$.

\begin{table}[t!]
\centering
\begin{tabular}{|c|c|c|c|c|} \hline  \hline
Parameters & $\mu_i\pm1\sigma$ & Parameters & $\mu_i\pm1\sigma$ & $3\sigma$ region \\ \hline
$m_{t}/\text{GeV}$ & $89.213 \pm 2.219$ &  $m_{\tau}/\text{GeV}$ & $1.293\pm 0.007$ & --\\
  $m_{u}/m_{c}$ & $0.00193 \pm 0.00060$ & $m_{e}/m_{\mu}$ & $0.00474 \pm 0.00004$ & --\\
  $m_{c}/m_{t}$ & $0.00280 \pm 0.00012$ & $m_{\mu}/m_{\tau}$ & $0.0588 \pm 0.0005$ & --\\
  $m_{b}/\text{GeV}$ & $0.965 \pm 0.011$ & $\Delta m_{21}^{2}/(10^{-5}\text{eV}^{2})$ & $7.41_{-0.20}^{+0.21}$ &$[6.82, 8.03]$  \\
  $m_{d}/m_{s}$ & $0.0505 \pm 0.0062$ & $r\equiv\Delta m_{21}^{2}/\Delta m_{31}^{2}$ & $0.02956 \pm 0.00084$ & $[0.02704, 0.03207]$\\
  $m_{s}/m_{b}$ & $0.0182 \pm 0.0010$ & $\sin^{2}\theta_{12}^{l}$ & $0.303\pm 0.012$ & $[0.270, 0.341]$\\
  $\theta_{12}^{q}$ & $0.2274 \pm 0.0007$ & $\sin^{2}\theta_{23}^{l}$ & $0.451_{-0.016}^{+0.019}$ & $[0.408, 0.603]$\\
 $\theta_{13}^{q}$ & $0.00349 \pm 0.00013$ & $\sin^{2}\theta_{13}^{l}$ & $0.02225_{-0.00059}^{+0.00056}$ & $[0.02052, 0.02398]$\\
  $\theta_{23}^{q}$ & $0.0400 \pm 0.0006 $ & $\delta_{CP}^{l}/^{\circ}$ & $232_{-26}^{+36}$ & $[144, 350]$\\
  $\delta_{CP}^{q}/^{\circ}$ & $69.21 \pm 3.11 $ & & &\\
\hline \hline
\end{tabular}
\caption{\label{Tab:parameter_values} The best fit values $\mu_i$ and $1\sigma$ uncertainties of the quark and lepton parameters when evolved to the GUT scale as calculated in~\cite{Antusch:2013jca}, with the SUSY breaking scale $M_{\text{SUSY}}=10$ TeV and $\tan\beta=5$, where the error widths represent $1\sigma$ intervals. The parameter $r\equiv \Delta m_{21}^{2}/ \Delta m_{31}^{2}$ is the ratio of neutrino mass-squared differences. The values of lepton mixing angles, leptonic Dirac CP violation phases $\delta^{l}_{CP}$ and the neutrino mass squared difference are taken from NuFIT 5.2~\cite{Esteban:2020cvm} for normal ordering with neutrino masses with Super-Kamiokande atmospheric data.
}
\end{table}

We use the minimization package \textbf{TMinuit} to find the minima of $\chi^{2}$ function, the best fit values of the input parameters as well as the predictions for fermion observables. The $\chi^{2}$ function is highly non-linear and possesses many local minima. It is possible that the minima found by \textbf{TMinuit} is a local minimum but not global minimum. To increase the quality of the minima, we start the minimization algorithm many times with different initial parameters and choose the lowest out of the many local minima. The parameter space of the inputs are restricted in the following regions:
\begin{eqnarray}
\text{couplings:} &&\quad g_{i}\in [-10^{6},10^{6}]\,,\\
\text{modulus:}&&\quad \tau\in \mathcal{F}:~|\text{Re}\tau|\leq \frac{1}{2}\,,~\text{Im}\tau>0\,,~|\tau|\geq 1\,,
\end{eqnarray}
where $\mathcal{F}$ is the fundamental domain of $\Gamma$. A fermion model is said to be compatible with experimental data if the predictions of neutrino masses and mixing parameters are within their experimental $3\sigma$ regions given in table~\ref{Tab:parameter_values}. For the charged-fermion masses and quark mixing angles and CP phase, we require that their best-fit values should not deviate from the experimental central values by more than 3 times of the corresponding $1\sigma$ region -- see table~\ref{Tab:parameter_values}. As a result of the analysis we obtain:

{\bf Lepton models:} For lepton models, we have considered two scenarios that the neutrino masses are generated by the Weinberg operator or the type-I seesaw mechanism. If light neutrino masses are described by the Weinberg operator, by performing $\chi^{2}$ analysis, we find that there are hundreds of lepton models which are compatible with experimental data. The minimal model contains $7$ free real input parameters. For the case that neutrino masses are generated through the type-I seesaw mechanism, we find a minimal model which can only use $6$ free parameters to explain all $9$ measured lepton observables. In addition, we obtain a lot of phenomenologically viable lepton models with the number of input parameters varying from $7$ to $9$.

{\bf Quark models:} For quark models, we scan the parameter space of quark models containing no more than $10$ input parameters one by one. In order to explain the experimental data given in table~\ref{Tab:parameter_values}, the quark models are found to contain at least $9$ parameters. In the following, we will give a representative quark model with $9$ parameters. There is no quark model with $8$ or less parameters that can accommodate the experimental at $3\sigma$ confidence level. However, some models with $8$ parameters can achieve a relatively small $\chi^{2}$, and only the quark mass ratio $m_{s}/m_{b}$ and $\theta^q_{13}$ slightly lie outside the $3\sigma$ region. It is reasonable to regard these models as a good leading order approximation, we will give such an example below.

{\bf Unified models:} In the above, we have searched the phenomenologically viable models in lepton sector and quark sector separately. Now we want to combine the viable lepton models with the viable quark models to give a unified description of both leptons and quarks with a common modulus $\tau$. By performing $\chi^{2}$ analysis of unified fermion models, we find the minimal viable unified fermion model can use 14 real parameters to describe the $22$ masses and mixing parameters of quarks and leptons. The detail of the model will be present in following.

\section{Benchmark models}
By performing $\chi^{2}$ analysis to the constructed fermions models, we have obtained thousands of phenomenologically viable models. It is impossible to show all viable models in detail, consequently we will provide some benchmark models where the quality of the results can be appropriate for leptons and quarks. The benchmark models are chosen from viable models which contain minimum number of free real input parameters.

\subsection{Lepton models\label{sec:lepton_models}}
For lepton models, we present one representative lepton model for the case of Majorana neutrino masses generated via the type-I seesaw mechanism and one for the case of the Weinberg operator.

\subsubsection{Lepton model with $6$ parameters for seesaw mechanism \label{sec:lepton_model_6para_SS}}
Among all phenomenologically viable lepton models for the case of Majorana neutrino masses generated via the type-I seesaw mechanism, the minimal lepton model contains only $6$ real input parameters. We have identified a single viable lepton model with six parameters that is highly predictive, as it employs these parameters to account for nine measured lepton observables. Moreover this model predicts absolute neutrino masses and Majorana CP violation phases.
For the sake of clarity, we list the details of the model in the following. The representation assignments and the modular weights of the lepton fields are,
  \begin{eqnarray}\nonumber
    &&L\sim \mathbf{3}\,,~~E_{D}^{c}\equiv (e^{c},\mu^{c}) \sim \mathbf{\widehat{2}'}\,,~~\tau^{c}\sim \mathbf{1'}\,,~~N^{c}\sim \mathbf{3}\,,\\
    &&k_{L}=-1\,,~~k_{E_{D}^{c}}=6\,,~~k_{\tau^{c}}=5\,,~~k_{N^{c}}=1\,.
  \end{eqnarray}
The corresponding modular invariant lepton superpotentials are given by
\begin{eqnarray}\nonumber
  \mathcal{W}_{E}&=&g^{E}_{1} \left( E_{D}^{c}L\right)_{\mathbf{\widehat{2}'}}Y^{(5)}_{\mathbf{\widehat{2}'}}H_{d}+g^{E}_{2} \left( E_{D}^{c}L\right)_{\mathbf{\widehat{4}}}Y^{(5)}_{\mathbf{\widehat{4}}}H_{d} + g^{E}_{3} \left( \tau^{c}L \right)_{\mathbf{3'}}Y^{(4)}_{\mathbf{3'}}H_{d}\,,\\
  \mathcal{W}_{\nu}&=&g^{D}H_{u}(N^{c}L)_{\mathbf{1}}+ \Lambda  \left(N^{c}N^{c}\right)_{\mathbf{2}}Y^{(2)}_{\mathbf{2}}\,.
\end{eqnarray}
Notice that there are $6$ free real input parameters including $3$ coupling constants $g^{E}_{1,2,3}$ in the charged lepton sector, $1$ effective coupling constant $(g^{D})^{2}/\Lambda$ in the neutrino sector and a complex modulus $\tau$. The corresponding charged lepton and neutrino mass matrices read as
\begin{eqnarray} \nonumber
  M_{E}&=&\left( \begin{array}{ccc}
                   -g^{E}_{1} Y^{(5)}_{\mathbf{\widehat{2}'},2}-\sqrt{2}g^{E}_{2}Y^{(5)}_{\mathbf{\widehat{4}},3} ~&  \sqrt{3}g^{E}_{2}Y^{(5)}_{\mathbf{\widehat{4}},1} & ~\sqrt{2}g^{E}_{1}Y^{(5)}_{\mathbf{\widehat{2}'},1}+g^{E}_{2}Y^{(5)}_{\mathbf{\widehat{4}},4}  \\
                   -g^{E}_{1} Y^{(5)}_{\mathbf{\widehat{2}'},1}+\sqrt{2}g^{E}_{2}Y^{(5)}_{\mathbf{\widehat{4}},4} ~& -\sqrt{2}g^{E}_{1}Y^{(5)}_{\mathbf{\widehat{2}'},2}+g^{E}_{2}Y^{(5)}_{\mathbf{\widehat{4}},3} & ~-\sqrt{3}g^{E}_{2}Y^{(5)}_{\mathbf{\widehat{4}},2}  \\
                  g^{E}_{3} Y^{(4)}_{\mathbf{3'},1}~& g^{E}_{3} Y^{(4)}_{\mathbf{3'},3} & ~g^{E}_{3} Y^{(4)}_{\mathbf{3'},2}  \\
                 \end{array} \right) v_{d}\,,\\
  M_D &=& g^{D}\begin{pmatrix}
1 & ~0~ & 0 \\
0 & ~0~ & 1 \\
0 & ~1~ &0 \\
\end{pmatrix}v_{u} \,,\quad
  M_N =\begin{pmatrix}
  -2Y^{(2)}_{\mathbf{2},1} ~&~ 0 ~&~0 \\
 0 ~&~ \sqrt{3}Y^{(2)}_{\mathbf{2},2}  ~&~ Y^{(2)}_{\mathbf{2},1}  \\
 0 ~&~ Y^{(2)}_{\mathbf{2},1} ~&~\sqrt{3}Y^{(2)}_{\mathbf{2},2} \\
\end{pmatrix} \Lambda\,.
\end{eqnarray}
Using the seesaw formula given in Eq.~\eqref{eq:seesaw_formula}, we can obtain the effective light neutrino mass matrix
\begin{equation}
M_{\nu}=\frac{(g^{D}v_{u})^{2}}{\Lambda}\left( \begin{array}{ccc} \frac{1}{2Y^{(2)}_{\mathbf{2},1}} ~& 0 ~& 0 \\
0 ~& \frac{\sqrt{3}Y^{(2)}_{\mathbf{2},2}}{Y^{(2)2}_{\mathbf{2},1}-3Y^{(2)2}_{\mathbf{2},2}} ~& -\frac{Y^{(2)}_{\mathbf{2},1}}{Y^{(2)2}_{\mathbf{2},1}-3Y^{(2)2}_{\mathbf{2},2}} \\
0 ~& -\frac{Y^{(2)}_{\mathbf{2},1}}{Y^{(2)2}_{\mathbf{2},1}-3Y^{(2)2}_{\mathbf{2},2}} ~& \frac{\sqrt{3}Y^{(2)}_{\mathbf{2},2}}{Y^{(2)2}_{\mathbf{2},1}-3Y^{(2)2}_{\mathbf{2},2}} \end{array} \right)\,,
\end{equation}
and the three light neutrino masses are given as
\begin{equation}
m_{1}=\frac{1}{|2Y^{(2)}_{\mathbf{2},1}|}\frac{(g^{D}v_{u})^{2}}{\Lambda}\,,\quad
m_{2}=\frac{1}{|Y^{(2)}_{\mathbf{2},1}-\sqrt{3}Y^{(2)}_{\mathbf{2},2}|}\frac{(g^{D}v_{u})^{2}}{\Lambda}\,,\quad
m_{3}=\frac{1}{|Y^{(2)}_{\mathbf{2},1}+\sqrt{3}Y^{(2)}_{\mathbf{2},2}|}\frac{(g^{D}v_{u})^{2}}{\Lambda}\,.
\end{equation}
\begin{figure}[t!]
\centering
\includegraphics[width=0.6\textwidth]{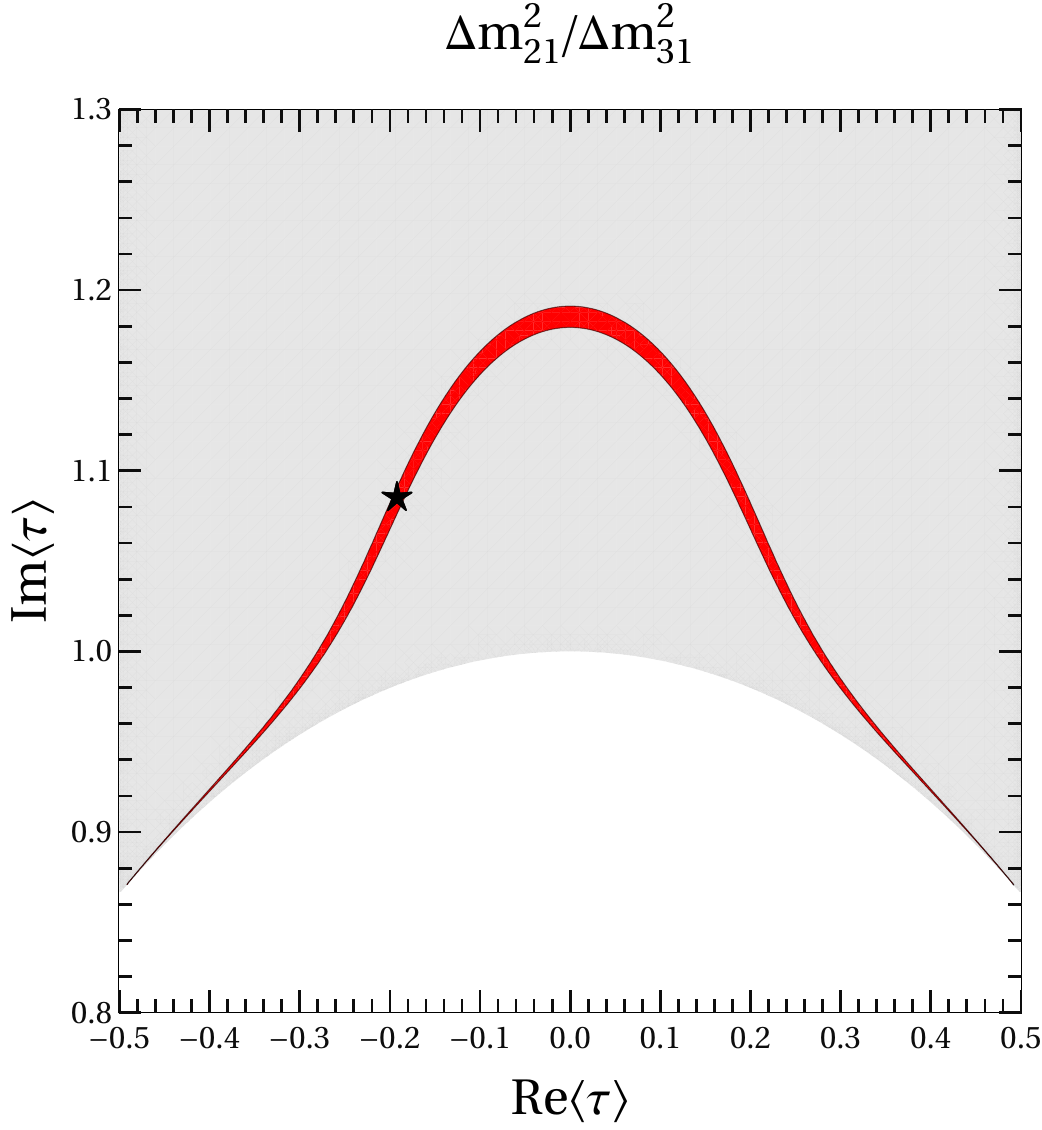}
\caption{The region of modulus $\braket{\tau}$ compatible with experimental data. The gray region is the fundamental domain of $\tau$, and the red area represents the viable range of $\braket{\tau}$ compatible with the experimental region of $\Delta m_{21}^{2}/\Delta m_{31}^{2}$ of the ratio of neutrino mass squared difference taken from~\cite{Esteban:2020cvm}. The black star represents the best fitting point.
}
\label{fig:mass_ratio}
\end{figure}
One can find that the neutrino masses only depend on the VEV of the modulus $\tau$ and an overall parameter $\frac{(g^{D}v_{u})^{2}}{ \Lambda}$, thus we can use the experimental result of the masses ratio $\Delta m_{21}^{2}/\Delta m_{31}^{2}$ to constrain the allowed region of $\braket{\tau}$. The corresponding result is shown in red region of figure~\ref{fig:mass_ratio}. To find out if or not that the model can explain the experimental results of leptons, we perform a global fit to the lepton experimental data for normally-ordered neutrino masses. The values of the $6$ input parameters at the best-fit point are
\begin{figure}[t!]
\centering
\includegraphics[width=6.5in]{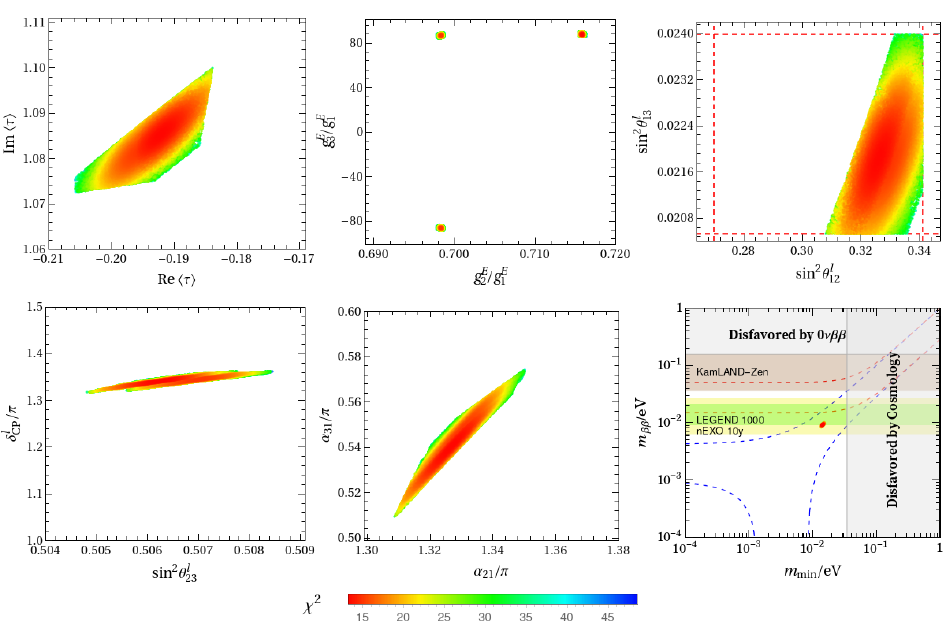}
\caption{\label{fig:model_lepton_6para_SS} The predictions for the correlations among the input free parameters, neutrino mixing angles, CP violation phases and neutrino masses in the lepton model with $6$ real input parameters. }
\end{figure}
\begin{eqnarray}
\nonumber
  &&\braket{\tau}=-0.19205 + 1.08536i\,,\quad g_{2}^{E}/g_{1}^{E}=0.71588\,,\quad g_{3}^{E}/g_{1}^{E}=87.44709\,,\\
  &&g_{1}^{E}v_{d}=0.07189~\text{GeV}\,,\quad \frac{\left(g^{D}v_{u}\right)^{2}}{ \Lambda}=0.02881~\text{eV}\,, \label{eq:bf_lepton_6para}
\end{eqnarray}
being the corresponding lepton observables
\begin{eqnarray}
\nonumber&& \sin^{2}\theta^{l}_{12}=0.3261\,,\quad \sin^{2}\theta^{l}_{13}=0.02182\,,\quad \sin^{2}\theta^{l}_{23}=0.5063\,,\quad \delta_{CP}^{l}=241.2^{\circ}\,,\\
  \nonumber&& \alpha_{21}=1.3268\pi\,,\quad \alpha_{31}=0.5401\pi\,,\quad m_e/m_{\mu}=0.004737\,,\quad m_{\mu}/m_{\tau}=0.05876\,,\\ \nonumber
  &&\frac{\Delta m_{21}^{2}}{\Delta m_{31}^{2}} = 0.03009\,,\quad  m_1=14.27~\text{meV}\,,\quad m_2=16.67~\text{meV}\,,\quad m_3=51.64~\text{meV}\,,\\ && m_{\beta\beta}=9.17~\text{meV}\,, \label{eq:ob_lepton_6para}
\end{eqnarray}
with $\chi^{2}_{\rm min}=13.17$. Here $\alpha_{21}$ and $\alpha_{31}$ are two Majorana CP-violation phases in the standard parametrization and $m_{\beta\beta}$ is the effective Majorana mass of neutrinoless double beta ($0\nu\beta\beta$) decay experiment. The predictions of all measured lepton observables lie in the corresponding $3\sigma$ experimental regions as given in table~\ref{Tab:parameter_values}. It is notable that only normal ordering neutrino mass spectrum can accommodate the experimental data and inverted ordering is disfavored in this model. Moreover, we use the sampler \textbf{MultiNest}~\cite{Feroz:2007kg,Feroz:2008xx} to scan the parameter space of the model, while all measured lepton observables are demanded to lie in the experimentally allowed $3\sigma$ regions. The correlations among the input free parameters, neutrino mixing angles, CP violation phases and neutrino masses are shown in figure~\ref{fig:model_lepton_6para_SS}. We can find both the free parameters and the lepton observables are limited in narrow ranges. The atmospherical neutrino mixing angle $\sin^{2}\theta^{l}_{23}$ is located in the second octant as $\sin^{2}\theta^{l}_{23}\in [0.504,0.508]$. Besides, there are $\sin^{2}\theta^{l}_{12}\in [0.308,0.341]$, $\delta_{CP}^{l}\in [1.316\pi,1.360\pi]$, $\alpha_{21}\in [1.309\pi,1.347\pi]$ and $\alpha_{31}\in [0.509\pi,0.569\pi]$.  The mass of the lightest neutrino $m_{1}$ is predicted to be in $[13.47\text{meV},15.05\text{meV}]$ and $\sum_{i}m_{i}$ is lies in $[79.41\text{meV},86.69\text{meV}]$. Cosmological data shows that the most stringent bound on the sum of neutrino masses is $\sum_{i}m_{i}<120~\text{meV}$ at $95\%$~C.L. from the Planck Collaboration results~\cite{Planck:2018vyg}. Thus, the present model is compatible with the Planck data as well.

In terms of light neutrino masses and the lepton mixing parameters, the effective Majorana mass $m_{\beta\beta}$ which is defined in the amplitude of the $0\nu\beta\beta$-decay can be expressed as
\begin{equation}
  m_{\beta\beta}=|m_{1}\cos^{2}\theta_{12}^{l}\cos^{2}\theta_{13}^{l}+m_{2}\sin^{2}\theta_{12}^{l}\cos^{2}\theta_{13}^{l}e^{i\alpha_{21}}+m_{3}\sin^{2}\theta_{13}e^{i(\alpha_{31}-2\delta_{CP}^{l})}|\,.
\end{equation}
In this model, $m_{\beta\beta}$ is predicted to be limited in $[8.55\text{meV},9.86\text{meV}]$ which respects the current experimental bound. According to the latest result of the KamLAND-Zen experiment,  the upper limit of $m_{\beta\beta}$ is $(36-156)~\text{meV}$ at $90\%$ C.L. ~\cite{KamLAND-Zen:2022tow}. The future large-scale $0\nu\beta\beta$-decay experiments expect to further improve the bound on $m_{\beta\beta}$. For instance, the SNO+ Phase II is expected to reach a sensitivity of $(19-46)$ meV~\cite{SNO:2015wyx}, the optimistic bound of LEGEND experiment is about $(15-50)$~meV~\cite{LEGEND:2017cdu}, and the nEXO will be able to probe $m_{\beta\beta}$ down to $(5.7-17.7)$~meV~\cite{nEXO:2017nam} which can check the prediction of this model.

\begin{figure}[t!]
\centering
\includegraphics[width=6.5in]{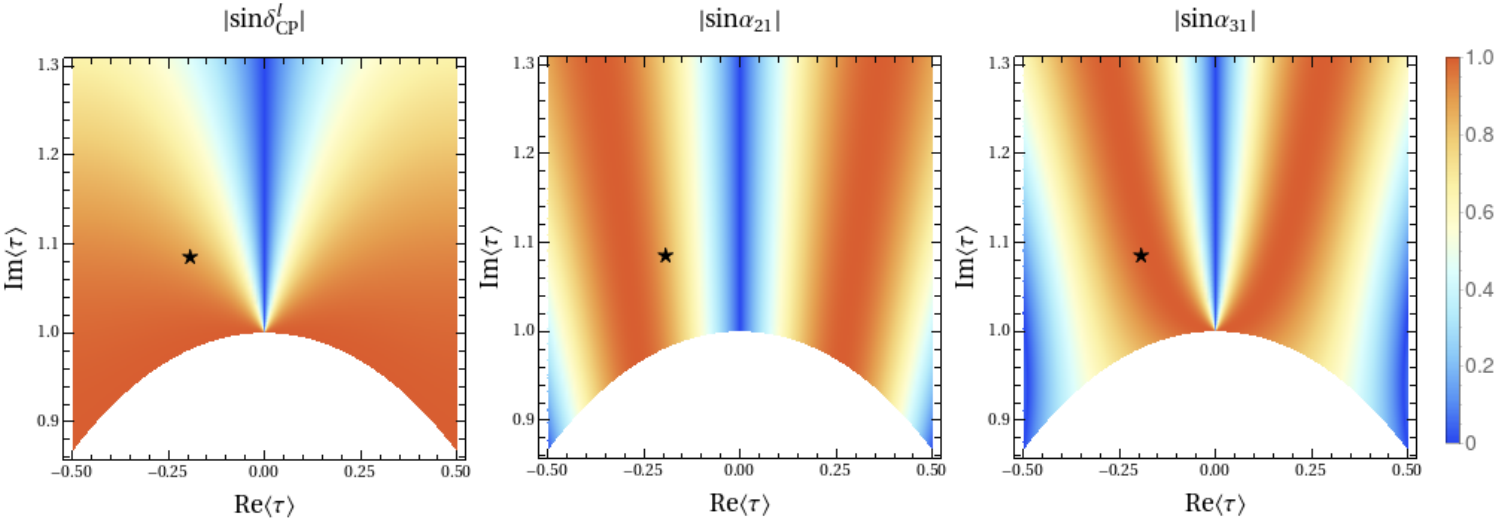}
\caption{The contour plots of the CP-violation phases on the $\text{Re}\braket{\tau}-\text{Im}\braket{\tau}$ plane. The black star refers to the best fitting point.}
\label{fig:phase}
\end{figure}

As mentioned in section~\ref{sec:fermion_models}, all couplings are real because of the gCP, and the VEV of $\tau$ is the unique source of CP violation. To find the correlations between $\braket{\tau}$ and CP-violation phases, we plot the contour regions of CP-violation phases in the plane $\text{Im}\braket{\tau}$ versus $\text{Re}\braket{\tau}$ in figure~\ref{fig:phase}. The coupling constants are fixed at their best fit values as given in Eq.~\eqref{eq:bf_lepton_6para}.

\subsubsection{Lepton model with $7$ parameters for Weinberg operator \label{sec:lepton_model_7para_WO}}

\begin{figure}[t!]
\centering
\includegraphics[width=6.5in]{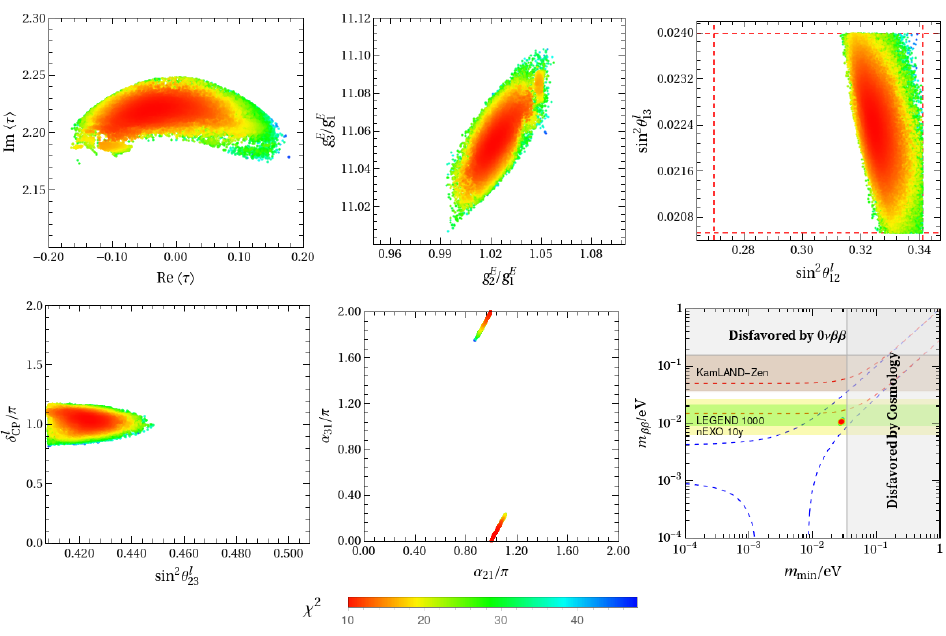}
\caption{\label{fig:model_lepton_7para_WO}The predictions for the correlations among the input free parameters, neutrino mixing angles, CP violation phases and neutrino masses in the lepton model with $7$ real input parameters.}
\end{figure}

In the case of neutrino masses generated via the Weinberg operator, we find that at least $7$ real parameters are need to explain the measured values of lepton observables. In this section, we give such a representative lepton model with $7$ input parameters. The details of the model are given as
  \begin{eqnarray}\nonumber
    &&L\sim \mathbf{3}\,,~~E_{D}^{c}\equiv (e^{c},\mu^{c}) \sim \mathbf{2}\,,~~\tau^{c}\sim \mathbf{1'}\,,\\
    &&k_{L}=1\,,~~k_{E_{D}^{c}}=5\,,~~k_{\tau^{c}}=3\,.
  \end{eqnarray}
The corresponding modular invariant lepton superpotentials are given by
\begin{eqnarray}\nonumber
  \mathcal{W}_{E}&=&g^{E}_{1} \left( E_{D}^{c}L\right)_{\mathbf{3}}Y^{(6)}_{\mathbf{3}I}H_{d} +g^{E}_{2} \left( E_{D}^{c}L\right)_{\mathbf{3}}Y^{(6)}_{\mathbf{3}II}H_{d}+g^{E}_{3} \left( E_{D}^{c}L\right)_{\mathbf{3'}}Y^{(6)}_{\mathbf{3'}}H_{d}+ g^{E}_{4} \left( \tau^{c}L \right)_{\mathbf{3'}}Y^{(4)}_{\mathbf{3'}}H_{d}\,,\\
  \mathcal{W}_{\nu}&=&\frac{1}{\Lambda}(LL)_{\mathbf{2}}Y^{(2)}_{\mathbf{2}}H_{u}H_{u}\,,
\end{eqnarray}
where $\Lambda$ is the new physical scale where lepton number is violated by two units. The corresponding charged lepton and neutrino mass matrices read as
{\footnotesize{
\begin{eqnarray} \nonumber
  M_{E}&=&\left(
\begin{array}{ccc}
 -2 g^{E}_{2} Y_{\mathbf{3}II,1}^{(6)}-2 g^{E}_{1}Y_{\mathbf{3}I,1}^{(6)} & g^{E}_{2} Y_{\mathbf{3}II,3}^{(6)}-\sqrt{3} g^{E}_{3} Y_{\mathbf{3'},2}^{(6)}+g^{E}_{1}Y_{\mathbf{3}I,3}^{(6)} & g^{E}_{2} Y_{\mathbf{3}II,2}^{(6)}-\sqrt{3} g^{E}_{3} Y_{\mathbf{3'},3}^{(6)}+g^{E}_{1}Y_{\mathbf{3}I,2}^{(6)} \\
 -2 g^{E}_{3} Y_{\mathbf{3'},1}^{(6)} & \sqrt{3} g^{E}_{2} Y_{\mathbf{3}II,2}^{(6)}+g^{E}_{3} Y_{\mathbf{3'},3}^{(6)}+\sqrt{3} g^{E}_{1}Y_{\mathbf{3}I,2}^{(6)} & \sqrt{3} g^{E}_{2} Y_{\mathbf{3}II,3}^{(6)}+g^{E}_{3} Y_{\mathbf{3'},2}^{(6)}+\sqrt{3} g^{E}_{1}Y_{\mathbf{3}I,3}^{(6)} \\
 g^{E}_{4} Y_{\mathbf{3'},1}^{(4)} & g^{E}_{4} Y_{\mathbf{3'},3}^{(4)} & g^{E}_{4} Y_{\mathbf{3'},2}^{(4)} \\
\end{array}
\right)v_{d}\,,\\
  M_\nu &=& \frac{v_{u}^{2}}{\Lambda}\left(
\begin{array}{ccc}
 -2 Y_{\mathbf{2},1}^{(2)} & 0 & 0 \\
 0 & \sqrt{3} Y_{\mathbf{2},2}^{(2)} & Y_{\mathbf{2},1}^{(2)} \\
 0 & Y_{\mathbf{2},1}^{(2)} & \sqrt{3} Y_{\mathbf{2},2}^{(2)} \\
\end{array}
\right)\,.
\end{eqnarray}}}
The agreement between the model predictions and the experimental data is optimized for the following values of the input parameters
\begin{eqnarray}\label{eq:bf_lepton_7para}\nonumber
  &&\braket{\tau}=-0.03966+ 2.21810i\,,\quad g_{2}^{E}/g_{1}^{E}=1.01858\,,\quad g_{3}^{E}/g_{1}^{E}=11.04561\,,\\
  && g^{E}_{4}/g^{E}_{1}=-0.01128\,,\quad g_{1}^{E}v_{d}=1.02083~\text{GeV}\,,\quad \frac{v_{u}^{2}}{\Lambda}=28.74885~\text{meV}\,.
\end{eqnarray}
The values of the masses and mixing parameters of leptons at the above best fit point are determined to be
\begin{eqnarray}
\nonumber&& \sin^{2}\theta^{l}_{12}=0.3263\,,\quad \sin^{2}\theta^{l}_{13}=0.02242\,,\quad \sin^{2}\theta^{l}_{23}=0.4296\,,\quad \delta_{CP}^{l}=187.5^{\circ}\,,\\
  \nonumber&& \alpha_{21}=1.028\pi\,,\quad \alpha_{31}=0.056\pi\,,\quad m_e/m_{\mu}=0.004800\,,\quad m_{\mu}/m_{\tau}=0.05988\,,\\ \nonumber
  &&\frac{\Delta m_{21}^{2}}{\Delta m_{31}^{2}} = 0.02945\,,\quad  m_1=28.11~\text{meV}\,,\quad m_2=29.39~\text{meV}\,,\quad m_3=57.50~\text{meV}\,,\\ && m_{\beta\beta}=10.49~\text{meV}\,,
\end{eqnarray}
with $\chi^{2}_{\rm min}=8.79$. The predictions are compatible with the experimental data at $3\sigma$ level--see table~\ref{Tab:parameter_values}. We display the allowed values of input parameters and the corresponding observables in figure~\ref{fig:model_lepton_7para_WO}. Unlike the predictions of the model in section~\ref{sec:lepton_model_6para_SS}, the prediction of $\sin^{2}\theta^{l}_{23}$ in this model is located in the first octant as $\sin^{2}\theta^{l}_{23}\in [0.408,0.448]$. The $3\sigma$ region of $\sin^{2}\theta^{l}_{13}$ can be achieved, while $\sin^{2}\theta_{12}^{l}\in[0.313,0.341]$. For the CP violation phases, $\delta_{CP}^{l}$ and $\alpha_{21}$ are located around $\pi$ as $\delta_{CP}^{l}\in[0.815\pi,1.171\pi]$ and $\alpha_{21}\in[0.873\pi,1.114\pi]$, the $\alpha_{31}$ varies around $0$ as $\alpha_{31}\in[0,0.232\pi]\cup[1.748\pi,2\pi]$. The lightest neutrino mass is $m_{1}\in[26.86\text{meV},29.40\text{meV}]$ and the sum of neutrino masses $\sum_{i}m_{i}\in [110.13\text{meV},120.00\text{meV}]$ which is close to the upper bound of the Planck Collaboration results~\cite{Planck:2018vyg}, $\sum_{i}m_{i}<120~\text{meV}$.
The allowed region of $m_{\beta\beta}$ is $[9.82~\text{meV},11.74~\text{meV}]$ which agrees with the current bound~\cite{KamLAND-Zen:2022tow} and can be checked by the future large-scale $0\nu\beta\beta$-decay experiments~\cite{nEXO:2017nam}.

\subsection{\label{sec:quark_models}Quark models}

We turn to the representative models of quarks in this section. The minimal phenomenologically viable quark model is found to contain $9$ real input parameters. As mentioned in section~\ref{sec:numerical_analysis}, some quark models with $8$ parameters can be regarded as leading order approximation. For these reasons, in the following, we present one representative quark model with $8$ parameters and another viable quark model with $9$ parameters.

\subsubsection{Quark model with $8$ input parameters\label{sec:quark_model_8para}}

The first benchmark quark model contains $8$ input parameters and it can not completely explain the experimental result of quarks.  The $2O$ representation and modular-weight assignments of quark fields are given as
\begin{eqnarray}\nonumber
  &&Q_{L}\sim \mathbf{3}\,,~~U_{D}^{c}\equiv (u^{c},c^{c})\sim \mathbf{\widehat{2}'}\,,~~t^{c}\sim \mathbf{1}\,,~~D^{c}\equiv (d^{c},s^{c},b^{c}) \sim \mathbf{3'}\,,\\
  && k_{Q_{L}}=3-k_{U_{D}^{c}}=6-k_{t^{c}}=4-k_{D^{c}}\,.
\end{eqnarray}
Then the Yukawa superpotentials for the quark masses are
\begin{eqnarray}\nonumber
  \mathcal{W}_{u}&=&g^{u}_{1}\left( U_{D}^{c}Q_{L} \right)_{\widehat{\mathbf{4}}}Y_{\widehat{\mathbf{4}}}^{(3)}H_{u} + g^{u}_{2}\left( t^{c}Q_{L}\right)_{\mathbf{3}}Y^{(6)}_{\mathbf{3}I}H_{u} + g^{u}_{3}\left( t^{c}Q_{L}\right)_{\mathbf{3}}Y^{(6)}_{\mathbf{3}II}H_{u} \,,\\
  \mathcal{W}_{d}&=&g^{d}_{1}\left( D^{c}Q_{L} \right)_{\mathbf{2}}Y^{(4)}_{\mathbf{2}}H_{d}+g^{d}_{2}\left( D^{c}Q_{L} \right)_{\mathbf{3}}Y^{(4)}_{\mathbf{3}}H_{d}+g^{d}_{3}\left( D^{c}Q_{L} \right)_{\mathbf{3'}}Y^{(4)}_{\mathbf{3'}}H_{d} \,,
\end{eqnarray}
where $g^{u}_{1,2,3}$ and $g^{d}_{1,2,3}$ are real coupling constants. We find the quark mass matrices take the following form
\begin{eqnarray}\nonumber
M_{u}&=&\left(\begin{array}{ccc}
-\sqrt{2}g^{u}_{1}Y^{(3)}_{\widehat{\mathbf{4}},3} ~& \sqrt{3}g^{u}_{1}Y^{(3)}_{\widehat{\mathbf{4}},1} &~ g^{u}_{1}Y^{(3)}_{\widehat{\mathbf{4}},4}  \\
\sqrt{2}g^{u}_{1}Y^{(3)}_{\widehat{\mathbf{4}},4} ~& g^{u}_{1}Y^{(3)}_{\widehat{\mathbf{4}},3} &~ -\sqrt{3}g^{u}_{1}Y^{(3)}_{\widehat{\mathbf{4}},2}  \\
g^{u}_{2}Y^{(6)}_{\mathbf{3}I,1}+g^{u}_{3}Y^{(6)}_{\mathbf{3}II,1} ~& g^{u}_{2}Y^{(6)}_{\mathbf{3}I,3}+g^{u}_{3}Y^{(6)}_{\mathbf{3}II,3} &~ g^{u}_{2}Y^{(6)}_{\mathbf{3}I,2}+g^{u}_{3}Y^{(6)}_{\mathbf{3}II,2}  \\ \end{array} \right)v_{u} \,,\\
M_{d}&=&\left(\begin{array}{ccc}
2g^{d}_{1}Y^{(4)}_{\mathbf{2},2} ~&-g^{d}_{2}Y^{(4)}_{\mathbf{3},2}+g^{d}_{3}Y^{(4)}_{\mathbf{3'},3}  &~g^{d}_{2}Y^{(4)}_{\mathbf{3},3}-g^{d}_{3}Y^{(4)}_{\mathbf{3'},2}  \\
-g^{d}_{2}Y^{(4)}_{\mathbf{3},2}-g^{d}_{3}Y^{(4)}_{\mathbf{3'},3}~&\sqrt{3}g^{d}_{1}Y^{(4)}_{\mathbf{2},1}-g^{d}_{2}Y^{(4)}_{\mathbf{3},1} &~-g^{d}_{1}Y^{(4)}_{\mathbf{2},2}+g^{d}_{3}Y^{(4)}_{\mathbf{3'},1}   \\
g^{d}_{2}Y^{(4)}_{\mathbf{3},3}+g^{d}_{3}Y^{(4)}_{\mathbf{3'},2}~& -g^{d}_{1}Y^{(4)}_{\mathbf{2},2}-g^{d}_{3}Y^{(4)}_{\mathbf{3'},1} &~ \sqrt{3}g^{d}_{1}Y^{(4)}_{\mathbf{2},1}+g^{d}_{2}Y^{(4)}_{\mathbf{3},1}  \\ \end{array} \right)v_{d} \,.
\end{eqnarray}
We perform a global fit to the quark experimental data, the values of the input parameters at the best-fit point are
\begin{eqnarray}
 &&\braket{\tau}=-0.30703 + 2.88560i\,,\quad g_{2}^{u}/g_{1}^{u}=777.66929\,,\quad g_{3}^{u}/g_{1}^{u}=551.39462\,,\\
\nonumber
 &&g_{2}^{d}/g_{1}^{d}=-1.75444\,,\quad g_{3}^{d}/g_{1}^{d}=4.23434\,,\quad g_{1}^{u}v_{u}=1.67475~\text{GeV}\,,\quad g_{1}^{d}v_{d}=0.27672~\text{GeV}\,,
\end{eqnarray}
being the corresponding quark observables
\begin{eqnarray}
\nonumber&& \theta^{q}_{12}=0.227\,,\quad \theta^{q}_{13}=0.00270\,,\quad \theta^{q}_{23}=0.0399\,,\quad \delta^{q}_{CP}=69.31^{\circ}\,,\\
&& m_{u}/m_{c}=0.00196\,,\quad m_{c}/m_{t}=0.00276\,,\quad m_{d}/m_{s}=0.0610\,,\quad m_{s}/m_{b}=0.0292\,.
\end{eqnarray}
Almost all observables lie in the $3\sigma$ experimental intervals, except $m_{s}/m_{b}$ and $\theta_{13}^{q}$ which are close to the upper and lower bounds respectively. The model can be regarded as a good leading order approximation.

\subsubsection{Quark model with $9$ input parameters\label{sec:quark_model_9para}}

In this section, we give an example of the minimal viable quark models with $9$ input parameters. The quark fields transform under the $2O$ modular symmetry as follow,
\begin{eqnarray}\nonumber
  &&Q_{D}\equiv (Q_{1},Q_{2})^{T}\sim \mathbf{2}\,,~~Q_{3}\sim \mathbf{1}\,,~~U_{D}^{c}\equiv (u^{c},c^{c})\sim \mathbf{2}\,,\\ \nonumber
  &&t^{c}\sim \mathbf{1}\,,~~D_{D}^{c}\equiv (d^{c},s^{c}) \sim \mathbf{\widehat{2}'}\,,~~b^{c}\sim \mathbf{1}\,,\\
  && k_{Q_{D}}=4-k_{U_{D}^{c}}=k_{Q_{3}}=-k_{t^{c}}=3-k_{D_{D}^{c}}=4-k_{b^{c}}\,.
\end{eqnarray}
Then the Yukawa superpotentials for the quark masses are
\begin{eqnarray}\nonumber
  \mathcal{W}_{u}&=&g^{u}_{1}\left( U_{D}^{c}Q_{D} \right)_{\mathbf{1}}Y_{\mathbf{1}}^{(4)}H_{u} + g^{u}_{2}\left( U_{D}^{c}Q_{D} \right)_{\mathbf{2}}Y_{\mathbf{2}}^{(4)}H_{u} + g^{u}_{3}\left( U_{D}^{c}Q_{3}\right)_{\mathbf{2}}Y^{(4)}_{\mathbf{2}}H_{u}+ g^{u}_{4}\left( t^{c}Q_{3}\right)_{\mathbf{1}}H_{u}\,,\\
  \mathcal{W}_{d}&=&g^{d}_{1}\left( D_{D}^{c}Q_{L} \right)_{\widehat{\mathbf{4}}}Y^{(3)}_{\widehat{\mathbf{4}}}H_{d}+g^{d}_{2}\left( b^{c}Q_{D} \right)_{\mathbf{2}}Y^{(4)}_{\mathbf{2}}H_{d}+g^{d}_{3}\left( b^{c}Q_{3} \right)_{\mathbf{1}}Y^{(4)}_{\mathbf{1}}H_{d} \,.
\end{eqnarray}
It can be found that this model depends only $9$ real free parameters including 7 real coupling constants $g^{u}_{1,2,3,4}$, $g^{d}_{1,2,3}$ and a complex $\tau$. The corresponding up- and down-type quark mass matrices take the following form
\begin{eqnarray}\nonumber
    M_{u}&=&\left(
\begin{array}{ccc}
 g^{u}_{1}Y_{\mathbf{1}}^{(4)}-g^{u}_{2} Y_{\mathbf{2},1}^{(4)} & g^{u}_{2} Y_{\mathbf{2},2}^{(4)} & g^{u}_{3} Y_{\mathbf{2},1}^{(4)} \\
 g^{u}_{2} Y_{\mathbf{2},2}^{(4)} & g^{u}_{2} Y_{\mathbf{2},1}^{(4)}+g^{u}_{1}Y_{\mathbf{1}}^{(4)} & g^{u}_{3} Y_{\mathbf{2},2}^{(4)} \\
 0 & 0 & g^{u}_{4} \\
\end{array}
\right)v_{u} \,,\\
   M_{d}&=&\left(
\begin{array}{ccc}
 g^{d}_{1}Y_{\widehat{\mathbf{4}},3}^{(3)} & -g^{d}_{1}Y_{\widehat{\mathbf{4}},2}^{(3)} & 0 \\
 g^{d}_{1}Y_{\widehat{\mathbf{4}},4}^{(3)} & g^{d}_{1}Y_{\widehat{\mathbf{4}},1}^{(3)} & 0 \\
 g^{d}_{2} Y_{\mathbf{2},1}^{(4)} & g^{d}_{2} Y_{\mathbf{2},2}^{(4)} & g^{d}_{3} Y_{\mathbf{1}}^{(4)} \\
\end{array}
\right)v_{d} \,.
\end{eqnarray}
We perform a global fit to the quark experimental data, the values of the input parameters at the best-fit point are
\begin{eqnarray}\nonumber
  &&\braket{\tau}=0.06845 + 2.20708i\,,\quad g_{2}^{u}/g_{1}^{u}=1.00535\,,\quad g_{3}^{u}/g_{1}^{u}=0.02844\,,\\ \nonumber
  &&g_{4}^{u}/g_{1}^{u}=-1.09856\times 10^{-5}\,,\quad g_{2}^{d}/g_{1}^{d}=-1.08833\,,\quad g_{3}^{d}/g_{1}^{d}=-0.00707\,,\\
  &&g_{1}^{u}v_{u}=44.47283~\text{GeV}\,,\quad g_{1}^{d}v_{d}=0.87481~\text{GeV}\,.
\end{eqnarray}
 The corresponding predictions of quark observables are
\begin{eqnarray}
\nonumber&& \theta^{q}_{12}=0.227\,,\quad \theta^{q}_{13}=0.00348\,,\quad \theta^{q}_{23}=0.0403\,,\quad \delta^{q}_{CP}=69.482^{\circ}\,,\\
&& m_{u}/m_{c}=0.00196\,,\quad m_{c}/m_{t}=0.00273\,,\quad m_{d}/m_{s}=0.0542\,,\quad m_{s}/m_{b}=0.0181\,,~~~
\end{eqnarray}
with $\chi^{2}=0.770$.
Obviously the above best-fit values are in very good agreement with the data, being all observables in the experimentally allowed $1\sigma$ intervals. By scanning the parameter space of this model, we find all the quark observables can take any values within the experimental $3\sigma$ regions.

\subsection{\label{sec:unified_model}Unified models of leptons and quarks}
In the above two sections, we have discussed the phenomenologically viable models in lepton sector and quark sector separately. Now we shall investigate whether the $2O$ modular symmetry can describe the flavor structures of quark and lepton simultaneously. In this section, we give an example of the unified model with 6 parameters in the lepton sector and 10 parameters in the quark sector. The complex modulus $\tau$ in the quark and lepton sectors is the same one, so the total number of real free parameters is $6+10-2=14$. This model is very predictive, since it uses 14 real free parameters to describe the 22 masses and mixing parameters of quarks and leptons. In the lepton sector, the model is chosen as same as the one which has been described in section~\ref{sec:lepton_models}. The assignments of quarks are
\begin{eqnarray}\nonumber
    &&Q_{D}\equiv (Q_{1},Q_{2})^{T}\sim \mathbf{2}\,,~~Q_{3}\sim \mathbf{1'}\,,~~U_{D}^{c}\equiv (u^{c},c^{c})\sim \mathbf{\widehat{2}'}\,,\\
\nonumber
    &&t^{c}\sim \mathbf{1'}\,,~~D_{D}^{c}\equiv (d^{c},s^{c}) \sim \mathbf{2}\,,~~b^{c}\sim \mathbf{1'}\,,\\
  && k_{Q_{D}}=3-k_{U_{D}^{c}}=k_{Q_{3}}=6-k_{t^{c}}=6-k_{D_{D}^{c}}=-k_{b^{c}}\,.
\end{eqnarray}
Then the Yukawa superpotentials for the quark masses are
\begin{eqnarray}\nonumber
  \mathcal{W}_{u}&=&g^{u}_{1}\left( U_{D}^{c}Q_{D} \right)_{\widehat{\mathbf{4}}}Y_{\widehat{\mathbf{4}}}^{(3)}H_{u} + g^{u}_{2}\left( t^{c}Q_{D}\right)_{\mathbf{2}}Y^{(6)}_{\mathbf{2}}H_{u}+ g^{u}_{3}\left( t^{c}Q_{3}\right)_{\mathbf{1}}Y^{(6)}_{\mathbf{1}}H_{u}\,,\\ \nonumber
  \mathcal{W}_{d}&=&g^{d}_{1}\left( D_{D}^{c}Q_{D} \right)_{\mathbf{1}}Y^{(6)}_{\mathbf{1}}H_{d}+g^{d}_{2}\left( D_{D}^{c}Q_{D} \right)_{\mathbf{1'}}Y^{(6)}_{\mathbf{1'}}H_{d}+g^{d}_{3}\left( D_{D}^{c}Q_{D} \right)_{\mathbf{2}}Y^{(6)}_{\mathbf{2}}H_{d}\\
  &~&+g^{d}_{4}\left( D_{D}^{c}Q_{3} \right)_{\mathbf{2}}Y^{(6)}_{\mathbf{2}}H_{d}+g^{d}_{5}\left( b^{c}Q_{3} \right)_{\mathbf{1}}H_{d} \,.
\end{eqnarray}
After symmetry breaking, the quark mass matrices take following forms
\begin{eqnarray}\nonumber
    M_{u}&=&\left(
\begin{array}{ccc}
 g^{u}_{1}Y_{\widehat{\mathbf{4}},3}^{(3)} & -g^{u}_{1}Y_{\widehat{\mathbf{4}},2}^{(3)} & 0 \\
 g^{u}_{1}Y_{\widehat{\mathbf{4}},4}^{(3)} & g^{u}_{1}Y_{\widehat{\mathbf{4}},1}^{(3)} & 0 \\
 -g^{u}_{2} Y_{\mathbf{2,2}}^{(6)} & g^{u}_{2} Y_{\mathbf{2,1}}^{(6)} & g^{u}_{3} Y_{\mathbf{1}}^{(6)} \\
\end{array}
\right)v_{u} \,,\\
   M_{d}&=&\left(
\begin{array}{ccc}
 g^{d}_{1}Y_{\mathbf{1}}^{(6)}-g^{d}_{3} Y_{\mathbf{2,1}}^{(6)} & g^{d}_{2} Y_{\mathbf{1'}}^{(6)}+g^{d}_{3} Y_{\mathbf{2,2}}^{(6)} & -g^{d}_{4} Y_{\mathbf{2,2}}^{(6)} \\
 g^{d}_{3} Y_{\mathbf{2,2}}^{(6)}-g^{d}_{2} Y_{\mathbf{1'}}^{(6)} & g^{d}_{3} Y_{\mathbf{2,1}}^{(6)}+g^{d}_{1}Y_{\mathbf{1}}^{(6)} & g^{d}_{4} Y_{\mathbf{2,1}}^{(6)} \\
 0 & 0 & g^{d}_{5} \\
\end{array}
\right)v_{d} \,.
\end{eqnarray}
By performing a $\chi^{2}$ analysis of the lepton sector and quark sector simultaneously, we find the best fit values of the input parameters are
\begin{eqnarray}\nonumber
  &&\braket{\tau}=-0.19991 + 1.07381i\,,\quad g_{2}^{E}/g_{1}^{E}=0.69822\,,\quad g_{3}^{E}/g_{1}^{E}=86.19360\,, \\
\nonumber&& g_{1}^{E}v_{d}=0.07132~\text{GeV}\,,\quad \frac{\left(g^{D}v_{u}\right)^{2}}{\Lambda}=30.27760~\text{meV}\,.\\
  \nonumber&& g_{2}^{u}/g_{1}^{u}=122.34700\,,\quad g_{3}^{u}/g_{1}^{u}=0.49606\,,\quad g_{2}^{d}/g_{1}^{d}=45.08310\,,\\
  \nonumber&&g_{3}^{d}/g_{1}^{d}=1.15319\,,\quad g_{4}^{d}/g_{1}^{d}=0.01338\,,\quad g_{5}^{d}/g_{1}^{d}=0.00362\,,\\
  &&g_{1}^{u}v_{u}=0.58474~\text{GeV}\,,\quad g_{1}^{d}v_{d}=0.29190~\text{GeV}\,.
\end{eqnarray}
With the above given values of input parameters, the predictions of fermion observables are
\begin{eqnarray}
\nonumber&& \sin^{2}\theta^{l}_{12}=0.3409\,,\quad \sin^{2}\theta^{l}_{13}=0.02271\,,\quad \sin^{2}\theta^{l}_{23}=0.5079\,,\quad \delta_{CP}^{l}=244.3^{\circ}\,,\\
\nonumber&& \alpha_{21}=1.3465 \pi\,,\quad \alpha_{31}=0.5694 \pi\,,\quad m_e/m_{\mu}=0.00474\,,\quad m_{\mu}/m_{\tau}=0.0588\,,\quad \\
  \nonumber && \frac{\Delta m_{21}^{2}}{\Delta m_{31}^{2}} = 0.02909\,,\quad m_1=15.00~\text{meV}\,,\quad m_2=17.30~\text{meV}\,,\quad m_3=52.69~\text{meV}\,,\quad \\
\nonumber&& m_{\beta\beta}=9.73~\text{meV}\,,\quad \theta^{q}_{12}=0.227\,,\quad \theta^{q}_{13}=0.00350\,,\quad \theta^{q}_{23}=0.0389\,,\quad \delta^{q}_{CP}=71.08^{\circ}\,,\\
&& m_{u}/m_{c}=0.00212\,,\quad m_{c}/m_{t}=0.00281\,,\quad m_{d}/m_{s}=0.0503\,,\quad m_{s}/m_{b}=0.0212\,,
\end{eqnarray}
with $\chi^{2}_{l}=20.0$, $\chi^{2}_{q}=12.6$ and $\chi^{2}_{\text{total}}=\chi^{2}_{l}+\chi^{2}_{q}=32.6$, being all observables in the experimentally allowed $3\sigma$ intervals.

In section~\ref{sec:lepton_models}, we have scanned the parameter space of lepton sector. Similarly, we scan the parameter space of the quark sector independently, and all the quark observables are constrained in the experimentally preferred $3\sigma$ regions.
\begin{figure}[t!]
\centering
\includegraphics[width=6.5in]{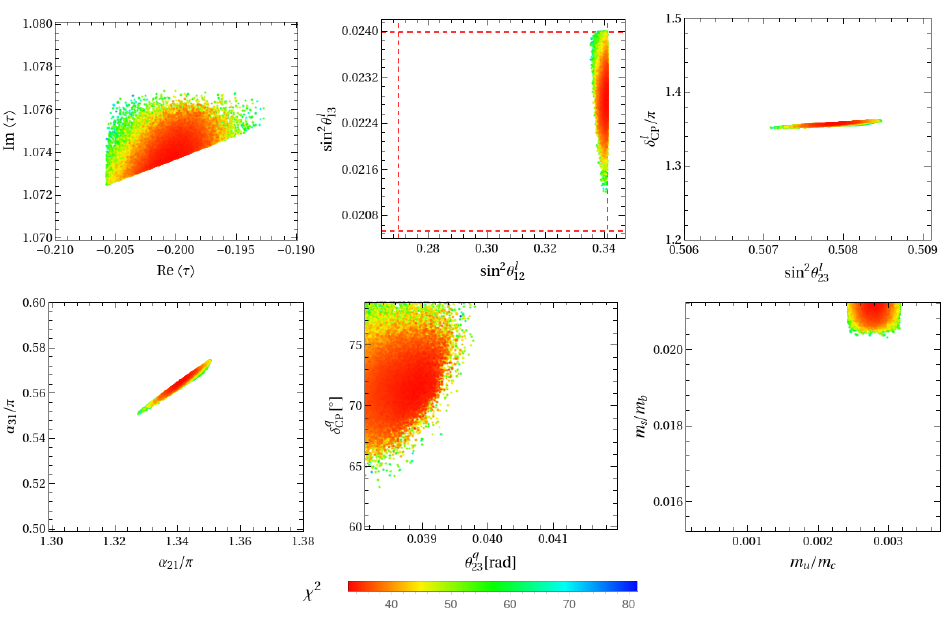}
\caption{\label{fig:model-lepton-quark-14par} The predictions for the correlations among the input parameters and the quark observables in the unified model. }
\end{figure}
We find that all quark observables can approximately take any values within their allowed $3\sigma$ ranges. The phenomenologically viable regions of $\tau$ obtained from quark and lepton sectors have an overlap, which is shown in figure~\ref{fig:model-lepton-quark-14par}. If we include the experimental constraints of both quarks and leptons, the allowed values of some lepton and quark observables can be constrained in narrower ranges as follow,
\begin{eqnarray}\nonumber
  &&\sin^{2}\theta^{l}_{12}\in [0.3353,0.3410]\,,\quad \sin^{2}\theta^{l}_{13}\in [0.02120,0.02398]\,,\quad \sin^{2}\theta^{l}_{23}\in [0.5071,0.5084]\,,\\ \nonumber
  &&\delta^{l}_{CP}\in [1.351\pi,1.361\pi]\,,\quad \alpha_{21}\in [1.328\pi,1.350\pi]\,,\quad \alpha_{31}\in [0.551\pi,0.574\pi]\,,\\ \nonumber
  &&m_{1}\in [14.66\text{meV},15.04\text{meV}]\,,\quad \sum_{i}m_{i}\in[81.98\text{meV},86.83\text{meV}]\,,\\ \nonumber
  &&m_{\beta\beta}\in [9.47\text{meV},9.90\text{meV}]\,,\quad \theta^{q}_{23}\in [0.0381,0.0398]\,,\quad \delta_{CP}^{q}\in[63.31^{\circ},78.55^{\circ}]\,,\\
  &&m_{u}/m_{c}\in [0.00184,0.00240]\,,\quad m_{s}/m_{b}\in [0.0203,0.0212]\,.
\end{eqnarray}
The correlation among the fermion observables and the allowed region of $\tau$ are plotted in figure~\ref{fig:model-lepton-quark-14par}.

\section{\label{sec:conclusion}Conclusion}

The origin of the flavor structure of quarks and leptons is a big puzzle in particle physics, and modular symmetry is a promising approach to address this flavor puzzle. The inhomogeneous and homogeneous finite modular groups $\Gamma_N$ and $\Gamma^{\prime}_N$ with $N\leq7$ have been intensively studied in the flavor models for quarks and leptons. It is intriguing that $\Gamma_3\cong A_4$, $\Gamma_4\cong S_4$ and $\Gamma_5\cong A_5$ are the symmetry groups of the platonic solids tetrahedron, octahedron (hexahedron), icosahedron (dodecahedron) respectively. Moreover, the homogeneous finite modular groups $\Gamma'_3\cong T'$ and $\Gamma'_5$ are isomorphic to the binary tetrahedral group and the binary icosahedral group respectively. It turns out that the binary octahedral group $2O$ is also a finite modular symmetry group~\cite{Liu:2021gwa}, while it hasn't been studied in the literature so far.

In the present work, we have performed a systematic analysis of how the flavor structure of quarks and leptons can be explained in the framework of $2O$ modular symmetry with gCP. The representation matrices of both generators $S$ and $T$ are unitary and symmetric in our working basis, consequently the gCP symmetry reduces to the traditional CP symmetry and all the coupling constants are constrained to be real. We have considered all possible representation assignments of the matter fields under $2O$, and the Higgs doublets $H_{u}$ and $H_{d}$ are assumed to be trivial singlets with vanishing modular weight. For each independent representation assignment $(\rho_{\psi^{c}},\rho_{\psi})$ of Dirac fermions, the resulting sub-matrices of the fermion mass matrices are summarized in table~\ref{tab:fermion_submatrix}. Hence the explicit form of the fermion mass matrix can be straightforwardly read out from table~\ref{tab:fermion_submatrix} once the modular weights and the transformations of quark/lepton fields under $2O$ are specified. We assume that the neutrinos are Majorana particles and two distinct neutrino mass generation mechanisms including the effective Weinberg operator and the type-I seesaw mechanism are investigated. The Majorana neutrino mass matrix can also be obtained from table~\ref{tab:fermion_submatrix} with $\rho_{\psi}=\rho_{\psi^{c}}=\rho_{L/N^{c}}$.

In bottom-up approach of modular symmetry, one can freely assign both modular weights and representation of matter fields, therefore many possible models can be constructed. We intend to understand how the modular symmetry $2O$ can help to understand the flavor structure of quarks and leptons, and we aim to find out the phenomenologically viable models with the minimum number of free parameters. Requiring the number of the free real parameters in lepton sector is less than 10, we have obtained thousands of lepton models for both cases of neutrino masses generated by the Weinberg operator and the type-I see-saw mechanism. In quark sector, we demand the number of the free real parameters less than 11 and over thirty thousands quark models with the modular symmetry $2O$ are found. In order to quantitatively estimate which models can accommodate experimental data, we have performed a thorough numerical analysis of them. We find hundreds of lepton models can explain the experimental data of lepton masses and the lepton mixing parameters. In the case of neutrino masses generated via the Weinberg operator, the phenomenologically viable lepton models contain at least $7$ free real parameters including the complex modulus $\tau$. For the type-I see-saw case, we find a minimal lepton model which can use only 6 real parameters to explain all measured lepton observables. To illustrate our findings, we presented two benchmark lepton models in section~\ref{sec:lepton_models}. The minimal phenomenologically viable lepton model with 6 real parameters is given in section~\ref{sec:lepton_model_6para_SS}, the best fit values of the input parameters and the corresponding predictions for neutrino masses and mixing parameters are give in Eq.~\eqref{eq:bf_lepton_6para} and Eq.~\eqref{eq:ob_lepton_6para} respectively. In section~\ref{sec:lepton_model_7para_WO}, we give a viable lepton model in the case of neutrino masses generated via the Weinberg operator, the best fit results and the interesting graphical correlations between lepton observables are provided. In quark sector, the results of $\chi^{2}$ analysis shows that thousands of quark models can accommodate the experimental data of quark masses, mixing angles and CP violation phase at $3\sigma$ confidence level. The minimal viable quark models contain $9$ free parameters, and one representative quark model with $9$ parameters is provided in section~\ref{sec:quark_model_9para} in detail. We have also presented a quark model with $8$ parameters in section~\ref{sec:quark_model_8para} as a good leading order approximation. This model can accommodate all the quark data except that the mass ratio $m_{s}/m_{b}$ is slightly larger. Furthermore, we have investigated whether the flavor structures of quark and lepton can be simultaneously described by the $2O$ modular symmetry with a common modulus. We give one predictive example of lepton-quark unification models in which the 22 masses and mixing parameters of quarks and leptons are described in terms of 12 free real coupling constants and a common complex modulus, see section~\ref{sec:unified_model}. This is the minimal modular invariant model for quarks and leptons with the smallest number of free parameters in the literature, as far as we know.

In the present work, we have analyzed quark and lepton models with the modular symmetry $2O$ in the limit of exact supersymmetry. The effects of supersymmetry breaking could lead to dependence of our results on other additional parameters such as the supersymmetry breaking scale $m_{\text{SUSY}}$, the messenger scale $M$, the cutoff scale $\Lambda$ and $\tan\beta$. Dimensional analysis shows that supersymmetry breaking contributions to Yukawa couplings and neutrino masses scale are $m_{\text{SUSY}}/M$~\cite{Criado:2018thu}. Hence the corrections induced by the SUSY breaking terms are negligible if the separation between $m_{\text{SUSY}}$ and $M$ is sufficiently large. This mild requirement can be satisfied for a very wide range of the effective SUSY breaking scale $m_{\text{SUSY}}$.

In the top-down approach of modular symmetry, it is found that the modular symmetry is always accompanied by a traditional flavor group. The nontrivial product of modular and traditional flavor symmetries gives rise to the eclectic flavor group~\cite{Nilles:2020nnc,Nilles:2020kgo}.  The eclectic flavor group is more predictive than the consideration of modular symmetry or traditional flavor symmetry alone. Terms allowed by the modular symmetry could be forbidden by the traditional flavor symmetry. In particular, the interplay of traditional flavor symmetry and modular symmetry can restrict the K\"ahler potential so that the minimal K\"ahler potential could be leading order term~\cite{Chen:2021prl,Baur:2022hma,Ding:2023ynd}. The eclectic extension of the traditional flavor group is limited to the finite modular groups $\Gamma_N$ and $\Gamma'_N$ in the present scheme of eclectic flavor symmetry. It is known that the eclectic flavor group $\Omega(1)$ is a nontrivial combination of the traditional
flavor group $\Delta(54)$ (or $\Delta(27)$) and the finite modular group $\Gamma'_3\cong T'$~\cite{Nilles:2020nnc}, the modular binary octahedral group $2O$ potentially gives rise to new candidate of eclectic flavor group by combining with certain traditional flavor symmetry. This is left for future work.

In conclusion, the modular binary octahedral group $2O$ provides a simple and economical framework to understand the flavor structures of the quarks and leptons. The minimal models identified in above are very predictive and they are expected to be tested at future neutrino oscillation facilities and $0\nu\beta\beta$-decay experiments. The modular symmetry $2O$ also opens up new possibility for the eclectic flavor group.

\subsection*{Acknowledgements}
GJD is supported by the National Natural Science Foundation of China under Grant Nos.~11975224, 11835013. XGL is supported by U.S. National Science Foundation under Grant No.~PHY-2210283. JNL is supported by the Grants No. NSFC-12147110 and the China Post-doctoral Science Foundation under Grant No. 2021M70. GJD would like to thank the School of Physics and Astronomy at the
University of Southampton for hospitality.

\newpage
\section*{Appendix}

\setcounter{equation}{0}
\renewcommand{\theequation}{\thesection.\arabic{equation}}

\begin{appendix}

\section{\label{app:VVMFs}Some results of VVMFs}
\subsection{\label{app:123dVVMF}VVMFs of minimal weight in $d\leq 3$ dimension}
In this section, we briefly describe the general results about 1-d, 2-d, and 3-d VVMFs of $\mathrm{SL}(2,\mathbb{Z})$~\cite{Liu:2021gwa}. They can immediately be used to construct VVMFs in the $1$ to $3$ dimensional irreps of group $2O$.

The MLDE of Eq.~\eqref{eq:MLDE} satisfied by 1-d VVMFs can be solved by an even power of the eta function. Consequently, all the one-dimensional VVMFs of minimal weight $k_0$ can be expressed as the following eta products
\begin{equation}
\mathbf{1}_{p}: \qquad Y(\tau)=\eta^{2k_0}(\tau)\,,
\end{equation}
where $p=0,1,\dots,11$ denotes the twelve 1-d irreps of $\mathrm{SL}(2,\mathbb{Z})$, and the minimal weight $k_0=p=\frac{6}{i\pi}\log\rho_{\mathbf{1}_{p}}(T)$.

The MLDE of Eq.~\eqref{eq:MLDE} satisfied by 2-d or 3-d VVMFs can be always transformed into the generalized hypergeometric equation. Thus, all the 2-d or 3-d VVMFs of minimal weight $k_0$ can be expressed as the generalized hypergeometric series:
\begin{equation}
Y(\tau)=\left(Y_1(\tau)\,,\dots \,, Y_n(\tau)\right)^T\,,
\end{equation}
where the component is~\footnote{It is important to note that in order to align with the corresponding modular transformation matrix $\rho(S)$, each component here needs to be accompanied by an appropriate constant factor.}
\begin{equation}
\label{eq:solHyperEq}
Y_i(\tau)=\eta^{2k_0}\left(\frac{K}{1728}\right)^{1-\beta_i} {}_nF_{n-1}(1+\alpha_1-\beta_i,\dots,1+\alpha_n-\beta_i;1+\beta_1-\beta_i,\check{\dots} ,1+\beta_n-\beta_i; K)\,,
\end{equation}
where $\check{}$ denotes omission of $1+\beta_i-\beta_i$, and the generalized hypergeometric series ${}_nF_{n-1}$ is defined by
\begin{equation}
{}_nF_{n-1}(a_1,\dots,a_n;b_1,\dots, b_{n-1}; z) = \sum_{m \geq 0}^{\infty} \dfrac{\prod_{j=1}^{n}a^{(m)}_j}{\prod_{k=1}^{n-1}b^{(m)}_k}\dfrac{z^m}{m!}\,,
\label{eq:HyperSeries}
\end{equation}
with $m\in \mathbb{N}$, $a,b \in \mathbb{C}$ and notation $a^{(m)}$ represents the rising factorial (also known as Pochhammer symbol). The function $K$ is the inverse of Klein $j$-invariant: $K(\tau)=1728/j(\tau)$.~\footnote{Pochhammer symbol $a^{(m)}$ is defined as $a^{(m)}=a(a+1)\dots(a+m-1)$ for $m\geq 1$, and $a^{(0)}=1$. The Klein $j$-invariant can be expressed by Eisenstein series as $j(\tau)=1728E^3_4(\tau)/(E_4^3(\tau)-E_6^2(\tau))$.}

For the two-dimensional case, $n=2, i=1,2$, the parameters $\beta_{1,2}$, $\alpha_{1,2}$ and minimal weight $k_0$ solely rely on the eigenvalues $e^{2\pi i r_{1,2}}$ of the corresponding 2-d irrep matrix $\rho(T)$:
\begin{equation}
\beta_1=\frac{r_2-r_1}{2}+\frac{11}{12}\,,\quad \beta_2=\frac{r_1-r_2}{2}+\frac{11}{12}\,,\quad
\alpha_1=0\,,\quad \alpha_2=\frac{1}{3}\,,\quad k_0=6(r_1+r_2)-1\,.
\end{equation}

For the three-dimensional case, $n=3, i=1,2,3$, the parameter $\beta_{1,2,3}$, $\alpha_{1,2,3}$ and minimal weight $k_0$ also depend only on the eigenvalues $e^{2\pi i r_{1,2,3}}$ of the corresponding 3-d irrep matrix $\rho(T)$:
\begin{align}
\nonumber&
\beta_1=\frac{5}{6}-\frac{1}{6}(4r_1-2r_2-2r_3)\,,~
\beta_2=\frac{5}{6}-\frac{1}{6}(4r_2-2r_3-2r_1)\,,~
\beta_3=\frac{5}{6}-\frac{1}{6}(4r_3-2r_1-2r_2)\,,\\
&\alpha_1=0\,,\quad \alpha_2=\frac{1}{3}\,,\quad \alpha_2=\frac{2}{3}\,,\quad k_0=4(r_1+r_2+r_3)-2\,.
\end{align}
\subsection{\label{app:4dVVMF}4-d VVMF of minimal weight for $2O$}
In this section, we provide a brief description of the construction of the 4-d VVMF in the irrep $\widehat{\mathbf{4}}$ of binary octahedral group $2O$. From the 4-d irrep matrix $\rho_{\widehat{\mathbf{4}}}(T)$ as depicted in Eq.~\eqref{eq:irr_reps}, we can determine the exponent $r_i$ associated with its eigenvalues $e^{2\pi i r_i}$:
\begin{equation}
(r_1,r_2,r_3,r_4)=(5/8,~3/8,~7/8,~1/8)\,.
\end{equation}
Because $\rho(S^2)=-1=-(-1)^{3(r_1+r_2+r_3+r_4)\mod 2}$, this 4-d irrep is the cyclic type~\cite{Liu:2021gwa}. Specifically, the VVMF module $\mathcal{M}(\widehat{\mathbf{4}})$ has the following basis
\begin{equation}
\mathcal{M}(\widehat{\mathbf{4}})=\langle F, ~D_{k_0}F,~D_{k_0}^2F,~D_{k_0}^3F\rangle\,.
\end{equation}
Here, $F$ is the 4-d VVMF of minimal weight $k_0$ where $k_0=3(r_1+r_2+r_3+r_4)-3=3$.
Consequently, we can express the VVMF $D_{k_0}^4 F$ of weight $k_0+8$ as a linear combination of these bases over the ring $\mathbb{C}[E_4,E_6]$:
\begin{equation}
\label{eq:4dMLDE}
\left(D^4_{k_0}+aE_4D^2_{k_0}+bE_6D_{k_0}+cE^2_4\right)F=0\,.
\end{equation}
For convenience, we simplify the above weight-$k_0$ MLDE by transforming it into a weight-zero MLDE through the rescaling of $F$ to $\widetilde{F}=\eta^{-2k_0}F$:
\begin{equation}
\label{eq:Reduced4dMLDE}
\left(D^4_{0}+aE_4D^2_{0}+bE_6D_{0}+cE^2_4\right)\widetilde{F}=0\,.
\end{equation}
As shown in Ref.~\cite{franc2016hypergeometric}, by introducing new variable $K(\tau)$ and notation $\theta_K$:
\begin{align}
K(\tau)=1728/j(\tau)\,,~~~~\qquad \theta_K=K\frac{d}{dK}\,,
\end{align}
the weight-zero MLDE mentioned above can be transformed into the following form
{\scriptsize
\begin{equation}
\label{eq:New4dMLDE}
\left(\theta_K^4 -\frac{2K+1}{1-K}\theta_K^3+\frac{44K^2-(36a+28)K+36a+11}{36(1-K)^2} \theta_K^2 +\frac{8K^2-(12a+36b+4)K-6a+36b-1}{36(1-K)^2}\theta_K+\frac{c}{(1-K)^2}\right)\widetilde{F}=0\,.
\end{equation}
}
From the $q$-expansion of $K(\tau)$: $K=1728q(1-744q+\dots)$, we can deduce the $K$-expansion of $\widetilde{F}$:
\begin{equation}
\label{eq:K-series}
\widetilde{F}=\eta^{-2k_0}F=\begin{pmatrix}
K^{r_1-\frac{k_0}{12}}\sum_{n\geq 0}a_1(n)K^n \\
K^{r_2-\frac{k_0}{12}}\sum_{n\geq 0}a_2(n)K^n \\
K^{r_3-\frac{k_0}{12}}\sum_{n\geq 0}a_3(n)K^n \\
K^{r_4-\frac{k_0}{12}}\sum_{n\geq 0}a_4(n)K^n
\end{pmatrix}\,.
\end{equation}
By substituting it into Equation~\eqref{eq:New4dMLDE} mentioned earlier, and considering only the leading-order terms of the series $K$, we can derive the indicial equation near $K=0$:
\begin{equation}
r^4-r^3+(a+11/36)r^2-(a/6-b+1/36)r+c=\prod_{j=1}^4(r-r'_j)=0\,,
\end{equation}
with
\begin{equation}
\label{eq:rprime}
r'_1=r_1-\frac{k_0}{12}=\frac{3}{8},~~r'_2=r_2-\frac{k_0}{12}=\frac{1}{8},~~r'_3=r_3-\frac{k_0}{12}=\frac{5}{8},~~r'_4=r_4-\frac{k_0}{12}=-\frac{1}{8}.
\end{equation}
Consequently, we can determine the values of the unknown parameters $a,b,c$ as follows:
\begin{equation}
a=-25/288\,,\quad b=25/864\,,\quad c=-15/4096\,.
\end{equation}
So Eq.~\eqref{eq:New4dMLDE} becomes
{\small
\begin{align}
\nonumber
&\Big[\left(K^4-2K^5+K^6\right)\frac{d^4}{d K^4}+\left(5K^3-13K^4+8K^5\right)\frac{d^3}{d K^3}+\left(\frac{135}{32}K^2-\frac{5095}{288}K^3+\frac{128}{9}K^4\right)\frac{d^2}{d K^2} \\
&~+\left(\frac{15}{64}K-\frac{365}{96}K^2+\frac{40}{9}K^3\right)\frac{d}{d K}-\frac{15}{4096}\Big]\widetilde{F}=0\,.
\end{align}}
Regrettably, it is not possible to translate it into the generalized hypergeometric equation. Therefore, we can only solve it by the power series method and get the $K$-series for $\widetilde{F}$.
More specifically, we plug the Eq.~\eqref{eq:K-series} into the aforementioned differential equation, and proceed to compare the terms of each order of $K$. This process yields the following recursive formulas for the coefficient $a_i(n)$:
\begin{align}
\label{eq:Recursion}
\nonumber
0&=\left[ (r'_i+m)_4 + 5(r'_i+m)_3 +\frac{135}{32}(r'_i+m)_2 +\frac{15}{64}(r'_i+m)_1 -\frac{15}{4096} \right] a_i(m) \\
\nonumber
&-\left[2(r'_i+m-1)_4+13(r'_i+m-1)_3 + \frac{5096}{288}(r'_i+m-1)_2+\frac{365}{96}(r'_i+m-1)_1 \right]a_i(m-1) \\
&+\left[(r'_i+m-2)_4+8(r'_i+m-2)_3+\frac{128}{9}(r'_i+m-2)_2+\frac{40}{9}(r'_i+m-2)_1\right]a_i(m-2)
\end{align}
for $m\geq2$, and
\begin{align}
\label{eq:1orderRecursion}
\nonumber
0&=\left[ (r'_i+1)_4 + 5(r'_i+1)_3 +\frac{135}{32}(r'_i+1)_2 +\frac{15}{64}(r'_i+1)_1 -\frac{15}{4096} \right] a_i(1) \\
&-\left[2(r'_i)_4+13(r'_i)_3 + \frac{5096}{288}(r'_i)_2+\frac{365}{96}(r'_i)_1 \right]a_i(0)
\end{align}
for $m=1$, and
\begin{align}
\label{eq:0orderRecursion}
0&=\left[ (r'_i)_4 + 5(r'_i)_3 +\frac{135}{32}(r'_i)_2 +\frac{15}{64}(r'_i)_1 -\frac{15}{4096} \right] a_i(0)
\end{align}
for $m=0$.
Here $r'_i$ can be found in Eq.~\eqref{eq:rprime}, and the falling factorial $(x)_{s}$ is defined by $(x)_{s}=x(x-1)\dots(x-s+1)$ for $s\geq 1$ and $(x)_{0}=1$.

Notice that the first recursion Eq.~\eqref{eq:0orderRecursion} is automatically satisfied as we expected, so we can take $a_i(0)=(1/1728)^{r'_i}$ for convenience, then from second recursion Eq.~\eqref{eq:1orderRecursion} we can obtain $a_i(1)$. Subsequently, by employing Eq.~\eqref{eq:Recursion}, we can iteratively obtain all the values of $a_i(m)$ for $m\geq 2$.

Finally, we change the variable of the $K$-series solution of $F$ from $K$ to $q=e^{2\pi i\tau}$, and subsequently revert $\widetilde{F}$ back to $F$. The $q$-expansions of $F_i$ are given by
\begin{align}
\nonumber
F_1(\tau)&=q^{5/8} \left(1-q-4 q^2+3 q^3+q^4+3 q^5+13 q^6-13 q^7-12 q^8-4 q^9 +\dots\right) \,,\\
\nonumber
F_2(\tau)&=q^{3/8} \left(1+q-7 q^2-6 q^3+16q^4+9 q^5-6 q^6+9 q^7-23 q^8-23 q^9+\dots\right) \,,\\
\nonumber
F_3(\tau)&=q^{7/8} \left(1-3q+ 3q^2-4q^3+3q^4+6q^5-3q^6+3q^7-15q^8+2q^9+\dots\right)\,, \\
F_4(\tau)&=q^{1/8} \left(1+3q-6q^2-23q^3+12q^4+66q^5-15q^6-84q^7+48q^8+58q^9+\dots\right) \,.
\end{align}
From the representation matrix $\rho_{\widehat{\mathbf{4}}}(S)$ as shown in Eq.~\eqref{eq:irr_reps}, we can organize these components into a VVMF with respect to irrep $\widehat{\mathbf{4}}$ of $2O$:
\begin{equation}
Y^{(3)}_{\widehat{\mathbf{4}}}(\tau) =(4\sqrt{3}F_1(\tau)\,,~~ 2\sqrt{3}F_2(\tau)\,,~~ -8F_3(\tau)\,,~~ F_4(\tau))^T\,.
\end{equation}

\section{\label{app:2O_group}Binary octahedral group $2O$}
The binary octahedral group $2O$ is the preimage of the octahedral group $O\cong S_4$ under the $2:1$ covering homomorphism $\mathrm{SU}(2) \to \mathrm{SO}(3)$. Alternatively, it can be understood as the Schur cover of the permutation group $S_4$ of ``$-$'' type. The group $2O$ can be generated by the modular generators $S$ and $T$ satisfying the relations~\cite{Liu:2021gwa}:
\begin{equation}
\label{eq:2Ogroup}
S^2=T^4=R,~~~ RT=TR,~~~R^2=(ST)^3=1\,,
\end{equation}
or equivalently
\begin{equation}
\label{eq:2Ogroup}
S^4=(ST)^3=S^2T^4=1,\quad S^2T=TS^2\,.
\end{equation}
It has 48 elements and the group ID is $[48,28]$ in \texttt{GAP}~\cite{GAP}. The cyclic group $Z^{R}_2\equiv\left\{1, R\right\}$ is the center. Notice that $2O$ is not a semidirect product of $Z^{R}_2$ by $S_4$, and there is no subgroup of $2O$ isomorphic to $S_4$. Interestingly, however, it contains the binary tetrahedral group $2T$ and the quaternion group $Q_8$ as its normal subgroups.

In addition to the representations of $S_4$: $\bm{1}$, $\bm{1}'$, $\bm{2}$, $\bm{3}$ and $\bm{3}'$, $2O$ possesses two doublet representations $\widehat{\bm{2}},\widehat{\bm{2}}'$ and one quartet representation $\widehat{\bm{4}}$. The explicit forms of the representation matrices $\rho(S)$ and $\rho(T)$ in each of the irreps are given in the following,
\begin{eqnarray}
\label{eq:irr_reps} \begin{array}{ccc}
\mathbf{1:} ~& S=1, ~&~ T=1\,, \\
\mathbf{1}':~& S=-1, ~&~ T=-1\,, \\
\mathbf{2:} ~&~ S=-\frac{1}{2}
\begin{pmatrix}
1  ~&~  \sqrt{3}  \\
\sqrt{3}  ~&~  -1
\end{pmatrix}, & T=\left(\begin{array}{cc}
1 ~&~ 0 \\
0 ~&~ -1 \\
\end{array}\right) \,,\\
\mathbf{\widehat{2}}: ~&~ S=\frac{i}{\sqrt{2}}
\begin{pmatrix}
1  ~&~  1  \\
1  ~&~  -1
\end{pmatrix}, & T=\left(\begin{array}{cc}
\xi^{3} ~&~ 0 \\
0 ~&~ \xi^{5} \\
\end{array}\right) \,,\\
\mathbf{\widehat{2}}': ~&~ S=\frac{i}{\sqrt{2}}
\begin{pmatrix}
1  ~&~  1  \\
1  ~&~  -1
\end{pmatrix}, & T=\left(\begin{array}{cc}
\xi ~&~ 0 \\
0 ~&~ \xi^{7} \\
\end{array}\right) \,,\\
\mathbf{3:} ~& S=\frac{1}{2}\left(\begin{array}{ccc}
0~&\sqrt{2}~&\sqrt{2}\\
\sqrt{2}~&-1~&1\\
\sqrt{2}~&1~&-1
\end{array}\right),
 ~&~ T=\left(
\begin{array}{ccc}
1~&0~&0\\
0~&\xi^{2}~&0\\
0~&0~&\xi^6
\end{array}\right) \,,\\
\mathbf{3'}: ~& S=-\frac{1}{2}\left(\begin{array}{ccc}
0~&\sqrt{2}~&\sqrt{2}\\
\sqrt{2}~&-1~&1\\
\sqrt{2}~&1~&-1
\end{array}\right),
 ~&~ T=-\left(
\begin{array}{ccc}
1~&0~&0\\
0~&\xi^{2}~&0\\
0~&0~&\xi^6
\end{array}\right) \,,\\
\mathbf{\widehat{4}}: ~& S=\dfrac{i}{2 \sqrt{2}}
\left(\begin{array}{cccc}
1 ~& 1 ~& \sqrt{3} ~& -\sqrt{3} \\
1 ~& -1 ~& -\sqrt{3} ~& -\sqrt{3} \\
\sqrt{3} ~& -\sqrt{3} ~& 1 ~& 1 \\
-\sqrt{3} ~& -\sqrt{3} ~& 1 ~& -1 \\
\end{array}\right),
 ~&~ T=\left(
\begin{array}{cccc}
\xi^{5}~&0~&0&~0\\
0~&\xi^{3}~&0&~0\\
0~&0~&\xi^7&~0\\
0~&0~&0&~\xi\\
\end{array}\right) \,,
\end{array}
\end{eqnarray}
with $\xi=e^{i\pi/4}$. Notice that the two 1-d irreps $\bm{1}$ and $\bm{1}'$ correspond to $\mathbf{1}_0$ and $\mathbf{1}_6$ of $\mathrm{SL}(2,\mathbb{Z})$, these three 2-d irreps $\mathbf{2}$, $\widehat{\mathbf{2}}$ and $\widehat{\mathbf{2}}'$ are the doublet representations $\mathbf{2}_{(0,\frac{1}{2})}$, $\mathbf{2}_{(\frac{3}{8},\frac{5}{8})}$ and $\mathbf{2}_{(\frac{1}{8},\frac{7}{8})}$ of $\mathrm{SL}(2,\mathbb{Z})$, and the two 3-d irreps $\mathbf{3}$ and $\mathbf{3}'$ correspond to $\mathbf{3}_{(0,\frac{1}{4}, \frac{3}{4})}$ and $\mathbf{3}_{(\frac{1}{2}, \frac{3}{4}, \frac{1}{4})}$ of $\mathrm{SL}(2,\mathbb{Z})$. We see the representation matrix $\rho(R)=1$ in the unhatted irreps $\bm{1}$, $\bm{1}'$, $\bm{2}$, $\bm{3}, \bm{3}'$ and $\rho(R)=-1$ in the hatted irreps $\widehat{\bm{2}}$, $\widehat{\bm{2}}'$, $\widehat{\bm{4}}$. Therefore, it is impossible to differentiate between the group $2O$ and $S_4$ when considering the irreps $\bm{1}$, $\bm{1}'$, $\bm{2}$, $\bm{3}$, and $\bm{3}'$. The 48 elements of $2O$ can be divided into 8 conjugacy classes, and their character table is shown in table~\ref{tab:character}, which is obtained by taking the trace of the explicit representation matrices.

\begin{table}[th!]
\begin{center}
\renewcommand{\tabcolsep}{2.8mm}
\renewcommand{\arraystretch}{1.3}
\begin{tabular}{|c|c|c|c|c|c|c|c|c|c|c|c|c|c|c|c|c|c|c|c|c|c|c|}\hline\hline
\text{Classes} & $1C_1$ & $1C_2$ & $6C_8$ & $6C'_8$ & $6C_4$ & $8C_6$ & $8C_3$ & $12C'_4$  \\ \hline
\text{Representative} & $1$ & $S^2$ & $T$ & $ST^2$ & $(ST^2)^2$ & $S^2TS$ & $S^2(ST)^2T^2$ & $S^3T^2ST$ \\ \hline
$\bm{1}$ & $1$ & $1$ & $1$ & $1$ & $1$ & $1$ & $1$ & $1$\\ \hline
$\bm{1'}$ & $1$ & $1$ & $-1$ & $-1$ & $1$ & $1$ & $1$ & $-1$\\ \hline
$\bm{2}$ & $2$ & $2$ & $0$ & $0$ & $2$ & $-1$ & $-1$ & $0$\\ \hline
$\widehat{\bm{2}}$ & $2$ & $-2$ & $-\sqrt{2}$ & $\sqrt{2}$ & $0$ & $1$ & $-1$ & $0$\\ \hline
$\widehat{\bm{2}}'$ & $2$ & $-2$ & $\sqrt{2}$ & $-\sqrt{2}$ & $0$ & $1$ & $-1$ & $0$\\ \hline
$\bm{3}$ & $3$ & $3$ & $1$ & $1$ & $-1$ & $0$ & $0$ & $-1$\\ \hline
$\bm{3'}$ & $3$ & $3$ & $-1$ & $-1$ & $-1$ & $0$ & $0$ & $1$\\ \hline
$\widehat{\bm{4}}$ & $4$ & $-4$ & $0$ & $0$ & $0$ & $-1$ & $1$ & $0$\\ \hline\hline
\end{tabular}
\caption{\label{tab:character} The character table of the binary octahedral group $2O$.}
\end{center}
\end{table}

We present the decompositions of tensor products of different $2O$ irreps and the corresponding Clebsch-Gordan coefficients in our basis. We use $\alpha_i$ to indicate the elements of the first representation of the product and $\beta_i$ to indicate those of the second representation.
The results are summarized in table~\ref{tab:2O_CG-1st}.

\begin{center}
\renewcommand{\arraystretch}{1.15}
\begin{small}
\setlength\LTcapwidth{\textwidth}
\setlength\LTleft{-0.12in}
\setlength\LTright{0pt}
 \begin{longtable}{|c|c|c|c|c|c|c|c|c|c|c|c|}
\caption{\label{tab:2O_CG-1st}
Multiplication rules of different irreps of $2O$ and the corresponding Clebsch-Gordan coefficients. Note that $\mathbf{1}\otimes \mathbf{r}=\mathbf{r}\sim \alpha \beta$.}\\
\midrule
\specialrule{0em}{1.0pt}{1.0pt}
\midrule

\endfirsthead

\midrule
\specialrule{0em}{1.0pt}{1.0pt}
\midrule

\endhead

\midrule
\specialrule{0em}{1.0pt}{1.0pt}
\midrule

\caption[]{continues on next page}\\
\endfoot

\endlastfoot

\multicolumn{3}{|c}{~~~~~~$\bm{1}' \otimes \bm{2} = \bm{2}$~~~~~~} & \multicolumn{3}{|c}{$\bm{1}' \otimes \widehat{\bm{2}} = \widehat{\bm{2}}'$} & \multicolumn{3}{|c}{~~~~~~$\bm{1}' \otimes \widehat{\bm{2}}'
        =\widehat{\bm{2}}$~~~~~~}&\multicolumn{3}{|c|}{$\bm{1}' \otimes \widehat{\bm{4}} = \widehat{\bm{4}}$}  \\
\multicolumn{3}{|c}{$\mathbf{2} \sim    \alpha_1\begin{pmatrix}
                 \beta_2 \\
                - \beta_1 \\
\end{pmatrix} $} &
\multicolumn{3}{|c}{$\widehat{\mathbf{2}}' \sim \alpha_1\begin{pmatrix}
         \beta_2 \\
        - \beta_1 \\
\end{pmatrix}$} &
\multicolumn{3}{|c}{$\widehat{\mathbf{2}} \sim \alpha_1\begin{pmatrix}
                 \beta_2 \\
                -\beta_1\\
        \end{pmatrix}$} &
\multicolumn{3}{|c|}{$\widehat{\mathbf{4}} \sim \alpha_1\begin{pmatrix}
        \beta_4 \\
        - \beta_3 \\
         \beta_2 \\
        - \beta_1 \\
\end{pmatrix}$}  \\ \hline

\multicolumn{6}{|c}{~~~~~~~~~~~~~~~~$\mathbf{1} \otimes \mathbf{3} = \mathbf{1}' \otimes \mathbf{3}' = \mathbf{3}$~~~~~~~~~~~~~~~} & \multicolumn{6}{|c|}{$\mathbf{1} \otimes \mathbf{3}' = \mathbf{1}' \otimes \mathbf{3} = \mathbf{3}'$} \\
\multicolumn{6}{|c}{$\mathbf{3}\sim\alpha_1\begin{pmatrix}
\beta_1 \\
\beta_2 \\
\beta_3 \\
\end{pmatrix} $}&
\multicolumn{6}{|c|}{$ \mathbf{3'}\sim \alpha_1\begin{pmatrix}
\beta_1 \\
\beta_2 \\
\beta_3 \\
\end{pmatrix} $} \\ \hline

\multicolumn{4}{|c}{$\mathbf{2} \otimes \mathbf{2} = \mathbf{1_s} \oplus \mathbf{1'_a} \oplus \mathbf{2_s}$} & \multicolumn{4}{|c}{$\mathbf{2} \otimes \widehat{\mathbf{2}} = \widehat{\mathbf{4}}$} & \multicolumn{4}{|c|}{$\mathbf{2} \otimes \widehat{\mathbf{2}}' = \widehat{\mathbf{4}}$}   \\
\multicolumn{4}{|c}{  $ \begin{array}{l}
\mathbf{1_s}\sim \alpha_1 \beta_1+\alpha_2 \beta_2 \\
\mathbf{1'_a}\sim \alpha_2 \beta_1-\alpha_1 \beta_2 \\
\mathbf{2_s}\sim\begin{pmatrix}
        \alpha_2 \beta_2-\alpha_1 \beta_1 \\
        \alpha_2 \beta_1+\alpha_1 \beta_2 \\
\end{pmatrix} \\
\end{array} $ } &
\multicolumn{4}{|c}{ $\widehat{\mathbf{4}} \sim \begin{pmatrix}
         -\alpha_1 \beta_2 \\
        \alpha_1 \beta_1 \\
        \alpha_2 \beta_1 \\
        \alpha_2 \beta_2 \\
        \end{pmatrix}$} &
\multicolumn{4}{|c|}{ $\widehat{\mathbf{4}} \sim \begin{pmatrix}
        \alpha_2 \beta_1 \\
        \alpha_2 \beta_2 \\
        -\alpha_1 \beta_2 \\
        \alpha_1 \beta_1 \\
        \end{pmatrix} $}  \\ \hline

\multicolumn{4}{|c}{$\widehat{\mathbf{2}} \otimes \widehat{\mathbf{2}} = \mathbf{1_a} \oplus \mathbf{3_s} $} & \multicolumn{4}{|c}{$\widehat{\mathbf{2}} \otimes \widehat{\mathbf{2}}' = \mathbf{1'} \oplus \mathbf{3'}$} & \multicolumn{4}{|c|}{$\widehat{\mathbf{2}}' \otimes \widehat{\mathbf{2}}' = \mathbf{1_a} \oplus \mathbf{3_s}$}  \\
\multicolumn{4}{|c}{  $\begin{array}{l}
\mathbf{1_a}\sim \alpha_2 \beta_1-\alpha_1 \beta_2 \\
\mathbf{3_s}\sim\begin{pmatrix}
\alpha_2 \beta_1+\alpha_1 \beta_2 \\
\sqrt{2} \alpha_2 \beta_2 \\
-\sqrt{2} \alpha_1 \beta_1 \\
\end{pmatrix}\\
\end{array}$ } &
\multicolumn{4}{|c}{ $ \begin{array}{l}
\mathbf{1'}\sim \alpha_1 \beta_1+\alpha_2 \beta_2\\
\mathbf{3'}\sim\begin{pmatrix}
\alpha_1 \beta_1-\alpha_2 \beta_2 \\
\sqrt{2} \alpha_2 \beta_1 \\
\sqrt{2} \alpha_1 \beta_2 \\
\end{pmatrix}\\
\end{array} $} &
\multicolumn{4}{|c|}{ $\begin{array}{l}
\mathbf{1_a}\sim \alpha_2 \beta_1-\alpha_1 \beta_2 \\
\mathbf{3_s}\sim\begin{pmatrix}
\alpha_2 \beta_1+\alpha_1 \beta_2 \\
-\sqrt{2} \alpha_1 \beta_1 \\
\sqrt{2} \alpha_2 \beta_2 \\
\end{pmatrix}\\
\end{array} $}  \\ \hline

\multicolumn{4}{|c}{$\mathbf{2} \otimes \mathbf{3} = \mathbf{3} \oplus \mathbf{3'}$} & \multicolumn{4}{|c}{$\mathbf{2} \otimes \mathbf{3'} = \mathbf{3} \oplus \mathbf{3'}$} & \multicolumn{4}{|c|}{$\widehat{\mathbf{2}} \otimes \mathbf{3} = \widehat{\mathbf{2}} \oplus \widehat{\mathbf{4}}$}  \\
\multicolumn{4}{|c}{  $ \begin{array}{l}
\mathbf{3} \sim \begin{pmatrix}
-2 \alpha_1 \beta_1 \\
\alpha_1 \beta_2+\sqrt{3} \alpha_2 \beta_3 \\
\sqrt{3} \alpha_2 \beta_2+\alpha_1 \beta_3 \\
\end{pmatrix} \\
\mathbf{3'} \sim \begin{pmatrix}
-2 \alpha_2 \beta_1 \\
\alpha_2 \beta_2-\sqrt{3} \alpha_1 \beta_3 \\
\alpha_2 \beta_3-\sqrt{3} \alpha_1 \beta_2 \\
\end{pmatrix}\\
\end{array} $ } &
\multicolumn{4}{|c}{ $ \begin{array}{l}
\mathbf{3} \sim \begin{pmatrix}
-2 \alpha_2 \beta_1 \\
\alpha_2 \beta_2-\sqrt{3} \alpha_1 \beta_3 \\
\alpha_2 \beta_3-\sqrt{3} \alpha_1 \beta_2 \\
\end{pmatrix} \\
\mathbf{3'} \sim \begin{pmatrix}
-2 \alpha_1 \beta_1 \\
\alpha_1 \beta_2+\sqrt{3} \alpha_2 \beta_3 \\
\sqrt{3} \alpha_2 \beta_2+\alpha_1 \beta_3 \\
\end{pmatrix}\\
\end{array} $} &
\multicolumn{4}{|c|}{ $\begin{array}{l}
\widehat{\mathbf{2}} \sim \begin{pmatrix}
 \alpha_1 \beta_1+\sqrt{2} \alpha_2 \beta_3 \\
\sqrt{2} \alpha_1 \beta_2-\alpha_2 \beta_1 \\
\end{pmatrix} \\
\widehat{\mathbf{4}} \sim \begin{pmatrix}
 -\sqrt{2} \alpha_2 \beta_1-\alpha_1 \beta_2 \\
\alpha_2 \beta_3-\sqrt{2} \alpha_1 \beta_1 \\
\sqrt{3} \alpha_2 \beta_2 \\
\sqrt{3} \alpha_1 \beta_3 \\
\end{pmatrix}\\
\end{array} $}  \\ \hline

\multicolumn{4}{|c}{$\widehat{\mathbf{2}} \otimes \mathbf{3'} = \widehat{\mathbf{2}}' \oplus \widehat{\mathbf{4}}$} & \multicolumn{4}{|c}{$\widehat{\mathbf{2}}' \otimes \mathbf{3} = \widehat{\mathbf{2}}' \oplus\widehat{\mathbf{4}}$} & \multicolumn{4}{|c|}{$\widehat{\mathbf{2}}' \otimes \mathbf{3'} = \widehat{\mathbf{2}} \oplus \widehat{\mathbf{4}}$}  \\
\multicolumn{4}{|c}{  $ \begin{array}{l}
\widehat{\mathbf{2}}' \sim \begin{pmatrix}
\alpha_2 \beta_1-\sqrt{2} \alpha_1 \beta_2 \\
\alpha_1 \beta_1+\sqrt{2} \alpha_2 \beta_3 \\
\end{pmatrix} \\
\widehat{\mathbf{4}} \sim \begin{pmatrix}
 \sqrt{3} \alpha_1 \beta_3 \\
-\sqrt{3} \alpha_2 \beta_2 \\
\alpha_2 \beta_3-\sqrt{2} \alpha_1 \beta_1 \\
\sqrt{2} \alpha_2 \beta_1+\alpha_1 \beta_2 \\
\end{pmatrix}\\
\end{array} $ } &
\multicolumn{4}{|c}{ $ \begin{array}{l}
\widehat{\mathbf{2}}' \sim \begin{pmatrix}
\sqrt{2} \alpha_2 \beta_2-\alpha_1 \beta_1 \\
\alpha_2 \beta_1+\sqrt{2} \alpha_1 \beta_3 \\
\end{pmatrix}\\
\widehat{\mathbf{4}} \sim \begin{pmatrix}
\sqrt{2} \alpha_2 \beta_1-\alpha_1 \beta_3 \\
\sqrt{3} \alpha_2 \beta_3 \\
\sqrt{2} \alpha_1 \beta_1+\alpha_2 \beta_2 \\
\sqrt{3} \alpha_1 \beta_2 \\
\end{pmatrix}\\
\end{array} $} &
\multicolumn{4}{|c|}{ $\begin{array}{l}
\widehat{\mathbf{2}} \sim \begin{pmatrix}
\alpha_2 \beta_1-\sqrt{2} \alpha_1 \beta_3 \\
\alpha_1 \beta_1+\sqrt{2} \alpha_2 \beta_2 \\
\end{pmatrix} \\
\widehat{\mathbf{4}} \sim \begin{pmatrix}
 \alpha_2 \beta_2-\sqrt{2} \alpha_1 \beta_1 \\
\sqrt{2} \alpha_2 \beta_1+\alpha_1 \beta_3 \\
\sqrt{3} \alpha_1 \beta_2 \\
-\sqrt{3} \alpha_2 \beta_3 \\
\end{pmatrix} \\
\end{array} $}  \\ \hline

\multicolumn{4}{|c}{$\mathbf{2} \otimes \widehat{\mathbf{4}} = \widehat{\mathbf{2}} \oplus \widehat{\mathbf{2}}' \oplus \widehat{\mathbf{4}}$} & \multicolumn{4}{|c}{$\widehat{\mathbf{2}} \otimes \widehat{\mathbf{4}} = \mathbf{2} \oplus \mathbf{3} \oplus \mathbf{3'}$} & \multicolumn{4}{|c|}{$\widehat{\mathbf{2}}' \otimes \widehat{\mathbf{4}} = \mathbf{2} \oplus \mathbf{3} \oplus \mathbf{3'}$}  \\
\multicolumn{4}{|c}{  $ \begin{array}{l}
\widehat{\mathbf{2}} \sim \begin{pmatrix}
\alpha_1 \beta_2+\alpha_2 \beta_3 \\
\alpha_2 \beta_4-\alpha_1 \beta_1 \\
\end{pmatrix} \\
\widehat{\mathbf{2}}' = \begin{pmatrix}
\alpha_2 \beta_1+\alpha_1 \beta_4 \\
\alpha_2 \beta_2-\alpha_1 \beta_3 \\
\end{pmatrix} \\
\widehat{\mathbf{4}} \sim \begin{pmatrix}
\alpha_1 \beta_1+\alpha_2 \beta_4 \\
\alpha_1 \beta_2-\alpha_2 \beta_3 \\
-\alpha_2 \beta_2-\alpha_1 \beta_3 \\
\alpha_2 \beta_1-\alpha_1 \beta_4 \\
\end{pmatrix} \\
\end{array} $ } &
\multicolumn{4}{|c}{ $ \begin{array}{l}
 \mathbf{2} \sim \begin{pmatrix}
\alpha_1 \beta_1+\alpha_2 \beta_2 \\
\alpha_2 \beta_3-\alpha_1 \beta_4 \\
\end{pmatrix}\\
\mathbf{3} \sim \begin{pmatrix}
\sqrt{2} \alpha_2 \beta_2-\sqrt{2} \alpha_1 \beta_1 \\
\alpha_2 \beta_1+\sqrt{3} \alpha_1 \beta_3 \\
\alpha_1 \beta_2-\sqrt{3} \alpha_2 \beta_4 \\
\end{pmatrix}\\
\mathbf{3'} \sim \begin{pmatrix}
-\sqrt{2} \alpha_2 \beta_3-\sqrt{2} \alpha_1 \beta_4 \\
\sqrt{3} \alpha_1 \beta_2+\alpha_2 \beta_4 \\
\sqrt{3} \alpha_2 \beta_1-\alpha_1 \beta_3 \\
\end{pmatrix}\\
        \end{array} $} &
\multicolumn{4}{|c|}{ $\begin{array}{l}
\mathbf{2} \sim \begin{pmatrix}
\alpha_1 \beta_3+\alpha_2 \beta_4 \\
\alpha_2 \beta_1-\alpha_1 \beta_2 \\
\end{pmatrix} \\
\mathbf{3} \sim \begin{pmatrix}
\sqrt{2} \alpha_2 \beta_4-\sqrt{2} \alpha_1 \beta_3 \\
\alpha_1 \beta_4-\sqrt{3} \alpha_2 \beta_2 \\
\sqrt{3} \alpha_1 \beta_1+\alpha_2 \beta_3 \\
\end{pmatrix} \\
\mathbf{3'} \sim \begin{pmatrix}
\sqrt{2} \alpha_2 \beta_1+\sqrt{2} \alpha_1 \beta_2 \\
\alpha_1 \beta_1-\sqrt{3} \alpha_2 \beta_3 \\
-\alpha_2 \beta_2-\sqrt{3} \alpha_1 \beta_4 \\
\end{pmatrix} \\
        \end{array} $}  \\ \hline

\multicolumn{6}{|c}{ $\mathbf{3} \otimes \mathbf{3} = \mathbf{3'} \otimes \mathbf{3'} = \mathbf{1_s} \oplus \mathbf{2_s} \oplus \mathbf{3_a} \oplus \mathbf{3'_s}$} & \multicolumn{6}{|c|}{ $\mathbf{3} \otimes \mathbf{3'} = \mathbf{1'} \oplus \mathbf{2} \oplus \mathbf{3} \oplus \mathbf{3'}$} \\
\multicolumn{6}{|c}{ $\begin{array}{l}
\mathbf{1_s} \sim \alpha_1 \beta_1+\alpha_3 \beta_2+\alpha_2 \beta_3 \\
\mathbf{2_s} \sim \begin{pmatrix}
        \alpha_3 \beta_2+\alpha_2 \beta_3-2 \alpha_1 \beta_1 \\
        \sqrt{3} \alpha_2 \beta_2+\sqrt{3} \alpha_3 \beta_3 \\
\end{pmatrix} \\
\mathbf{3_a} \sim \begin{pmatrix}
        \alpha_3 \beta_2-\alpha_2 \beta_3 \\
        \alpha_2 \beta_1-\alpha_1 \beta_2 \\
        \alpha_1 \beta_3-\alpha_3 \beta_1 \\
\end{pmatrix}  \\
\mathbf{3'_s} \sim \begin{pmatrix}
        \alpha_3 \beta_3-\alpha_2 \beta_2 \\
        \alpha_3 \beta_1+\alpha_1 \beta_3 \\
        -\alpha_2 \beta_1-\alpha_1 \beta_2 \\
\end{pmatrix} \\
\end{array} $} &
\multicolumn{6}{|c|}{ $\begin{array}{l}
\mathbf{1'} \sim \alpha_1 \beta_1+\alpha_3 \beta_2+\alpha_2 \beta_3 \\
\mathbf{2} \sim \begin{pmatrix}
        \sqrt{3} \alpha_2 \beta_2+\sqrt{3} \alpha_3 \beta_3 \\
        2 \alpha_1 \beta_1-\alpha_3 \beta_2-\alpha_2 \beta_3 \\
\end{pmatrix} \\
\mathbf{3} \sim \begin{pmatrix}
        \alpha_3 \beta_3-\alpha_2 \beta_2 \\
        \alpha_3 \beta_1+\alpha_1 \beta_3 \\
        -\alpha_2 \beta_1-\alpha_1 \beta_2 \\
\end{pmatrix}  \\
\mathbf{3'} \sim \begin{pmatrix}
        \alpha_3 \beta_2-\alpha_2 \beta_3 \\
        \alpha_2 \beta_1-\alpha_1 \beta_2 \\
        \alpha_1 \beta_3-\alpha_3 \beta_1 \\
\end{pmatrix} \\
\end{array} $}  \\ \hline

\multicolumn{6}{|c}{ $\mathbf{3} \otimes \widehat{\mathbf{4}} = \widehat{\mathbf{2}} \oplus \widehat{\mathbf{2}}' \oplus \widehat{\mathbf{4}}_{1} \oplus \widehat{\mathbf{4}}_{2}$} & \multicolumn{6}{|c|}{ $\mathbf{3'} \otimes \widehat{\mathbf{4}} = \widehat{\mathbf{2}} \oplus \widehat{\mathbf{2}}' \oplus \widehat{\mathbf{4}}_{1} \oplus \widehat{\mathbf{4}}_{2}$} \\
\multicolumn{6}{|c}{ $\begin{array}{l}
\widehat{\mathbf{2}} \sim \begin{pmatrix}
\sqrt{3} \alpha_2 \beta_4-\alpha_3 \beta_1-\sqrt{2} \alpha_1 \beta_2 \\
\alpha_2 \beta_2+\sqrt{3} \alpha_3 \beta_3-\sqrt{2} \alpha_1 \beta_1 \\
\end{pmatrix} \\
\widehat{\mathbf{2}}' \sim \begin{pmatrix}
\sqrt{3} \alpha_3 \beta_2-\alpha_2 \beta_3-\sqrt{2} \alpha_1 \beta_4 \\
\sqrt{3} \alpha_2 \beta_1+\alpha_3 \beta_4-\sqrt{2} \alpha_1 \beta_3 \\
\end{pmatrix} \\
\widehat{\mathbf{4}}_{1} \sim \begin{pmatrix}
\alpha_1 \beta_1+\sqrt{6} \alpha_3 \beta_3-2 \sqrt{2} \alpha_2 \
\beta_2 \\
-2 \sqrt{2} \alpha_3 \beta_1-\alpha_1 \beta_2-\sqrt{6} \alpha_2 \
\beta_4 \\
\sqrt{6} \alpha_2 \beta_1+3 \alpha_1 \beta_3 \\
-\sqrt{6} \alpha_3 \beta_2-3 \alpha_1 \beta_4 \\
\end{pmatrix} \\
\widehat{\mathbf{4}}_{2} \sim \begin{pmatrix}
-\alpha_1 \beta_1-\sqrt{2} \alpha_2 \beta_2 \\
\alpha_1 \beta_2-\sqrt{2} \alpha_3 \beta_1 \\
\alpha_1 \beta_3+\sqrt{2} \alpha_3 \beta_4 \\
\sqrt{2} \alpha_2 \beta_3-\alpha_1 \beta_4 \\
\end{pmatrix} \\
\end{array} $} &
\multicolumn{6}{|c|}{ $\begin{array}{l}
\widehat{\mathbf{2}} \sim \begin{pmatrix}
\sqrt{3} \alpha_2 \beta_1+\alpha_3 \beta_4-\sqrt{2} \alpha_1 \beta_3 \\
\sqrt{2} \alpha_1 \beta_4+\alpha_2 \beta_3-\sqrt{3} \alpha_3 \beta_2 \\
\end{pmatrix} \\
\widehat{\mathbf{2}}' \sim \begin{pmatrix}
\alpha_2 \beta_2+\sqrt{3} \alpha_3 \beta_3-\sqrt{2} \alpha_1 \beta_1 \\
\alpha_3 \beta_1+\sqrt{2} \alpha_1 \beta_2-\sqrt{3} \alpha_2 \beta_4 \\
\end{pmatrix} \\
\widehat{\mathbf{4}}_{1} \sim \begin{pmatrix}
\alpha_1 \beta_4-\sqrt{2} \alpha_2 \beta_3 \\
\alpha_1 \beta_3+\sqrt{2} \alpha_3 \beta_4 \\
\sqrt{2} \alpha_3 \beta_1-\alpha_1 \beta_2 \\
-\alpha_1 \beta_1-\sqrt{2} \alpha_2 \beta_2 \\
\end{pmatrix} \\
\widehat{\mathbf{4}}_{2} \sim \begin{pmatrix}
\sqrt{6} \alpha_3 \beta_2+3 \alpha_1 \beta_4 \\
\sqrt{6} \alpha_2 \beta_1+3 \alpha_1 \beta_3 \\
2 \sqrt{2} \alpha_3 \beta_1+\alpha_1 \beta_2+\sqrt{6} \alpha_2 \beta_4 \\
\alpha_1 \beta_1+\sqrt{6} \alpha_3 \beta_3-2 \sqrt{2} \alpha_2 \beta_2 \\
\end{pmatrix}\\
\end{array} $} \\ \hline
\multicolumn{12}{|c|}{ $\widehat{\mathbf{4}} \otimes \widehat{\mathbf{4}} = \mathbf{1_{a}} \oplus \mathbf{1'_s} \oplus \mathbf{2_a} \oplus \mathbf{3_{1,s}} \oplus\mathbf{3_{2,s}} \oplus \mathbf{3'_{1,a}} \oplus \mathbf{3'_{2,s}}$ } \\
\multicolumn{12}{|c|}{ $\begin{array}{l}
\mathbf{1_a} \sim \alpha_2 \beta_1-\alpha_1 \beta_2+\alpha_4 \beta_3-\alpha_3 \beta_4 \\
\mathbf{1'_s} \sim \alpha_3 \beta_1+\alpha_1 \beta_3+\alpha_4 \beta_2+\alpha_2 \beta_4 \\
\mathbf{2_a} \sim \begin{pmatrix}
\alpha_1 \beta_2-\alpha_2 \beta_1+\alpha_4 \beta_3-\alpha_3 \beta_4 \\
\alpha_3 \beta_1-\alpha_1 \beta_3+\alpha_4 \beta_2-\alpha_2 \beta_4 \\
\end{pmatrix}\\
\mathbf{3_{1,s}} \sim \begin{pmatrix}
\alpha_4 \beta_3+\alpha_3 \beta_4-\alpha_2 \beta_1-\alpha_1 \beta_2 \\
\sqrt{2} \alpha_1 \beta_1+\sqrt{2} \alpha_4 \beta_4 \\
-\sqrt{2} \alpha_2 \beta_2-\sqrt{2} \alpha_3 \beta_3 \\
\end{pmatrix} \\
\mathbf{3_{2,s}} \sim \begin{pmatrix}
 2 \sqrt{2} \alpha_2 \beta_1+2 \sqrt{2} \alpha_1 \beta_2 \\
\sqrt{3} \alpha_3 \beta_2+\sqrt{3} \alpha_2 \beta_3-\alpha_1 \beta_1-3 \alpha_4 \beta_4 \\
\alpha_2 \beta_2+3 \alpha_3 \beta_3+\sqrt{3} \alpha_4 \beta_1+\sqrt{3} \alpha_1 \beta_4 \\
\end{pmatrix} \\
\mathbf{3'_{1,a}}\sim \begin{pmatrix}
 \alpha_4 \beta_2-\alpha_2 \beta_4+\alpha_1 \beta_3-\alpha_3 \beta_1 \\
\sqrt{2} \alpha_1 \beta_4-\sqrt{2} \alpha_4 \beta_1 \\
\sqrt{2} \alpha_2 \beta_3-\sqrt{2} \alpha_3 \beta_2 \\
\end{pmatrix} \\
\mathbf{3'_{2,s}} \sim \begin{pmatrix}
\sqrt{2} \alpha_4 \beta_2+\sqrt{2} \alpha_2 \beta_4-\sqrt{2} \alpha_3 \beta_1-\sqrt{2} \alpha_1 \beta_3 \\
\sqrt{3} \alpha_2 \beta_2+\alpha_4 \beta_1+\alpha_1 \beta_4-\sqrt{3} \alpha_3 \beta_3 \\
\sqrt{3} \alpha_4 \beta_4+\alpha_3 \beta_2+\alpha_2 \beta_3-\sqrt{3} \alpha_1 \beta_1 \\
\end{pmatrix} \\
\end{array} $} \\
\midrule
\specialrule{0em}{1.0pt}{1.0pt}
\midrule
\end{longtable}
\end{small}
\end{center}

\section{\label{app:fixed-point}Fermion mass hierarchy from deviation of $\tau$ from the fixed points }

It is known that there are three inequivalent modular symmetry fixed points $\tau_0=i\infty, i, \omega\equiv e^{2\pi i/3}$ in the fundamental domain of $\mathrm{SL}(2,\mathbb{Z})$~\cite{Novichkov:2018ovf,Ding:2019gof,Novichkov:2020eep}. At these special values of $\tau$, modular symmetry is only partially broken and certain residual symmetry is preserved. Modulo a possible $\mathbb{Z}^{R}_2$ factor with $R=S^2$, the three fixed points $\tau_0=i\infty$, $\tau_0=i$ and $\tau=\omega$ are invariant under the action of $T$, $S$ and $ST$ respectively, and therefore the residual symmetries $\mathbb{Z}^{T}_N$,
$\mathbb{Z}^{S}_4$ and $\mathbb{Z}^{ST}_3$ are preserved respectively where $N$ refers to the level satisfying $T^N=1$~\cite{Novichkov:2018ovf,Ding:2019gof,Novichkov:2020eep}. Exactly at the fixed points, the residual symmetries would enforce the presence of multiple zeros in the fermion mass matrices. When the modulus $\tau$ deviates from any fixed point $\tau_0$ given above, the zero entries in the fermion mass matrix would become non-zero. It turns out that their magnitudes would be determined by the size of the departure $\epsilon$ from $\tau_0$ and by the field transformation properties under the residual symmetry group. As a consequence, the fermion mass hierarchies could be naturally produced in the vicinity of $\tau_0$~\cite{Okada:2020ukr,Feruglio:2021dte,Novichkov:2021evw,Feruglio:2023bav,Feruglio:2023mii}. The hierarchical structures in the vicinity of symmetric points have been systematically analyzed for the finite modular groups $\Gamma_N$ and $\Gamma'_N$~\cite{Novichkov:2021evw}.  In the following, we will follow Ref.~\cite{Novichkov:2021evw} and investigate the possible mass hierarchies arising solely due to the proximity of the modulus $\tau$ to the fixed points $\tau_0$ in the context of modular symmetry $2O$. Firstly we recapitulate method and results of this approach. After electroweak symmetry breaking, the fermion mass terms in modular invariant theory can be generally written as
\begin{equation}
\mathcal{L}_m=\psi^{c}_iM_{ij}(\tau)\psi_j\,,
\end{equation}
where $i, j=1, 2, 3$ are flavor indices, and each matrix element $M_{ij}(\tau)$ is a modular form. The modular symmetry transformations of the superfields $\psi$ and $\psi^c$ are given by
\begin{eqnarray}
\psi\stackrel{\gamma}{\rightarrow}(c\tau+d)^{-k}\rho(\gamma)\psi,
~~~\psi^c\stackrel{\gamma}{\rightarrow}(c\tau+d)^{-k^c}\rho^c(\gamma)\psi^c\,,
\end{eqnarray}
where $-k$ and $-k^c$ denote the modular weights of $\psi$ and $\psi^c$ respectively, both $\rho(\gamma)$ and $\rho^c(\gamma)$ are unitary representations of the finite modular groups. Modular invariance requires the fermion mass matrix $M(\tau)$ transforms as
\begin{equation}
\label{eq:mass-matrix-modular-invariance}M(\tau)\stackrel{\gamma}{\rightarrow}M(\gamma\tau)=(c\tau+d)^{k_{tot}}\left[\rho^{c}(\gamma)\right]^{*}M(\tau)\rho^{\dagger}(\gamma)\,,
\end{equation}
where $k_{tot}=k+k^c$ denotes the total modular weight. Taking $\gamma$ to the residual symmetry generator and $\tau$ in the vicinity of modular fixed points, one can use the transformation rule of Eq.~\eqref{eq:mass-matrix-modular-invariance} to constrain the form of the mass matrix $M(\tau)$ as well as the fermion mass hierarchies. It is remarkable that the magnitude of each matrix element can be estimated as some power of the small deviation $\epsilon$ from fixed points, and the power exponent only depends on how the representations of $\psi$ and $\psi^c$ decompose under the residual symmetry group. We list the order of magnitude of the matrix elements in the vicinity of the three fixed points in the following.
\begin{itemize}

\item{$\tau_0=i\infty$}

In the $T$-diagonal basis $\rho^{(c)}(T)=\text{diag}(\rho^{(c)}_i)$, for $(\rho^c_i\rho_j)^{*}=\zeta^{p_{ij}}$ with $0\leq p_{ij}<N$ and $\zeta=e^{2\pi i/N}$, the entry $M_{ij}(\tau)$ can be expanded as
\begin{equation}
M_{ij}=a_0q^{p_{ij}}_N+a_1q^{p_{ij}+N}_N+a_2q^{p_{ij}+2N}_N+\ldots,~~~q_N\equiv e^{2\pi i\tau/N}\,,
\end{equation}
in the vicinity of the fixed point, where the expansion coefficients $a_0$, $a_1$, $a_2, \ldots$ are expected to be order one. Therefore the magnitude of the entry $M_{ij}$ is
\begin{equation}
M_{ij}\sim\mathcal{O}(\epsilon^{p_{ij}}),~~~\epsilon=|q_N|=e^{-2\pi\text{Im}\tau/N}\,.
\end{equation}

\item{$\tau_0=i$}

In this case, it is convenient to switch to the basis in which the generator $S$ is represented by a diagonal matrix with $\rho^{(c)}(S)=\text{diag}(\rho^{(c)}_i)$. For $(i^{k_{tot}}\rho^c_i\rho_j)^{*}=(-1)^{p_{ij}}$ with $p_{ij}=0, 1$, we have
\begin{equation}
M_{ij}\sim\mathcal{O}(\epsilon^{p_{ij}}),~~~\epsilon=|s|,~~~s\equiv\frac{\tau-i}{\tau+i}\,.
\end{equation}
Hence the mass matrix entry $M_{ij}$ is expected to be $\mathcal{O}(1)$ or $\mathcal{O}(\epsilon)$, and this is
insufficient to reproduce the observed mass hierarchies of charged leptons and quarks.

\item{$\tau_0=\omega$}

Analogous to previous cases, it is convenient to adopt a basis where $ST$ is represented by a diagonal matrix with $\rho^{(c)}(ST)=\text{diag}(\rho^{(c)}_i)$. For $\omega^{k_{tot}}\rho^c_i\rho_j=\omega^{p_{ij}}$ with $p_{ij}=0, 1, 2$, the magnitude of the matrix entry $M_{ij}$ is
\begin{equation}
M_{ij}\sim\mathcal{O}(\epsilon^{p_{ij}}),~~~\epsilon=|u|,~~~u\equiv\frac{\tau-\omega}{\tau-\omega^2}\,.
\end{equation}

\end{itemize}

In short, the mass matrix entries $M_{ij}$ are expected to be of order $\mathcal{O}(\epsilon^{p_{ij}})$ in the vicinity of the modular fixed point $\tau_0$, where $\epsilon$ parametrizes deviation of $\tau$ from $\tau_0$. The power exponent $p_{ij}$ depends on how the representations of $\psi$ and $\psi^c$ decompose under the residual symmetry group and on their respective weights $k$ and $k^c$, and the value of $p_{ij}$ can be extracted from products of factors  corresponding to representations of the residual symmetry group.

Since the complex modulus $\tau$ is invariant under the action of modular transformation $R=S^2$ with $R^2\tau=\tau$, the full residual symmetry groups of the fixed points $\tau_0=i\infty$, $\tau_0=i$ and $\tau_0=\omega$ are $\mathbb{Z}^T_N\times\mathbb{Z}^{R}_2$, $\mathbb{Z}^S_4$ and $\mathbb{Z}^{ST}_3\times\mathbb{Z}^{R}_2$ respectively. Notice that $\mathbb{Z}^{R}_2$ is hidden in the residual group $\mathbb{Z}^S_4$. A generic field multiplet, whose modular transformation is characterized by the modular weight $k$ and the representation $\rho$ of the finite modular group, would decompose into one-dimensional representations of the
residual symmetry group. As explained in above, one can straightforwardly find the residual symmetry representations for any multiplet of a finite modular group. We give the decompositions of different $2O$ multiplets under the residual groups in table~\ref{tab:decomposition}.

\begin{table}[hptb!]
\renewcommand{\arraystretch}{1.5}
\centering
\small
\begin{tabular}{|c|c|c|c|}
\hline  \hline

$\rho$
&$\mathbb{Z}_{4}^{S}~~(\tau=i)$
&$\mathbb{Z}_{3}^{ST}\times \mathbb{Z}_{2}^{R}~~(\tau=\omega)$
&$\mathbb{Z}_{8}^{T}\times \mathbb{Z}_{2}^{R}~~(\tau=i\infty)$  \\\hline

$\bm{1}$ ~&~ $\bm{1}_{k}$ ~&~ $\bm{1}_{k}^{\pm}$ ~&~ $\bm{1}_{0}^{\pm}$\\\hline

$\bm{1}^{\prime}$ ~&~ $\bm{1}_{k+2}$ ~&~ $\bm{1}_{k}^{\pm}$ ~&~ $\bm{1}_{4}^{\pm}$\\ \hline

$\bm{2}$ ~&~ $\bm{1}_{k}\oplus\bm{1}_{k+2}$ ~&~ $\bm{1}_{k+1}^{\pm}\oplus\bm{1}_{k+2}^{\pm}$ ~&~ $\bm{1}_{0}^{\pm}\oplus\bm{1}_{4}^{\pm}$\\\hline

$\bm{{\widehat2}}$ ~&~ $\bm{1}_{k+1}\oplus\bm{1}_{k+3}$ ~&~ $\bm{1}_{k+1}^{\mp}\oplus\bm{1}_{k+2}^{\mp}$ ~&~ $\bm{1}_{3}^{\mp}\oplus\bm{1}_{5}^{\mp}$\\\hline

$\bm{{\widehat2}}^{\prime}$ ~&~ $\bm{1}_{k+1}\oplus\bm{1}_{k+3}$ ~&~ $\bm{1}_{k+1}^{\mp}\oplus\bm{1}_{k+2}^{\mp}$ ~&~ $\bm{1}_{1}^{\mp}\oplus\bm{1}_{7}^{\mp}$\\\hline

$\bm{3}$ ~&~ $\bm{1}_{k}\oplus\bm{1}_{k+2}\oplus\bm{1}_{k+2}$ ~&~ $\bm{1}_{k}^{\pm}\oplus\bm{1}_{k+1}^{\pm}\oplus\bm{1}_{k+2}^{\pm}$ ~&~ $\bm{1}_{0}^{\pm}\oplus\bm{1}_{2}^{\pm}\oplus\bm{1}_{6}^{\pm}$\\\hline

$\bm{3}^{\prime}$ ~&~ $\bm{1}_{k}\oplus\bm{1}_{k}\oplus\bm{1}_{k+2}$ ~&~ $\bm{1}_{k}^{\pm}\oplus\bm{1}_{k+1}^{\pm}\oplus\bm{1}_{k+2}^{\pm}$ ~&~$\bm{1}_{2}^{\pm}\oplus\bm{1}_{4}^{\pm}\oplus\bm{1}_{6}^{\pm}$\\\hline

$\bm{{\widehat4}}$ ~&~$\bm{1}_{k+1}\oplus\bm{1}_{k+1}\oplus\bm{1}_{k+3}\oplus\bm{1}_{k+3}$ ~&~ $\bm{1}_{k}^{\mp}\oplus\bm{1}_{k}^{\mp}\oplus\bm{1}_{k+1}^{\mp}\oplus\bm{1}_{k+2}^{\mp}$ ~&~ $\bm{1}_{1}^{\mp}\oplus\bm{1}_{3}^{\mp}\oplus\bm{1}_{5}^{\mp}\oplus\bm{1}_{7}^{\mp}$\\\hline \hline
\end{tabular}
\caption{\label{tab:decomposition}
Decomposition of the $2O$ multiplets of modular weight $k$ in the representation $\rho$ under the residual symmetry groups. The irrep subscripts should be understood modulo 4 and 3 in the second and third columns respectively. The upper (lower) signs in the superscript correspond to even (odd) values of $k$. }
\end{table}

After identifying the decompositions of fields under residual symmetry groups, one can apply the above general results to construct the hierarchical mass matrix in the vicinity of a fixed point in terms of powers of $\epsilon$, in the appropriate basis. Subsequently one can extract the hierarchical pattern of the fermion masses which is the singular values of the mass matrix $M(\tau)$. In table \ref{tab:mass-pattern}, we list the hierarchical patterns which can arise from the proximity of $\tau$ to the fixed points $\tau_0=i\infty$ and $\tau_0=\omega$. We find that hierarchical mass patterns can be obtained from some representation assignments, the promising cases are summarized in table~\ref{tab:heirarchy-summary}. If the large lepton mixing angles are taken into account further,  only the cases in which lepton fields are singlets of $2O$ or the lepton masses are vanishing in the limit $\epsilon\rightarrow 0$, can survive.

\begin{center}
\renewcommand{\arraystretch}{1.2}
\LTcapwidth=\textwidth
{\begin{longtable}[!]{|c|c|ccc|c|}
\caption{\label{tab:mass-pattern}The mass spectrum patterns $(m_{1}, m_{2}, m_{3})$ of the bilinear $\psi^{c}\psi$ in the vicinity of the fixed points $\tau_0=\omega$ and $\tau_0=i\infty$. Here $\psi$ and $\psi^c$ transform in the representations $\rho$ and $\rho^c$ of the finite modular group $2O$ respectively, and their modular weights are denoted by $k$ and $k^c$ with $k_{tot}\equiv k+k^c$. The mass spectrums are invariant if the modular transformations of $\psi$ and $\psi^c$ are interchanged.  } \\
\midrule
\specialrule{0em}{1.0pt}{1.0pt}
\midrule
\multirow{2}{*}{$\rho$}
& \multirow{2}{*}{$\rho^c$}
& & $\tau\simeq\omega$~&
& \multirow{2}{*}{$\tau\simeq i\infty$} \\
& & $k_{tot}=0\,(\text{mod}\,3)$ & $k_{tot}=1\,(\text{mod}\,3)$ &
$k_{tot}=2\,(\text{mod}\,3)$ & \\
\hline
\endfirsthead
\midrule
\specialrule{0em}{1.0pt}{1.0pt}
\midrule
\multirow{2}{*}{$\rho$}
& \multirow{2}{*}{$\rho^c$}
& & $\tau\simeq\omega$ &
& \multirow{2}{*}{$\tau\simeq i\infty$} \\
& & $k_{tot}=0\,(\text{mod}\,3)$ & $k_{tot}=1\,(\text{mod}\,3)$ &
$k_{tot}=2\,(\text{mod}\,3)$ & \\
\hline
\endhead
\midrule
\specialrule{0em}{1.0pt}{1.0pt}
\midrule
\caption[]{continues on next page}\\
\endfoot
\endlastfoot

$\bm{3}$
&$\bm{1}\oplus\bm{1}\oplus\bm{1}$
&$(\epsilon^{2},\epsilon,1)$&$(\epsilon^{2},\epsilon,1)$
&$(\epsilon^{2},\epsilon,1)$
&$(\epsilon^{6},\epsilon^{2},1)$\\
$\bm{3}$
&$\bm{1}\oplus\bm{1}\oplus\bm{1}^{\prime}$
&$(\epsilon^{2},\epsilon,1)$&$(\epsilon^{2},\epsilon,1)$
&$(\epsilon^{2},\epsilon,1)$
&$(\epsilon^{2},\epsilon^{2},1)$\\
$\bm{3}$
&$\bm{1}\oplus\bm{1}^{\prime}\oplus\bm{1}^{\prime}$
&$(\epsilon^{2},\epsilon,1)$
&$(\epsilon^{2},\epsilon,1)$&$(\epsilon^{2},\epsilon,1)$
&$(\epsilon^{6},\epsilon^{2},1)$\\
$\bm{3}$
&$\bm{1}^{\prime}\oplus\bm{1}^{\prime}\oplus\bm{1}^{\prime}$
&$(\epsilon^{2},\epsilon,1)$
&$(\epsilon^{2},\epsilon,1)$&$(\epsilon^{2},\epsilon,1)$
&$(\epsilon^{6},\epsilon^{4},\epsilon^{2})$\\
$\bm{3}$
&$\bm{2}\oplus\bm{1}$
&$(1,1,1)$&$(1,1,1)$&$(1,1,1)$
&$(\epsilon^{2},\epsilon^{2},1)$\\
$\bm{3}$
&$\bm{2}\oplus\bm{1}^{\prime}$
&$(1,1,1)$&$(1,1,1)$&$(1,1,1)$
&$(\epsilon^{6},\epsilon^{2},1)$\\
$\bm{3}$
&$\bm{3}$
&$(1,1,1)$&$(1,1,1)$&$(1,1,1)$
&$(1,1,1)$\\
$\bm{3}$
&$\bm{3}^{\prime}$
&$(1,1,1)$&$(1,1,1)$&$(1,1,1)$
&$(\epsilon^{4},1,1)$\\
$\bm{3}^{\prime}$
&$\bm{1}\oplus\bm{1}\oplus\bm{1}$
&$(\epsilon^{2},\epsilon,1)$
&$(\epsilon^{2},\epsilon,1)$&$(\epsilon^{2},\epsilon,1)$
&$(\epsilon^{6},\epsilon^{4},\epsilon^{2})$\\
$\bm{3}^{\prime}$
&$\bm{1}\oplus\bm{1}\oplus\bm{1}^{\prime}$
&$(\epsilon^{2},\epsilon,1)$
&$(\epsilon^{2},\epsilon,1)$&$(\epsilon^{2},\epsilon,1)$
&$(\epsilon^{6},\epsilon^{2},1)$\\
$\bm{3}^{\prime}$
&$\bm{1}\oplus\bm{1}^{\prime}\oplus\bm{1}^{\prime}$
&$(\epsilon^{2},\epsilon,1)$
&$(\epsilon^{2},\epsilon,1)$&$(\epsilon^{2},\epsilon,1)$
&$(\epsilon^{2},\epsilon^{2},1)$\\
$\bm{3}^{\prime}$
&$\bm{1}^{\prime}\oplus\bm{1}^{\prime}\oplus\bm{1}^{\prime}$
&$(\epsilon^{2},\epsilon,1)$
&$(\epsilon^{2},\epsilon,1)$&$(\epsilon^{2},\epsilon,1)$
&$(\epsilon^{6},\epsilon^{2},1)$\\
$\bm{3}^{\prime}$
&$\bm{2}\oplus\bm{1}$
&$(1,1,1)$&$(1,1,1)$&$(1,1,1)$
&$(\epsilon^{6},\epsilon^{2},1)$\\
$\bm{3}^{\prime}$
&$\bm{2}\oplus\bm{1}^{\prime}$
&$(1,1,1)$&$(1,1,1)$&$(1,1,1)$
&$(\epsilon^{2},\epsilon^{2},1)$\\
$\bm{3}^{\prime}$
&$\bm{3}^{\prime}$
&$(1,1,1)$&$(1,1,1)$&$(1,1,1)$
&$(1,1,1)$\\
$\bm{2}\oplus\bm{1}$
&$\bm{2}\oplus\bm{1}$
&$(1,1,1)$&$(1,1,1)$&$(1,1,1)$
&$(1,1,1)$\\
$\bm{2}\oplus\bm{1}$
&$\bm{2}\oplus\bm{1}^{\prime}$
&$(1,1,1)$&$(1,1,1)$&$(1,1,1)$
&$(\epsilon^{4},1,1)$\\
$\bm{2}\oplus\bm{1}^{\prime}$
&$\bm{2}\oplus\bm{1}^{\prime}$
&$(1,1,1)$&$(1,1,1)$&$(1,1,1)$
&$(1,1,1)$\\
$\bm{2}\oplus\bm{1}$
&$\bm{1}\oplus\bm{1}\oplus\bm{1}$
&$(\epsilon^{2},\epsilon,1)$
&$(\epsilon^{2},\epsilon,1)$&$(\epsilon^{2},\epsilon,1)$
&$(\epsilon^{4},1,1)$\\
$\bm{2}\oplus\bm{1}$
&$\bm{1}\oplus\bm{1}\oplus\bm{1}^{\prime}$
&$(\epsilon^{2},\epsilon,1)$
&$(\epsilon^{2},\epsilon,1)$&$(\epsilon^{2},\epsilon,1)$
&$(1,1,1)$\\
$\bm{2}\oplus\bm{1}$
&$\bm{1}\oplus\bm{1}^{\prime}\oplus\bm{1}^{\prime}$
&$(\epsilon^{2},\epsilon,1)$
&$(\epsilon^{2},\epsilon,1)$&$(\epsilon^{2},\epsilon,1)$
&$(\epsilon^{4},1,1)$\\
$\bm{2}\oplus\bm{1}$
&$\bm{1}^{\prime}\oplus\bm{1}^{\prime}\oplus\bm{1}^{\prime}$
&$(\epsilon^{2},\epsilon,1)$
&$(\epsilon^{2},\epsilon,1)$&$(\epsilon^{2},\epsilon,1)$
&$(\epsilon^{4},\epsilon^{4},1)$\\
$\bm{2}\oplus\bm{1}^{\prime}$
&$\bm{1}\oplus\bm{1}\oplus\bm{1}$
&$(\epsilon^{2},\epsilon,1)$&$(\epsilon^{2},\epsilon,1)$
&$(\epsilon^{2},\epsilon,1)$
&$(\epsilon^{4},\epsilon^{4},1)$\\
$\bm{2}\oplus\bm{1}^{\prime}$
&$\bm{1}\oplus\bm{1}\oplus\bm{1}^{\prime}$
&$(\epsilon^{2},\epsilon,1)$
&$(\epsilon^{2},\epsilon,1)$&$(\epsilon^{2},\epsilon,1)$
&$(\epsilon^{4},1,1)$\\
$\bm{2}\oplus\bm{1}^{\prime}$
&$\bm{1}\oplus\bm{1}^{\prime}\oplus\bm{1}^{\prime}$
&$(\epsilon^{2},\epsilon,1)$
&$(\epsilon^{2},\epsilon,1)$&$(\epsilon^{2},\epsilon,1)$
&$(1,1,1)$\\
$\bm{2}\oplus\bm{1}^{\prime}$
&$\bm{1}^{\prime}\oplus\bm{1}^{\prime}\oplus\bm{1}^{\prime}$
&$(\epsilon^{2},\epsilon,1)$
&$(\epsilon^{2},\epsilon,1)$&$(\epsilon^{2},\epsilon,1)$
&$(\epsilon^{4},1,1)$\\
$\bm{1}\oplus\bm{1}\oplus\bm{1}$
&$\bm{1}\oplus\bm{1}\oplus\bm{1}$
&$(1,1,1)$&$(\epsilon^{2},\epsilon^{2},\epsilon^{2})$
&$(\epsilon,\epsilon,\epsilon)$
&$(1,1,1)$\\
$\bm{1}\oplus\bm{1}\oplus\bm{1}$
&$\bm{1}\oplus\bm{1}\oplus\bm{1}^{\prime}$
&$(1,1,1)$&$(\epsilon^{2},\epsilon^{2},\epsilon^{2})$
&$(\epsilon,\epsilon,\epsilon)$
&$(\epsilon^{4},1,1)$\\
$\bm{1}\oplus\bm{1}\oplus\bm{1}$
&$\bm{1}\oplus\bm{1}^{\prime}\oplus\bm{1}^{\prime}$
&$(1,1,1)$&$(\epsilon^{2},\epsilon^{2},\epsilon^{2})$
&$(\epsilon,\epsilon,\epsilon)$
&$(\epsilon^{4},\epsilon^{4},1)$\\
$\bm{1}\oplus\bm{1}\oplus\bm{1}$
&$\bm{1}^{\prime}\oplus\bm{1}^{\prime}\oplus\bm{1}^{\prime}$
&$(1,1,1)$&$(\epsilon^{2},\epsilon^{2},\epsilon^{2})$
&$(\epsilon,\epsilon,\epsilon)$
&$(\epsilon^{4},\epsilon^{4},\epsilon^{4})$\\
$\bm{1}\oplus\bm{1}\oplus\bm{1}^{\prime}$
&$\bm{1}\oplus\bm{1}\oplus\bm{1}^{\prime}$
&$(1,1,1)$&$(\epsilon^{2},\epsilon^{2},\epsilon^{2})$
&$(\epsilon,\epsilon,\epsilon)$
&$(1,1,1)$\\
$\bm{1}\oplus\bm{1}\oplus\bm{1}^{\prime}$
&$\bm{1}\oplus\bm{1}^{\prime}\oplus\bm{1}^{\prime}$
&$(1,1,1)$&$(\epsilon^{2},\epsilon^{2},\epsilon^{2})$
&$(\epsilon,\epsilon,\epsilon)$
&$(\epsilon^{4},1,1)$\\
$\bm{1}\oplus\bm{1}\oplus\bm{1}^{\prime}$
&$\bm{1}^{\prime}\oplus\bm{1}^{\prime}\oplus\bm{1}^{\prime}$
&$(1,1,1)$&$(\epsilon^{2},\epsilon^{2},\epsilon^{2})$
&$(\epsilon,\epsilon,\epsilon)$
&$(\epsilon^{4},\epsilon^{4},1)$\\
$\bm{1}\oplus\bm{1}^{\prime}\oplus\bm{1}^{\prime}$
&$\bm{1}\oplus\bm{1}^{\prime}\oplus\bm{1}^{\prime}$
&$(1,1,1)$&$(\epsilon^{2},\epsilon^{2},\epsilon^{2})$
&$(\epsilon,\epsilon,\epsilon)$
&$(1,1,1)$\\
$\bm{1}\oplus\bm{1}^{\prime}\oplus\bm{1}^{\prime}$
&$\bm{1}^{\prime}\oplus\bm{1}^{\prime}\oplus\bm{1}^{\prime}$
&$(1,1,1)$&$(\epsilon^{2},\epsilon^{2},\epsilon^{2})$
&$(\epsilon,\epsilon,\epsilon)$
&$(\epsilon^{4},1,1)$\\
$\bm{1}^{\prime}\oplus\bm{1}^{\prime}\oplus\bm{1}^{\prime}$
&$\bm{1}^{\prime}\oplus\bm{1}^{\prime}\oplus\bm{1}^{\prime}$
&$(1,1,1)$&$(\epsilon^{2},\epsilon^{2},\epsilon^{2})$
&$(\epsilon,\epsilon,\epsilon)$
&$(1,1,1)$\\
$\bm{3}$
&$\bm{{\widehat2}}\oplus\bm{1}$
&$(1,1,1)$&$(1,1,1)$&$(1,1,1)$
&$(\epsilon^{7},\epsilon,1)$\\
$\bm{3}$
&$\bm{{\widehat2}}\oplus\bm{1}^{\prime}$
&$(1,1,1)$&$(1,1,1)$&$(1,1,1)$
&$(\epsilon^{7},\epsilon^{4},\epsilon)$\\
$\bm{3}$
&$\bm{{\widehat2}}^{\prime}\oplus\bm{1}$
&$(1,1,1)$&$(1,1,1)$&$(1,1,1)$
&$(\epsilon^{7},\epsilon,1)$\\
$\bm{3}$
&$\bm{{\widehat2}}^{\prime}\oplus\bm{1}^{\prime}$
&$(1,1,1)$&$(1,1,1)$&$(1,1,1)$
&$(\epsilon^{2},\epsilon,\epsilon)$\\
$\bm{3}^{\prime}$
&$\bm{{\widehat2}}\oplus\bm{1}$
&$(1,1,1)$&$(1,1,1)$&$(1,1,1)$
&$(\epsilon^{2},\epsilon,\epsilon)$\\
$\bm{3}^{\prime}$
&$\bm{{\widehat2}}\oplus\bm{1}^{\prime}$
&$(1,1,1)$&$(1,1,1)$&$(1,1,1)$
&$(\epsilon^{7},\epsilon,1)$\\
$\bm{3}^{\prime}$
&$\bm{{\widehat2}}^{\prime}\oplus\bm{1}$
&$(1,1,1)$&$(1,1,1)$&$(1,1,1)$
&$(\epsilon^{7},\epsilon^{4},\epsilon)$\\
$\bm{3}^{\prime}$
&$\bm{{\widehat2}}^{\prime}\oplus\bm{1}^{\prime}$
&$(1,1,1)$&$(1,1,1)$&$(1,1,1)$
&$(\epsilon^{7},\epsilon,1)$\\
$\bm{{\widehat2}}\oplus\bm{1}$
&$\bm{2}\oplus\bm{1}$
&$(1,1,1)$&$(1,1,1)$&$(1,1,1)$
&$(\epsilon^{3},\epsilon,1)$\\
$\bm{{\widehat2}}\oplus\bm{1}$
&$\bm{2}\oplus\bm{1}^{\prime}$
&$(1,1,1)$&$(1,1,1)$&$(1,1,1)$
&$(\epsilon^{3},\epsilon,1)$\\
$\bm{{\widehat2}}\oplus\bm{1}^{\prime}$
&$\bm{2}\oplus\bm{1}^{\prime}$
&$(1,1,1)$&$(1,1,1)$&$(1,1,1)$
&$(\epsilon^{3},\epsilon,1)$\\
$\bm{{\widehat2}}\oplus\bm{1}$
&$\bm{1}\oplus\bm{1}\oplus\bm{1}$
&$(\epsilon^{2},\epsilon,1)$
&$(\epsilon^{2},\epsilon,1)$&$(\epsilon^{2},\epsilon,1)$
&$(\epsilon^{5},\epsilon^{3},1)$\\
$\bm{{\widehat2}}\oplus\bm{1}$
&$\bm{1}\oplus\bm{1}\oplus\bm{1}^{\prime}$
&$(\epsilon^{2},\epsilon,1)$
&$(\epsilon^{2},\epsilon,1)$&$(\epsilon^{2},\epsilon,1)$
&$(\epsilon^{3},\epsilon,1)$\\
$\bm{{\widehat2}}\oplus\bm{1}$
&$\bm{1}\oplus\bm{1}^{\prime}\oplus\bm{1}^{\prime}$
&$(\epsilon^{2},\epsilon,1)$
&$(\epsilon^{2},\epsilon,1)$&$(\epsilon^{2},\epsilon,1)$
&$(\epsilon^{7},\epsilon,1)$\\
$\bm{{\widehat2}}\oplus\bm{1}$
&$\bm{1}^{\prime}\oplus\bm{1}^{\prime}\oplus\bm{1}^{\prime}$
&$(\epsilon^{2},\epsilon,1)$
&$(\epsilon^{2},\epsilon,1)$&$(\epsilon^{2},\epsilon,1)$
&$(\epsilon^{7},\epsilon^{4},\epsilon)$\\
$\bm{{\widehat2}}\oplus\bm{1}^{\prime}$
&$\bm{1}\oplus\bm{1}\oplus\bm{1}$
&$(\epsilon^{2},\epsilon,1)$
&$(\epsilon^{2},\epsilon,1)$&$(\epsilon^{2},\epsilon,1)$
&$(\epsilon^{5},\epsilon^{4},\epsilon^{3})$\\
$\bm{{\widehat2}}\oplus\bm{1}^{\prime}$
&$\bm{1}\oplus\bm{1}\oplus\bm{1}^{\prime}$
&$(\epsilon^{2},\epsilon,1)$
&$(\epsilon^{2},\epsilon,1)$&$(\epsilon^{2},\epsilon,1)$
&$(\epsilon^{5},\epsilon^{3},1)$\\
$\bm{{\widehat2}}\oplus\bm{1}^{\prime}$
&$\bm{1}\oplus\bm{1}^{\prime}\oplus\bm{1}^{\prime}$
&$(\epsilon^{2},\epsilon,1)$
&$(\epsilon^{2},\epsilon,1)$&$(\epsilon^{2},\epsilon,1)$
&$(\epsilon^{3},\epsilon,1)$\\
$\bm{{\widehat2}}\oplus\bm{1}^{\prime}$
&$\bm{1}^{\prime}\oplus\bm{1}^{\prime}\oplus\bm{1}^{\prime}$
&$(\epsilon^{2},\epsilon,1)$
&$(\epsilon^{2},\epsilon,1)$&$(\epsilon^{2},\epsilon,1)$
&$(\epsilon^{7},\epsilon,1)$\\
$\bm{{\widehat2}}^{\prime}\oplus\bm{1}$
&$\bm{2}\oplus\bm{1}$
&$(1,1,1)$&$(1,1,1)$&$(1,1,1)$
&$(\epsilon^{3},\epsilon,1)$\\
$\bm{{\widehat2}}^{\prime}\oplus\bm{1}$
&$\bm{2}\oplus\bm{1}^{\prime}$
&$(1,1,1)$&$(1,1,1)$&$(1,1,1)$
&$(\epsilon^{5},\epsilon^{3},1)$\\
$\bm{{\widehat2}}^{\prime}\oplus\bm{1}^{\prime}$
&$\bm{2}\oplus\bm{1}^{\prime}$
&$(1,1,1)$&$(1,1,1)$&$(1,1,1)$
&$(\epsilon^{3},\epsilon,1)$\\
$\bm{{\widehat2}}^{\prime}\oplus\bm{1}$
&$\bm{1}\oplus\bm{1}\oplus\bm{1}$
&$(\epsilon^{2},\epsilon,1)$
&$(\epsilon^{2},\epsilon,1)$&$(\epsilon^{2},\epsilon,1)$
&$(\epsilon^{7},\epsilon,1)$\\
$\bm{{\widehat2}}^{\prime}\oplus\bm{1}$
&$\bm{1}\oplus\bm{1}\oplus\bm{1}^{\prime}$
&$(\epsilon^{2},\epsilon,1)$
&$(\epsilon^{2},\epsilon,1)$&$(\epsilon^{2},\epsilon,1)$
&$(\epsilon^{3},\epsilon,1)$\\
$\bm{{\widehat2}}^{\prime}\oplus\bm{1}$
&$\bm{1}\oplus\bm{1}^{\prime}\oplus\bm{1}^{\prime}$
&$(\epsilon^{2},\epsilon,1)$
&$(\epsilon^{2},\epsilon,1)$&$(\epsilon^{2},\epsilon,1)$
&$(\epsilon^{5},\epsilon^{3},1)$\\
$\bm{{\widehat2}}^{\prime}\oplus\bm{1}$
&$\bm{1}^{\prime}\oplus\bm{1}^{\prime}\oplus\bm{1}^{\prime}$
&$(\epsilon^{2},\epsilon,1)$
&$(\epsilon^{2},\epsilon,1)$&$(\epsilon^{2},\epsilon,1)$
&$(\epsilon^{5},\epsilon^{4},\epsilon^{3})$\\
$\bm{{\widehat2}}^{\prime}\oplus\bm{1}^{\prime}$
&$\bm{1}\oplus\bm{1}\oplus\bm{1}$
&$(\epsilon^{2},\epsilon,1)$
&$(\epsilon^{2},\epsilon,1)$&$(\epsilon^{2},\epsilon,1)$
&$(\epsilon^{7},\epsilon^{4},\epsilon)$\\
$\bm{{\widehat2}}^{\prime}\oplus\bm{1}^{\prime}$
&$\bm{1}\oplus\bm{1}\oplus\bm{1}^{\prime}$
&$(\epsilon^{2},\epsilon,1)$
&$(\epsilon^{2},\epsilon,1)$&$(\epsilon^{2},\epsilon,1)$
&$(\epsilon^{7},\epsilon,1)$\\
$\bm{{\widehat2}}^{\prime}\oplus\bm{1}^{\prime}$
&$\bm{1}\oplus\bm{1}^{\prime}\oplus\bm{1}^{\prime}$
&$(\epsilon^{2},\epsilon,1)$
&$(\epsilon^{2},\epsilon,1)$&$(\epsilon^{2},\epsilon,1)$
&$(\epsilon^{3},\epsilon,1)$\\
$\bm{{\widehat2}}^{\prime}\oplus\bm{1}^{\prime}$
&$\bm{1}^{\prime}\oplus\bm{1}^{\prime}\oplus\bm{1}^{\prime}$
&$(\epsilon^{2},\epsilon,1)$&$(\epsilon^{2},\epsilon,1)$
&$(\epsilon^{2},\epsilon,1)$
&$(\epsilon^{5},\epsilon^{3},1)$\\
$\bm{{\widehat2}}\oplus\bm{1}$
&$\bm{{\widehat2}}\oplus\bm{1}$
&$(1,1,1)$&$(1,1,1)$&$(1,1,1)$
&$(1,1,1)$\\
$\bm{{\widehat2}}\oplus\bm{1}$
&$\bm{{\widehat2}}\oplus\bm{1}^{\prime}$
&$(1,1,1)$&$(1,1,1)$&$(1,1,1)$
&$(\epsilon^{4},1,1)$\\
$\bm{{\widehat2}}\oplus\bm{1}^{\prime}$
&$\bm{{\widehat2}}\oplus\bm{1}^{\prime}$
&$(1,1,1)$&$(1,1,1)$&$(1,1,1)$
&$(1,1,1)$\\
$\bm{{\widehat2}}\oplus\bm{1}$
&$\bm{{\widehat2}}^{\prime}\oplus\bm{1}$
&$(1,1,1)$&$(1,1,1)$&$(1,1,1)$
&$(\epsilon^{6},\epsilon^{2},1)$\\
$\bm{{\widehat2}}\oplus\bm{1}$
&$\bm{{\widehat2}}^{\prime}\oplus\bm{1}^{\prime}$
&$(1,1,1)$&$(1,1,1)$&$(1,1,1)$
&$(\epsilon^{2},\epsilon,\epsilon)$\\
$\bm{{\widehat2}}^{\prime}\oplus\bm{1}$
&$\bm{{\widehat2}}^{\prime}\oplus\bm{1}$
&$(1,1,1)$&$(1,1,1)$&$(1,1,1)$
&$(1,1,1)$\\
$\bm{{\widehat2}}^{\prime}\oplus\bm{1}$
&$\bm{{\widehat2}}^{\prime}\oplus\bm{1}^{\prime}$
&$(1,1,1)$&$(1,1,1)$&$(1,1,1)$
&$(\epsilon^{4},1,1)$\\
$\bm{{\widehat2}}^{\prime}\oplus\bm{1}^{\prime}$
&$\bm{{\widehat2}}^{\prime}\oplus\bm{1}^{\prime}$
&$(1,1,1)$&$(1,1,1)$&$(1,1,1)$
&$(1,1,1)$\\
\midrule
\specialrule{0em}{1.0pt}{1.0pt}
\midrule
\end{longtable}
}
\end{center}

\begin{table}[hptb!]
\centering
\renewcommand{\arraystretch}{1.5}
\begin{tabular}{|c|c|c|}
\hline\hline
Fixed point&Pattern&Viable $\rho^c\otimes \rho$\\
\hline
$\tau\simeq\omega$ & $(\epsilon^{2}, \epsilon, 1)$
& $[\bm{1}^{(\prime)}\oplus\bm{1}^{(\prime)}
\oplus\bm{1}^{(\prime)}]
\otimes[\bm{3}^{(\prime)}$ or $\bm{{\widehat 2}}^{(\prime)}\oplus\bm{1}^{(\prime)}$ or $\bm{2}\oplus\bm{1}^{(\prime)}$]\\ \hline
$\tau\simeq i\infty$ & $(\epsilon^{5}, \epsilon^{4}, \epsilon^{3})$
 & $[\bm{1}^{\prime}\oplus\bm{1}^{\prime}
\oplus\bm{1}^{\prime}] \otimes[\bm{{\widehat 2}}^{\prime}\oplus\bm{1}]$,
$[\bm{1}\oplus\bm{1}\oplus\bm{1}]
\otimes[\bm{{\widehat 2}}\oplus\bm{1}^{\prime}]$\\ \hline

$\tau\simeq i\infty$ & $(\epsilon^{6}, \epsilon^{4}, \epsilon^{2})$
& $[\bm{1}^{\prime}\oplus\bm{1}^{\prime}
\oplus\bm{1}^{\prime}]\otimes\bm{3}$, $[\bm{1}\oplus\bm{1}\oplus\bm{1}]
\otimes\bm{3}^{\prime}$\\\hline

\multirow{2}{*}{$\tau\simeq i\infty$} & \multirow{2}{*}{$(\epsilon^{7}, \epsilon^{4}, \epsilon)$} & $[\bm{1}^{\prime}\oplus\bm{{\widehat 2}}]
\otimes\bm{3}$, $[\bm{{\widehat 2}}^{\prime}\oplus\bm{1}]
\otimes\bm{3}^{\prime}$,\\
 & & $[\bm{1}^{\prime}\oplus\bm{1}^{\prime}
 \oplus\bm{1}^{\prime}]
\otimes[\bm{{\widehat 2}}\oplus\bm{1}]$,
 $[\bm{1}\oplus\bm{1}\oplus\bm{1}]
\otimes[\bm{{\widehat 2}}^{\prime}
\oplus\bm{1}^{\prime}]$\\ \hline

\multirow{2}{*}{$\tau\simeq i\infty$} & \multirow{2}{*}{$(\epsilon^{6}, \epsilon^{2}, 1)$} & [$\bm{1}\oplus\bm{1}\oplus\bm{1}$ or $\bm{1}\oplus\bm{1}^{\prime}\oplus\bm{1}^{\prime}$ or $\bm{2}\oplus\bm{1}^{\prime}$]
$\otimes\bm{3}$, \\

&& [$\bm{1}\oplus\bm{1}\oplus\bm{1}^{\prime}$ or $\bm{1}^{\prime}\oplus\bm{1}^{\prime}\oplus\bm{1}^{\prime}$
 or $\bm{2}\oplus\bm{1}$]
$\otimes\bm{3}^{\prime}$,
$[\bm{{\widehat 2}}^{\prime}\oplus\bm{1}]\otimes[\bm{{\widehat 2}}\oplus\bm{1}]$\\ \hline

\multirow{2}{*}{$\tau\simeq i\infty$} & \multirow{2}{*}{$(\epsilon^{3}, \epsilon, 1)$} & $[\bm{2}\oplus\bm{1}]\otimes[\bm{{\widehat 2}}^{(\prime)}\oplus\bm{1}]$, $[\bm{2}\oplus\bm{1}^{\prime}]\otimes$[$\bm{{\widehat 2}}\oplus\bm{1}^{(\prime)}$
 or $\bm{{\widehat 2}}^{\prime}\oplus\bm{1}^{\prime}$] \\

&&$[\bm{1}\oplus\bm{1}\oplus\bm{1}^{\prime}] \otimes[\bm{{\widehat 2}}^{(\prime)}\oplus\bm{1}]$, $[\bm{1}\oplus\bm{1}^{\prime}\oplus\bm{1}^{\prime}]
\otimes [\bm{{\widehat2}}^{(\prime)}\oplus\bm{1}^{\prime}]$ \\
\hline\hline
\end{tabular}
\caption{\label{tab:heirarchy-summary}
The promising hierarchical mass patterns which could be realized due to the proximity of the modulus $\tau$ to the fixed points in the modular symmetry $2O$. These patterns are unchanged if one permute the modular transformations $\rho$ and $\rho^c$.  }
\end{table}

\subsection{Model realization}

Inspired by the general analysis in above, we shall present an example model in which the mass hierarchies of charged lepton arise from small departure of modulus $\tau$ from the fixed point $\tau_0=i\infty$. The representation assignment is $\rho=\bm{{\widehat 2}}'\oplus\bm{1}$ and $\rho^c=\bm{3}^{\prime}$, and the fermion mass pattern is predicted to be $(m_1, m_2, m_3)\sim (\epsilon^7, \epsilon^4, \epsilon)$, as can be seen from table~\ref{tab:heirarchy-summary}. Hence the three charged lepton masses are vanishing and lepton mixing matrix can not be fixed in the symmetric limit of $\tau=\tau_0$. In this model, the light neutrino masses are generated through the type-I seesaw mechanism with only two RH neutrinos. The Higgs fields $H_u$ and $H_d$ are trivial singlets of modular symmetry group $2O$ and their modular weights are vanishing. To be more specific, the modular transformations of the lepton fields are
\begin{eqnarray}
\nonumber
&&L_{D}\sim \mathbf{\widehat{2}'}\,,~~L_{3}\sim \mathbf{1}\,,~~E^{c}\sim \mathbf{3'}\,,~~N^{c}\sim \mathbf{2}\,,\\
&&k_{L_{D}}=0\,,~~k_{L_{3}}=3\,,~~k_{E^{c}}=5\,,~~k_{N^{c}}=3\,.
\end{eqnarray}
Thus the superpotential for the lepton masses read as
\begin{eqnarray}
\nonumber\mathcal{W}_{E}&=&\alpha_{1} \left( E^{c}L_{D}\right)_{\mathbf{\widehat{2}}}Y^{(5)}_{\mathbf{\widehat{2}}}H_{d}+\alpha_{2} \left( E^{c}L_{D}\right)_{\widehat{\mathbf{4}}}Y^{(5)}_{\widehat{\mathbf{4}}}H_{d}+\alpha_{3} \left( E^{c}L_{3}\right)_{\mathbf{3'}}Y^{(8)}_{\mathbf{3'}I}H_{d} + \alpha_{4} \left( E^{c}L_{3}\right)_{\mathbf{3'}}Y^{(8)}_{\mathbf{3'}II}H_{d}\,,\\
\mathcal{W}_{\nu}&=&\beta_{1}H_{u}(N^{c}L_{D})_{\mathbf{4}}Y^{(3)}_{\mathbf{4}}+\beta_{2}H_{u}(N^{c}L_{3})_{\mathbf{2}}Y^{(6)}_{\mathbf{2}}+ \Lambda  \left(N^{c}N^{c}\right)_{\mathbf{1}}Y^{(6)}_{\mathbf{1}}+ \beta_3\Lambda  \left(N^{c}N^{c}\right)_{\mathbf{2}}Y^{(6)}_{\mathbf{2}}\,,
\end{eqnarray}
which yield the following charged lepton and neutrino mass matrices
\begin{eqnarray}
\nonumber M_{E}&=&\left(
\begin{array}{ccc}
\sqrt{2} \alpha_{2} Y_{\widehat{\mathbf{4}},2}^{(5)}+\alpha_{1}Y_{\mathbf{\widehat{2}},1}^{(5)} ~&~ \sqrt{2} \alpha_{2} Y_{\widehat{\mathbf{4}},1}^{(5)}-\alpha_{1}Y_{\mathbf{\widehat{2}},2}^{(5)} ~&~ \alpha_{3} Y_{\mathbf{3'}I,1}^{(8)}+\alpha_{4} Y_{\mathbf{3'}II,1}^{(8)} \\
-\sqrt{3} \alpha_{2} Y_{\widehat{\mathbf{4}},4}^{(5)} ~&~ \sqrt{2} \alpha_{1}Y_{\mathbf{\widehat{2}},1}^{(5)}-\alpha_{2} Y_{\widehat{\mathbf{4}},2}^{(5)} ~&~ \alpha_{3} Y_{\mathbf{3'}I,3}^{(8)}+\alpha_{4} Y_{\mathbf{3'}II,3}^{(8)} \\
\alpha_{2} Y_{\widehat{\mathbf{4}},1}^{(5)}+\sqrt{2} \alpha_{1}Y_{\mathbf{\widehat{2}},2}^{(5)} ~&~ -\sqrt{3} \alpha_{2} Y_{\widehat{\mathbf{4}},3}^{(5)} ~&~ \alpha_{3} Y_{\mathbf{3'}I,2}^{(8)}+\alpha_{4} Y_{\mathbf{3'}II,2}^{(8)}
\end{array}
\right)v_{d}\,,\\
\nonumber M_D &=& \left(
\begin{array}{ccc}
\beta_{1}Y_{\widehat{\mathbf{4}},3}^{(3)} ~&~ \beta_{1}Y_{\widehat{\mathbf{4}},4}^{(3)} ~&~ \beta_{2} Y_{\mathbf{2},1}^{(6)} \\
-\beta_{1}Y_{\widehat{\mathbf{4}},2}^{(3)} ~&~ \beta_{1}Y_{\widehat{\mathbf{4}},1}^{(3)} ~&~ \beta_{2} Y_{\mathbf{2},2}^{(6)}
\end{array}
\right)v_{u} \,,\\
  M_N &=&\left(
\begin{array}{cc}
Y_{\mathbf{1}}^{(6)}-\beta_{3} Y_{\mathbf{2},1}^{(6)} & \beta_{3} Y_{\mathbf{2},2}^{(6)} \\
 \beta_{3} Y_{\mathbf{2},2}^{(6)} & Y_{\mathbf{1}}^{(6)}+\beta_{3} Y_{\mathbf{2},1}^{(6)}
\end{array}
\right) \Lambda\,.
\end{eqnarray}
We impose gCP symmetry on this model so that all couplings are real.

For the complex modulus $\tau$ with large imaginary part, the small expansion parameter is $q_8=e^{\pi i\tau/4}$. Using the $q$-expansion of the modular forms given in section~\ref{sec:VVMF}, we find that the charged lepton and neutrino mass matrices are approximately given by
\begin{eqnarray}
\nonumber M_{E}&\simeq&\left(
\begin{array}{ccc}
 \left(\alpha_{1}-2 \sqrt{6} \alpha_{2}\right) q_8^3 ~&~ -\left(2\alpha_{1}+12 \sqrt{6} \alpha_{2}\right)q_8^5 ~&~  2 \sqrt{2} \alpha_{3} q_8^4 \\
 -\sqrt{3} \alpha_{2} q_8 ~&~ \left(\sqrt{2}\alpha_{1}+2 \sqrt{3} \alpha_{2} \right)q_8^3 ~&~ -\left(\alpha_{3}+\alpha_{4}\right) q_8^2 \\
 \left(2 \sqrt{2} \alpha_{1}-12 \sqrt{3} \alpha_{2}\right) q_8^5 ~&~ -40 \sqrt{3} \alpha_{2} q_8^7 ~&~ \left(4 \alpha_{3} +20 \alpha_{4}\right) q_8^6 \\
\end{array}
\right)v_{d}\,,\\
\nonumber M_D &\simeq& \left(
\begin{array}{ccc}
 -8 \beta_{1}q_8^7 ~&~ \beta_{1}q_8 ~&~ \beta_{2} \\
 -2 \sqrt{3} \beta_{1}q_8^3 ~&~ 4 \sqrt{3} \beta_{1}q_8^5 ~&~ 8 \sqrt{3} \beta_{2} q_8^4 \\
\end{array}
\right)v_{u} \,,\\
  M_N &\simeq&\left(
\begin{array}{cc}
 1-\beta_{3} ~&~ 8 \sqrt{3} \beta_{3} q_8^4 \\
 8 \sqrt{3} \beta_{3} q_8^4 ~&~ 1+\beta_{3}
\end{array}
\right) \Lambda\,.
\end{eqnarray}

Then we can read off the approximate expressions of the charged lepton mass ratios as follow
\begin{eqnarray}
\nonumber &&\frac{m_{e}}{m_{\mu}}\simeq
\frac{24\sqrt{6}\,\left|\alpha_2\alpha_4(\alpha^2_1-2\sqrt{6}\alpha_1\alpha_2-48\alpha^2_2)\right|}
 {[\alpha_{1}\left(\alpha_{3}+\alpha_{4}\right)
-2\sqrt{6}\alpha_{2}\left(2\alpha_{3}+\alpha_{4}\right)]^{2}}\;|q_{8}|^{3},\\
 &&\frac{m_{\mu}}{m_{\tau}}\simeq
\frac{|\alpha_{1}\left(\alpha_{3}+\alpha_{4}\right)
-2\sqrt{6}\alpha_{2}\left(2\alpha_{3}+\alpha_{4}\right)|}{3|\alpha_{2}|^{2}}\;|q_{8}|^{3}\,.
\end{eqnarray}
This is in agreement with the results of the general analysis summarized in table~\ref{tab:heirarchy-summary}. The light neutrino mass matrix is given by the seesaw formula
\begin{eqnarray}
  M_{\nu}=-M_{D}^{T}M_{N}^{-1}M_{D}
  \simeq\left(\begin{array}{ccc}
 -\frac{12\beta_{1}^{2}q_{8}^{6}}{1+\beta_{3}}
 ~&~
 \frac{32\beta_{1}^{2}q_{8}^{8}}{\beta_{3}+1}
  ~&~
  \frac{8\beta_{1}\beta_{2}(11\beta_{3}-7)q_{8}^{7}}
  {\beta_{3}^{2}-1}
   \\
 \frac{32\beta_{1}^{2}q_{8}^{8}}{\beta_{3}+1}
 ~&~
\frac{\beta_{1}^{2}q_{8}^{2}}{\beta_{3}-1}
 ~&~
 \frac{\beta_{1}\beta_{2}q_{8}}{\beta_{3}-1}
   \\
 \frac{8\beta_{1}\beta_{2}(11\beta_{3}-7)q_{8}^{7}}
  {\beta_{3}^{2}-1}
 ~&~
 \frac{\beta_{1}\beta_{2}q_{8}}{\beta_{3}-1}
 ~&~
 \frac{\beta_{2}^{2}}{\beta_{3}-1}
\end{array}
\right) \frac{v_{u}^{2}}{\Lambda}\,.
\end{eqnarray}
It follows that the neutrino masses are
\begin{equation}
m_{1}=0,~~m_{2}=\frac{12|\beta_{1}|^{2}|q_{8}|^{3}}
  {|1+\beta_{3}|}\frac{v_{u}^{2}}{\Lambda},~~m_{3}=\frac{|\beta_{2}|^{2}}
  {1-|\beta_{3}|}\frac{v_{u}^{2}}{\Lambda}\,.
\end{equation}
Obviously the neutrino mass spectrum is normal ordering.

We perform a numerical analysis, and the best fit values of the free parameters are determined to be
\begin{eqnarray}
\label{eq:bf_lepton}
\nonumber&&\braket{\tau}=0.29247 + 2.31919i\,,~~ \alpha_{2}/\alpha_{1}=-0.18349\,,~~ \alpha_{3}/\alpha_{1}=-1.29357\,,~~ \alpha_{4}/\alpha_{1}=4.18393\,,\\
\nonumber&&\beta_{2}/\beta_{1}=0.24374\,,~~ \beta_{3}=-1.00101\,,~~ \alpha_{1}v_{d}=13.21417~\text{MeV}\,,~~ \frac{\left(\beta_{1}v_{u}\right)^{2}}{\Lambda}=0.21443~\text{eV}\,.\\
\end{eqnarray}
We see that the coupling constants are of the same order of magnitudes and the imaginary part of the modulus $\tau$ is approximately 2.319 which is large and the departure parameter $\epsilon=e^{-\pi\,\text{Im}\tau/4}\approx0.161$. The lepton masses and mixing parameters at the best-fit point are given by
\begin{eqnarray}
\nonumber&& \sin^{2}\theta^{l}_{12}=0.3030\,,\quad \sin^{2}\theta^{l}_{13}=0.02225\,,\quad \sin^{2}\theta^{l}_{23}=0.4509\,,\quad \delta_{CP}^{l}=189.6^{\circ}\,,\\
\nonumber&& \alpha_{21}=0.1225\pi\,,\quad \alpha_{31}=0.6699\pi\,,\quad m_e/m_{\mu}=0.004800\,,\quad m_{\mu}/m_{\tau}=0.05900\,,\\
&&m_1=0~\text{meV}\,,\quad m_2=8.61~\text{meV}\,,\quad m_3=50.06~\text{meV}\,,
\end{eqnarray}
with $\chi^{2}=5.1$.

\end{appendix}


\providecommand{\href}[2]{#2}\begingroup\raggedright\endgroup

\end{document}